\documentclass[]{aastex631}

\usepackage{graphicx}
\usepackage{float}
\usepackage[caption=false]{subfig}
\usepackage{enumitem}
\usepackage{amsmath}
\usepackage{comment}
\usepackage{makecell}
\usepackage[english]{babel}
\usepackage[T1]{fontenc}
\usepackage[utf8]{inputenc}

\newcommand{\degree}{^\circ}
\newcommand{\naone}{Na~D$_1$}
\newcommand{\natwo }{Na~D$_2$}
\newcommand{\planetMetallicity}{6$\times$ to 14$\times$}
\newcommand{\planetCtoO}{0.65 to 0.94}
\newcommand{\planetMetallicityScattering}{4$\times$ to 8$\times$}
\newcommand{\planetCtoOScattering}{0.26 to 0.58}
\newcommand{\redChisq}{$\tilde{\chi}^2$}
\newcommand{\planetGAlbedo}{0.64}

\defcitealias{lustig-yaeger2023lhs47b}{Lustig-Yaeger \& Fu et al. (2023)}
\defcitealias{moran2023gj486bwater}{Moran \& Stevenson et al. (2023)}
\accepted{May 31, 2024}

\shorttitle{A JWST Emission Spectrum of WASP-69 b Indicates Aerosols}
\shortauthors{Schlawin et al.}

\begin{document}

\title{Multiple Clues for Dayside Aerosols and Temperature Gradients in WASP-69 b from a Panchromatic JWST Emission Spectrum}

\correspondingauthor{Everett Schlawin}
\email{eas342 AT EMAIL Dot Arizona .edu}

\author[0000-0001-8291-6490]{Everett Schlawin}
\affiliation{Steward Observatory, 933 North Cherry Avenue, Tucson, AZ 85721, USA}

\author[0000-0003-1622-1302]{Sagnick Mukherjee}
\affiliation{Department of Astronomy and Astrophysics, University of California, Santa Cruz, CA, USA}

\author[0000-0003-3290-6758]{Kazumasa Ohno}
\affiliation{Division of Science, National Astronomical Observatory of Japan, Tokyo, Japan}
\affiliation{Department of Astronomy and Astrophysics, University of California, Santa Cruz, CA, USA}

\author[0000-0003-4177-2149]{Taylor Bell}
\affiliation{Bay Area Environmental Research Institute, NASA's Ames Research Center, Moffett Field, CA 94035, USA}
\affiliation{Space Science and Astrobiology Division, NASA's Ames Research Center, Moffett Field, CA 94035, USA}

\author[0000-0002-9539-4203]{Thomas G. Beatty}
\affiliation{Department of Astronomy, University of Wisconsin--Madison, Madison, WI 53703, USA}

\author[0000-0002-8963-8056]{Thomas P. Greene}
\affiliation{Space Science and Astrobiology Division, NASA’s Ames Research Center, Moffett Field, CA, USA}

\author[0000-0001-6247-8323]{Michael Line}
\affiliation{School of Earth and Space Exploration, Arizona State University, Tempe, AZ, USA}

\author[0000-0002-8211-6538]{Ryan C. Challener}
\affiliation{Department of Astronomy, Cornell University, 122 Sciences Drive, Ithaca, NY 14853, USA}

\author[0000-0001-9521-6258]{Vivien Parmentier}
\affiliation{Université de la Côte d’Azur, Observatoire de la Côte d’Azur, CNRS, Laboratoire Lagrange, France}

\author[0000-0002-9843-4354]{Jonathan J. Fortney}
\affiliation{Department of Astronomy and Astrophysics, University of California, Santa Cruz, CA, USA}

\author[0000-0003-3963-9672]{Emily Rauscher}
\affiliation{Department of Astronomy, University of Michigan, 1085 S. University Ave., Ann Arbor, MI 48109, USA}

\author[0000-0002-3295-1279]{Lindsey Wiser}
\affiliation{School of Earth and Space Exploration, Arizona State University, Tempe, AZ, USA}

\author[0000-0003-0156-4564]{Luis Welbanks}
\affiliation{School of Earth and Space Exploration, Arizona State University, Tempe, AZ, USA}

\author[0000-0002-8517-8857]{Matthew Murphy}
\affiliation{Steward Observatory, 933 North Cherry Avenue, Tucson, AZ 85721, USA}

\author[0000-0001-8745-2613]{Isaac Edelman}
\affiliation{Bay Area Environmental Research Institute, NASA's Ames Research Center, Moffett Field, CA 94035, USA}

\author[0000-0003-1240-6844]{Natasha Batalha}
\affiliation{Space Science and Astrobiology Division, NASA’s Ames Research Center, Moffett Field, CA, USA}

\author[0000-0002-6721-3284]{Sarah E. Moran}
\affiliation{Lunar and Planetary Laboratory, University of Arizona, Tucson, AZ 85721, USA}

 \author[0000-0001-6086-4175]{Nishil Mehta}
 \affiliation{Université de la Côte d’Azur, Observatoire de la Côte d’Azur, CNRS, Laboratoire Lagrange, France}

\author[0000-0002-7893-6170]{Marcia Rieke}
\affiliation{Steward Observatory, 933 North Cherry Avenue, Tucson, AZ 85721, USA}



\begin{abstract}

WASP-69 b is a hot, inflated, Saturn-mass planet (0.26 M$_{Jup}$) with a zero-albedo equilibrium temperature of 963 K.
Here, we report the JWST 2 to 12 \micron\ emission spectrum of the planet consisting of two eclipses observed with NIRCam grism time series and one eclipse observed with MIRI LRS.
The emission spectrum shows absorption features of water vapor, carbon dioxide and carbon monoxide, but no strong evidence for methane.
WASP-69 b's emission spectrum is poorly fit by cloud-free homogeneous models.
We find three possible model scenarios for the planet: 1) a Scattering Model that raises the brightness at short wavelengths with a free Geometric Albedo parameter 2) a Cloud Layer model that includes high altitude silicate aerosols to moderate long wavelength emission and 3) a Two-Region model that includes significant dayside inhomogeneity and cloud opacity with two different temperature-pressure profiles.
In all cases, aerosols are needed to fit the spectrum of the planet.
The Scattering model requires an unexpectedly high Geometric Albedo of \planetGAlbedo. 
 Our atmospheric retrievals indicate inefficient redistribution of heat and an inhomogeneous dayside distribution, which is tentatively supported by MIRI LRS broadband eclipse maps that show a central concentration of brightness.
Our more plausible models (2 and 3) retrieve chemical abundances enriched in heavy elements relative to solar composition by \planetMetallicity\ solar and a C/O ratio of \planetCtoO, whereas the less plausible highly reflective scenario (1) retrieves a slightly lower metallicity and lower C/O ratio.
\end{abstract}




\section{Introduction} \label{sec:intro}

\subsection{The Value of Spectra of Moderately Irradiated Hot Jupiters}
JWST is opening up a new window into giant irradiated planet atmospheres to reveal their chemical, physical and dynamical processes \citep[e.g.][]{ahrer2022WASP39bERS,alderson2022wasp395bG395H,rustamkulov2022nirspecPrism,feinstein2023wasp39bNIRISS,bean2023hd149026b,grant2023quartzWASP17b,xue2023hd209458b}.
Emission spectra and phase curves are particularly valuable for understanding the temperature structure, composition and 3D effects in the atmospheres of hot planets \citep{bell2023_ers,coulombe2023wasp18EmissionSpec}.
Most JWST studies thus far have probed either cooler, smaller planets (GJ 1214 b T$_{eq}$=600 K, \citet{kempton2023reflectiveMetalRichGJ1214,gao2023gj1214};  LHS 475 b T$_{eq}$= 590 K, \citetalias{lustig-yaeger2023lhs47b}; GJ 486 b T$_{eq}$=700 K, \citetalias{moran2023gj486bwater}) or more highly irradiated Jupiters (WASP-39 b T$_{eq}$=1170~K \citep{ers2023wasp39b_CO2}, WASP-43 b T$_{eq}$ =1400~K \citep{bell2023_ers}, HD 149026 b T$_{eq}$=1700 K \citep{bean2023hd149026b}, HD 209458 b T$_{eq}$=1450~K \citep{xue2023hd209458b}), WASP-77 A b T$_{eq}$=1700~K \citep{august2023subsolarMetallicityWASP77Ab}.
However, the study of giant planets in the 800-1000 K temperature range is potentially a valuable window in the chemical equilibrium of the methane (CH$_4$) and ammonia (NH$_3$) molecules in exoplanet atmospheres, due to both photochemistry and mixing from deeper layers \citep{fortney2020beyondTeq}.

Methane and ammonia are expected to be the dominant Carbon-bearing and Nitrogen-bearing molecules below 900 K to 1000 K for hot Jupiter atmospheres observed at 0.1 bar \citep{moses13} so their abundances are critical for evaluating the level of Carbon and Nitrogen in exoplanet atmospheres.
Recently methane has been definitively detected for the first time via low resolution space-based spectroscopy with JWST, which opens up this molecule as a tracer of dynamical and chemical processes in exoplanets.
Methane was found in the moderately-irradiated Jupiter, WASP-80 b ($T_{eq}=825 K$), \citep{bell2023_methane} as well as the lower temperature sub-Neptunes K2-18 b T$_{eq}=270 K$ \citep{madhusudhan2023k2d18} and TOI-270 d T$_{eq}= 354 $K \citep{benneke2024toi270dMiscibleMetalRich,holmberg2024toi270d}.
The moderately irradiated Jupiter HAT-P-18 b (T$_{eq}$=850 K) may also have methane, as found by retrieval modeling, but there are no strong, high signal-to-noise absorption features in the SOSS bandpass from 0.6~\micron\ to 2.8~\micron\ \citep{fu2022hatp18bJWST} and a subsequent re-analysis did not find evidence for methane in HAT-P-18 b with the same data \citep{fournier-tondreau2023NIRtransHATP18b}.
Additional spectra of irradiated giant planets in the 800-1000 K temperature regime will be valuable probes of the atmospheric physics.
This temperature regime efficiently covers the  transition between CH$_4$-dominated versus CO-dominated reservoirs of carbon in planet atmospheres.
The presence of CH$_4$ in planets below $\sim$950~K is also predicted to supply the building blocks for photochemical hazes in planets that can explain some of the muted atmospheric features below 950 K as observed by HST transmission spectra \citep{gao2020aerosolsSilicatesAndHazes}.

\subsection{WASP-69 b}
WASP-69 b is an inflated Saturn mass transiting planet (0.26 $M_{Jup}$, 1.06 $R_{Jup}$) orbiting a K-type star (4715 K, 0.83 M$_\odot$, 0.81 R$_\odot$, Age $\sim$ 2 Gyr), with a zero-albedo full redistribution equilibrium temperature of 963 K \citep{anderson2014wasp69wasp84wasp70Ab}.
The host star shows evidence for metal enrichment with metallicities reported from [Fe/H]= 0.15 $\pm$ 0.08 \citep{anderson2014wasp69wasp84wasp70Ab} to 0.29 $\pm$ 0.04 \citep{sousa2021sweetcat}.
The equilibrium temperature regime of WASP-69 b is where CH$_4$ can potentially become a dominant carbon-bearing molecule in chemical equilibrium \citep{moses13}, but the cooling history of a planet and vertical mixing can potentially alter the composition of the planet from equilibrium expectations \citep{fortney2020beyondTeq}.

Atoms and molecules have been previously detected in WASP-69 b's atmosphere and give initial insights into its composition and aerosols.
Searches for Na in the atmosphere of WASP-69 b with high resolution spectroscopy have yielded a range of results, but most of the studies find that Na  is significantly detected in the \natwo\ line but less significantly detected in the \naone\ line, potentially from obscuration by atmospheric aerosols \citep{casasayas-barris2017sodiumWASP69b,deibert2019highResSpec,langeveld2022surveyNaHighRes,khalafinejad2021wasp69b_lowResHighRes}.\footnote{
Specifically, \citet{casasayas-barris2017sodiumWASP69b} report that the \natwo\ line was detected at a 5$\sigma$, while the \naone\ was not detected.
\citet{casasayas-barris2017sodiumWASP69b} find that the \natwo\ has an excess transit depth of 0.53 $\pm$0.14\%\ for a a 1.5~\AA\ bandpass  and a line contrast of 5.8 $\pm$ 0.3\% from a Gaussian fit the transit profile.
The slightly different oscillator strengths of the two lines makes it possible for \naone\ to be more obscured \citep{huitson2012sodiumHD189}.
\citet{deibert2019highResSpec}, by contrast, do not find significant absorption by WASP-69 b with the GRACES instrument for either the \natwo\ or \naone\ lines.
\citet{langeveld2022surveyNaHighRes} find that WASP-69 b has a significant Na line contrast of 3.28 $\pm$ 0.70\% for the \natwo\ line and 1.26 $\pm$ 0.61 \% for the \naone\ line. 
Muted but significant detections of both the \natwo\ and \naone\ lines were detected by \citet{khalafinejad2021wasp69b_lowResHighRes}.
The ratio of the two lines \natwo / \naone\ of 2.5 $\pm$ 0.7 suggests that there is aerosol opacity that muted the \naone\ line  \citep{khalafinejad2021wasp69b_lowResHighRes}.
Thus, the majority of Na line absorption studies at high spectral resolution mostly suggest an atmosphere with significant sodium but the \naone\ line is at least partially obscured by aerosols as compared to \natwo.
WASP-69 b's Na detections have also been theorized to be potentially influenced by a volcanically active moon \citep{oza2019sodiumPotassiumVolanicSatellites}; so far no moons have been detected \citep{narang2023wasp69bnoRadioLoudMoon}.}

Other atmospheric features of WASP-69 b found with transmission spectroscopy include its scattering slope, tentative evidence for TiO, Carbon-bearing and Oxygen-bearing molecules and an escaping outflow, all described below.
\citet{murgas2020rayleighScatteringWASP69b} find a spectral slope in the GTC OSIRIS optical transmission spectrum that is consistent with Rayleigh scattering from hydrogen.
However, the spectral slope is also consistent with stellar activity and the transit light source effect \citep[e.g.][]{rackham2018transitSourceEffect}.
\citet{estrela2021aerosolsMicrobarWASP69b} find an even stronger spectral slope than \citet{murgas2020rayleighScatteringWASP69b} with HST's Space Telescope Imaging Spectrograph (STIS).
The large transit depths down to $\sim$0.4~\micron\ are best explained by aerosols at high altitudes ($\mu$bar pressure levels) \citep{estrela2021aerosolsMicrobarWASP69b}.
There is further evidence of Raleigh scattering in ground-based spectroscopy from 0.5 to 0.9~\micron\ with tentative evidence of TiO in the spectrum \citep{ouyang2023tentativeTiOwasp69b}.
Extended Helium has also been detected in WASP-69 b with a signal-to-noise ratio of 18 that is blue-shifted by several km/s \citep{nortmann2018heliumWASP69b}. This was confirmed with narrowband photometry, constraining the mass loss rate to be 10$^{-4}$ to 10$^{-3}$ M$_{Jup}$/Gyr \citep{vissapragada2020heliumWASP69bWASP52b}.
The envelope escaping WASP-69 b extends several radii and is confined to a tail like structure that extends up to 7.5 planetary radii behind the planet's direction of motion \citep{tyler2023escapingEnvelopeWASP69b}.
However, the inferred mass loss rate from the observed escaping tail ($\sim 10^{-3}$ M$_{Jup}$/Gyr) is not a significant fraction of the overall mass (0.26 M$_{Jup}$) to measurably alter the deeper atmosphere.

In transit transmission, 5 molecules were detected with high resolution cross correlation \citep{guilluy2022fiveMoleculesWASP69b}: CO, CH$_4$, NH$_3$, H$_2$O and C$_2$H$_2$.
In the same analysis, the detection of H$_2$O and CH$_4$ and the absence of CO$_2$ disfavored models larger than 10$\times$ solar enrichment of heavy elements.
If the atmosphere is assumed to be solar composition, the  presence of the 5 detected molecules favors models that have high C/O ratios, perhaps larger than 1.0.
The presence of C$_2$H$_2$ in the terminator also implies some disequilibrium processes.

\citet{baxter2020transitionHotUltrahot} and \citet{wallack2019atmCompositionCoolGasGiants} used secondary eclipse measurements at the Spitzer 3.6~\micron\ and 4.5~\micron\ bandpasses to constrain the composition and temperature profile of WASP-69~b.
\citet{baxter2020transitionHotUltrahot} found that WASP-69~b and other planets  below $\lesssim 1660 $K lack temperature inversions, ie. their gas temperatures decrease as a function of altitude.
The spectral slope of the Spitzer 3.6~\micron\ and 4.5~\micron\ eclipse depths was also fit with chemically equilibrated atmospheric models.
The low spectral slope observed in the planet ($\lesssim 20 ppm/\micron$) favors a high atmospheric metallicity of the planet above $\sim30\times$ solar \citep{wallack2019atmCompositionCoolGasGiants}.
However, the two photometric eclipse depths cannot uniquely constrain the composition of WASP-69 b due to the correlation with the planet's temperature-pressure profile, aerosols, abundance ratios and possible disequilibrium effects \citep{wallack2019atmCompositionCoolGasGiants}.
JWST observations covering a wide range of wavelengths that encompass many molecular features and probe different pressure levels can better constrain the abundances, dynamics, chemical processes and aerosols in WASP-69 b and similar planets' atmospheres.

\subsubsection{Paper Outline}\label{sec:introOutline}
In Section \ref{sec:obs}, we describe the observations of 3 eclipses spanning 2{~\micron} to 12~\micron.
In Section \ref{sec:wasp69analysis}, we describe three independent analyses to extract the spectrum of WASP-69 b from the raw data and also describe the absolute flux of the star as compared to a stellar model.
In Section \ref{sec:specGenProperties}, we present the planet's dayside spectrum and its overall properties.
We introduce several types of models, the results of atmospheric retrievals and why they must incorporate aerosols and 3D effects in Section \ref{sec:retrievals}.
We find some evidence for an inhomogeneous dayside temperature distribution, which we investigate independently with broadband eclipse mapping in Section \ref{sec:eclipseMapping}.
We discuss the energy budget of the planet's dayside and the properties of atmospheric aerosols and synthesize WASP-69 b's inferred composition across all of our models in Section \ref{sec:discussion}.
Finally, we conclude in Section \ref{sec:conclusions} that WASP-69 b is chemically enriched relative to solar composition, has dayside aerosols and temperature gradients away from the substellar point.

\section{Observations}\label{sec:obs}

\begin{figure*}
\gridline{\fig{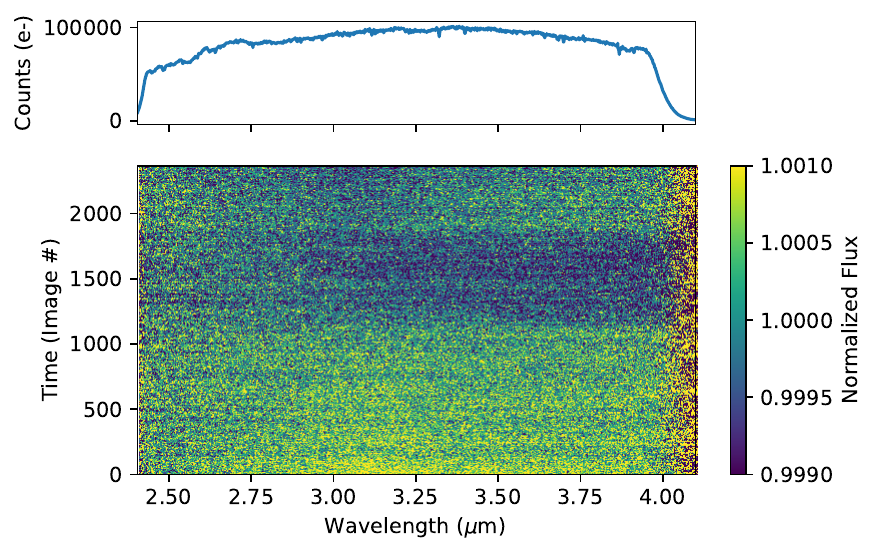}{0.49\textwidth}{Prog 1185 Obs 6: NIRCam - F322W2}\\
\fig{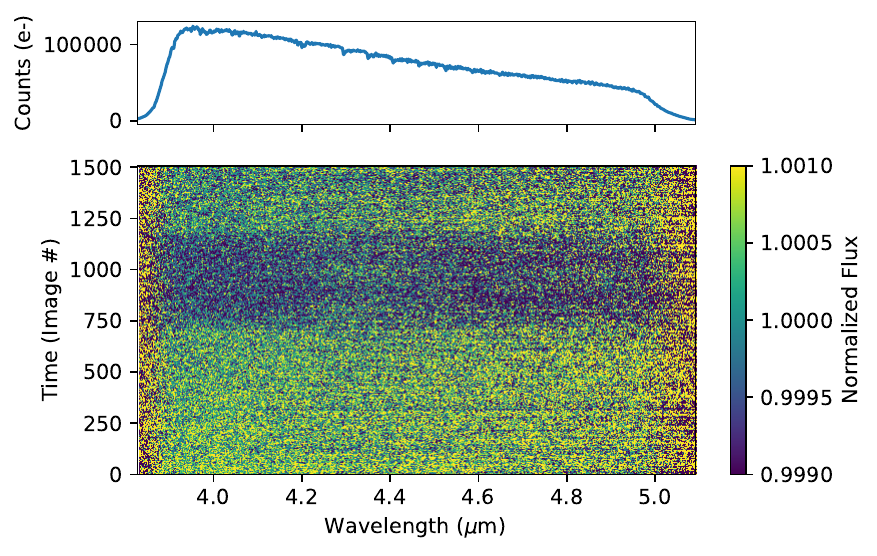}{0.49\textwidth}{Prog 1185 Obs 7: NIRCam - F444W}
          }
\gridline{\fig{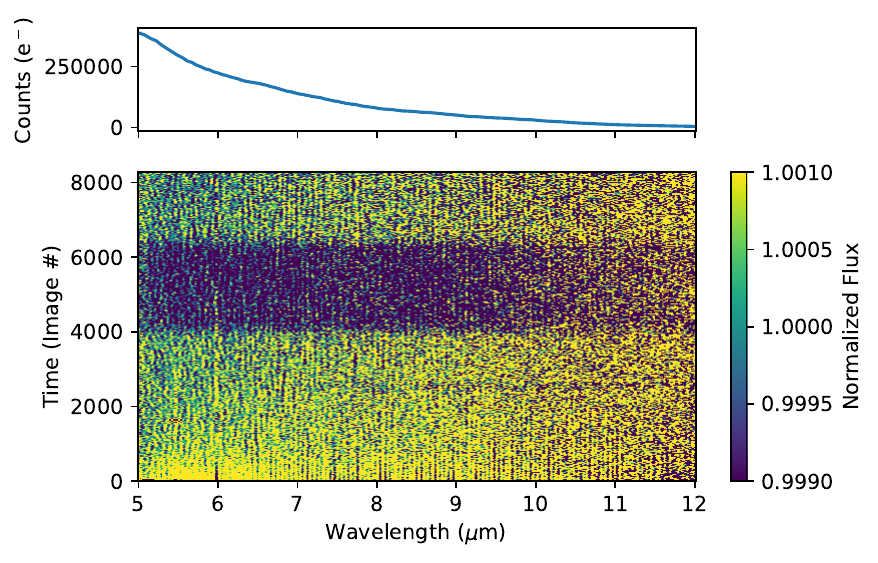}{0.49\textwidth}{Prog 1177 Obs 3: MIRI LRS}}
\caption{Dynamic Spectra for WASP-69 b's show the eclipse as a horizontal dark band.
The eclipse depth (as seen by the darkness of the band) increases from short wavelengths toward long wavelengths and also shows a dip where there is strong CO$_2$ atmospheric absorption near 4.3~\micron.
The dynamic spectra are the extracted spectra for each detector column (NIRCam) and row (MIRI) at full spectral and full time resolution without binning. The top panels of each dynamic spectrum show the average spectrum in electrons for one integration. \label{fig:dynamicSpec}}
\end{figure*}

WASP-69 b was observed as part of the MIRI And NIRCam Assay for the Transmission and Emission of Exoplanets (MANATEE), which combines observations from GTO program 1177 (Observation 3) and GTO program 1185 (Observation 6 and 7) to build up near-infrared to mid-infrared spectra on a wide variety of planets.
Table \ref{tab:observations} contains a summary of observations.

For NIRCam, we used the readout pattern with the most number of reads per integration to mitigate the effects of 1/f noise \citep{schlawin2020jwstNoiseFloorI} while still keeping data volume manageable\footnote{With BRIGHT2, our NIRCam observations had data volume excess above the sustainable rate (0.87 MB/s) with an excess of 7 (9) GB for the F322W (F444W) filters, respectively, but below the medium volume excess threshold of 15 GB. With the usually-favored RAPID readout pattern (1 frame per group), the observations would exceed the medium threshold with an excess of 24 GB.}, which led us to the BRIGHT2 readout pattern, which has no skipped frames and 2 co-added frames per group.
We selected the F210M filter and WLP8 pupil with the short wavelength arm of NIRCam to measure the the eclipse at 2.1~$\mu$m for both NIRCam observations.
For MIRI, we used the FASTR1 read pattern with the slit-less Low Resolution Spectrometer (LRS) mode.
The 3 separate observations spanned from 372 to 373 minutes in duration, or 2.8 times the eclipse duration for the planet.
This long baseline ensured significant detector settling time for persistence effects like charge trapping \citep{zhou2017chargeTrap} and MIRI detector upward and downward ramps \citep{bell2023firstLookWasp43,bouwman2023specPerformanceMIRI} to settle.
The observation details were specified before Cycle 1 measurements of these settling constants, but the on-orbit measurements of exoplanet lightcurves show that the time constants for the charge trapping ramp-up are short for the near-infrared instruments that employ HgCdTe H2RG detectors ($\lesssim$ 15 min) \citep[e.g.][]{ahrer2022WASP39bERS,feinstein2023wasp39bNIRISS,espinoza2022hatp14Spec,schlawin2023SWNIRCamPerformance}, so future near-infrared observations may succeed with shorter pre-transit or pre-eclipse baselines.

The pixel-level lightcurves for the three observations are shown in Figure \ref{fig:dynamicSpec} along with the detected counts in electrons per integration.
These dynamic spectra are normalized by the median spectrum per integration to show the change as a function of time and wavelength.
The eclipse as the planet goes behind the star is clearly detected in all three observations at this pixel-level resolution, as seen by the dark horizontal band across the dynamic spectra.
As expected, the eclipse depth rises from the shortest to longest wavelengths because the planet has a cooler temperature than the star.
Also visible in the dynamic spectra is the shallower eclipse depth near 4.3~\micron\ due to carbon dioxide in the planet's atmosphere.

Some systematic correlated noise is also visible in the dynamic spectra in both the NIRCam and MIRI data.
For NIRCam, horizontal stripes and bands are visible due to the 1/f noise present in the detector, which arises from the electronics that read the detector in the horizontal (wavelength) direction for NIRCam \citep{schlawin2020jwstNoiseFloorI,ahrer2022WASP39bERS}.
1/f noise appears as horizontal striping in the individual detector groups that manifests as horizontal striping in the dynamic spectra for the two NIRCam observations.
The 1/f noise was already mitigated by row-by-row subtraction using background pixels \citep[e.g.][]{schlawin2023SWNIRCamPerformance}, as described in Section \ref{sec:wasp69analysis}.
However, the row-by-row subtraction is only possible with background pixels of the right-most amplifier for the F322W2 NIRCam observation and the left two amplifiers of the F444W NIRCam observation, when they are oriented in the Data Management System orientation that is used in MAST.
1/f noise that is present in the middle two amplifiers can be mitigated by subtracting the median of rows of the other amplifiers, but this can only subtract 1/f noise that is common to the shared SIDECAR ASIC device and not noise that is unique to each amplifier's p-type field-effect transistor (PFET) \citep{schlawin2020jwstNoiseFloorI}.
Therefore, residuals 1/f noise is visible in Figure \ref{fig:dynamicSpec}, mostly at wavelengths shorter than 3.89~\micron\ for the the F322W2 observation 6 and longer than 4.09~\micron\ for observation 7.
Also noticeable in the dynamic spectra is a downward flux trend, most prominently in the NIRCam F322W2 spectrum near 3.1~\micron.
As will be described in Section \ref{sec:wasp69analysis}, we further mitigate and account for 1/f noise with co-variance weighted extraction, fitting excess noise and model the time-dependent trend with either a polynomial or the detector housing temperature.

For MIRI, there are also two systematic effects visible in Figure \ref{fig:dynamicSpec}.
There is a ramp visible in the first several hundred integrations which is typical of MIRI/LRS observations, especially at the shortest wavelengths \citep[e.g.,][]{bell2023firstLookWasp43,bell2024nightsideCloudsDisEqChemWASP43b,bouwman2023specPerformanceMIRI}.
There is also an odd-even effect in alternating spectral-axis pixels that appears as vertical striping in the dynamic spectrum; this is caused by the simultaneous resetting of pairs of rows (in MIRI's 90-degree rotated reference frame), and such odd-even striping is also typical of MIRI/LRS observations \citep[e.g.,][]{bell2024nightsideCloudsDisEqChemWASP43b}. This odd-even effect, the 390 Hz noise seen in the data from some MIRI subarrays (\citealt{bouwman2023specPerformanceMIRI,bell2024nightsideCloudsDisEqChemWASP43b}), and other still-unknown noise sources result in excessively noisy MIRI/LRS spectra at high-resolutions which is significantly improved by spectral binning \citep{bell2024nightsideCloudsDisEqChemWASP43b,welbanks2023wasp107b}. As a result, we used wider bin sizes (0.25\,$\mu$m full width) to average over these effects and give more reliable uncertainty estimates.

\begin{deluxetable*}{cccccccccc}[b!]
\tablecaption{Summary of WASP-69 b observations\label{tab:observations}}
\tablecolumns{6}
\tablewidth{0pt}
\tablehead{
\colhead{UT Exp Date} &
\colhead{Prog} &
\colhead{Obs \#} &
\colhead{Instrument} &
\colhead{SW Filter} &
\colhead{Spectroscopic} &
\colhead{Spec Wave Range} &
\colhead{N$_{int}$} &
\colhead{N$_{group}$} &
\colhead{Readout} \\
\colhead{YYYY-mm-dd} &
\colhead{} &
\colhead{} &
\colhead{} &
\colhead{} &
\colhead{Description} &
\colhead{\micron}  &
\colhead{} &
\colhead{} &
\colhead{Pattern}
}
\startdata
2022-10-25 & 1185 & 6 & NIRCam & F210M &  F322W2 & 2.4 - 4.01 & 2367 & 3 & BRIGHT2 \\
2023-06-06 & 1185 & 7 & NIRCam & F210M &  F444W & 3.88 - 4.98 & 1507 & 5 & BRIGHT2  \\
2023-05-02 & 1177 & 3 & MIRI & \nodata & LRS & 5.0 - 14  & 8276 & 16 & FASTR1
\enddata
\end{deluxetable*}

\begin{deluxetable*}{cccc}[b!]
\tablecaption{Summary of WASP-69 b Properties and Priors\label{tab:props}}
\tablecolumns{2}
\tablewidth{0pt}
\tablehead{
\colhead{Parameter} &
\colhead{Prior/Existing Value} &
\colhead{Prior/Existing Source} &
\colhead{TESS Posterior (this work)}}
\startdata
t$_0$ & 2455748.83344 $\pm$ 1.8e-4 & \citet{casasayas-barris2017sodiumWASP69b} &2455748.83345 $\pm$ 1.8e-4 \\
P & 3.8681382 $\pm$ 1.7e-6 d & \citet{casasayas-barris2017sodiumWASP69b} &  3.86813942 $\pm$ 1.8e-7 d \\
$i$ & 86.71$\degree$ $\pm$ 0.20 &  \citet{casasayas-barris2017sodiumWASP69b} & 86.631$\degree$ $\pm$ 0.029 \\
a/R$_*$ & 12.0 $\pm$ 0.46 &  \citet{casasayas-barris2017sodiumWASP69b} & 11.919 $\pm$ 0.05 \\
R$_p$/R$_*$ & 0.1336 $\pm$ 0.005 & \citet{casasayas-barris2017sodiumWASP69b} & 0.1285 $\pm$ 0.00025 \\
e & 0.0 & fixed & fixed \\
\hline
R$_*$ & 0.813 $\pm$ 0.028 R$_\odot$ & \citet{anderson2014wasp69wasp84wasp70Ab} & \nodata \\
log g$_*$ (cgs) & 4.535 $\pm$ 0.023  & \citet{anderson2014wasp69wasp84wasp70Ab} & \nodata \\
Distance & 50.29 $\pm$ 0.04 pc & \citet{gaia2023DR3} & \nodata \\
\enddata
\tablecomments{For the spectroscopic analysis, we use the prior values from \citet{casasayas-barris2017sodiumWASP69b}, which come from \citet{anderson2014wasp69wasp84wasp70Ab}. We also fit the publicly available TESS data and report the posteriors (right) to improve the orbital parameter precisions for eclipse mapping.}
\end{deluxetable*}

\section{Data Analysis and Extraction}\label{sec:wasp69analysis}

\subsection{Reduction and Lightcurve Extraction}\label{sec:reduction}

\subsubsection{\texttt{tshirt}}\label{sec:tshirtRed}
We extract spectra in a similar manner as in previous JWST data analyzed with \texttt{tshirt} and published \citep{ahrer2022WASP39bERS,bell2023_methane,welbanks2023wasp107b}, but with the exception that we allow for the small curvature and tilt of the trace (about 4 pixels across the spectrum with NIRCam, but negligibly for MIRI LRS).
For NIRCam data, we begin with the \texttt{\_uncal.fits} stage 0 data products from MAST and process them with a modified version of the \texttt{jwst} pipeline \citep{bushouse2023jwstPipeline1p10p2} version 1.10.2, except for the F210M photometry on Observation 6, which was extracted soon after observation with an older \texttt{jwst} version 1.6.0.
We ran the \texttt{jwst} pipeline with CRDS context \texttt{jwst\_1077.pmap} for Program 1185 observation 6's spectroscopy and CRDS context \texttt{jwst\_1093.pmap} for observation 7's spectroscopy, but checked that the used reference files were the same for both observations' contexts.
For the photometry, we used \texttt{jwst\_1009.pmap} for both observations.
We also used \texttt{tshirt} to extract the spectrum of the MIRI LRS data in Program 1177 observation 3.
For MIRI LRS data, we used SDP version 2023\_4a, \texttt{jwst} version 1.13.4, CRDS version 11.17.15 and CRDS context \texttt{jwst\_1225.pmap}.
For reference on different version numbers, we have found negligible differences in extracted \texttt{tshirt} photometry and spectroscopy $\lesssim$ 10 ppm depths between JWST version numbers and reference file contexts from JWST versions 1.8 to 1.13 and pmap contexts 1039 to 1188.

Our main modification to the \texttt{jwst} pipeline for NIRCam is to replace the reference pixel correction with a row-by-row, odd-even by amplifier (ROEBA) correction to mitigate 1/f noise in the photometry and spectroscopy \citep[e.g.][]{schlawin2023SWNIRCamPerformance}.
For MIRI LRS data, we do not use the ROEBA correction.
We also skip the dark current step for NIRCam because it adds noise for subarray modes.
We also set the jump sigma rejection threshold for cosmic rays to 6~$\sigma$ for NIRCam and 7~$\sigma$ for MIRI and stop processing the \texttt{jwst} pipeline after the \texttt{\_rateints.fits} products.
We manually divide by the NIRCam spectral images by the flat field reference file for imaging in CRDS (\texttt{jwst\_nircam\_flat\_0313.fits} for the F444W filter and \texttt{jwst\_nircam\_flat\_0266.fits} for the F322W2 filter).
The \texttt{jwst} pipeline was run on NIRCam data with the JWST Time Series Observation Wrapper (\texttt{jtow}) version 0.1.4 (\url{https://github.com/eas342/jtow.git}) and the spectra and photometry were extracted with the Time Series Helper and Integration Reduction Tool \texttt{tshirt} version 0.3 (\url{https://github.com/eas342/tshirt}).
For MIRI LRS data, we used an updated \texttt{tshirt} version 0.4 that includes wavelength calibration of MIRI LRS.
We do not apply a flat field correction to the MIRI LRS data.

For the photometric F210M data, we use a circular aperture of 92 pixels and a background annulus from 93 to 129 pixels, as shown in Appendix  \ref{sec:analysisDetails}.
For spectroscopy, we fit the traces of the spectrum for one image and use this to define the aperture and background for all images.
For the trace centroids, we fit the profile with a Gaussian in each column and then fit the centroids with a 3rd order polynomial with iterative 3~$\sigma$ clipping of outlier points.
For the long wavelength extraction, we next subtract the background using a linear fit along the Y (spatial) direction for all the pixels that are more than 7 pixels away from the source (rounded to the nearest whole pixel) and also located in a rectangular region from pixels Y=5 to 65.
We use a covariance-weighted extraction \citep{schlawin2020jwstNoiseFloorI} for the non-NaN pixels that are within 5 pixels from the source trace, assuming a read noise of 14 $e^-$ and a correlation of 0.08 between pixels.
We use the interpolated PSF to estimate the missing flux in the spatial direction for pixels that are marked as NaN or outliers more than 30$\sigma$ from a spline fit and multiply the flux to account for the missing pixels' fractional flux.
The extracted average spectrum in ($e^-$) and spectroscopic lightcurves at the pixel resolution are shown in Figure \ref{fig:dynamicSpec}.
We next bin the spectra in wavelength for lightcurve analysis to approximately 0.010~\micron\ wide wavelength bins, but rounded to the nearest whole pixel.

\subsubsection{\texttt{Eureka!}}\label{sec:eurekaRed}

Our \texttt{Eureka!} NIRCam and MIRI reductions used version 0.10 of the \texttt{Eureka!} pipeline \citep{bell2022eureka}; \texttt{jwst} package version 1.10.2 \citep{bushouse2023jwstPipeline1p10p2}; CRDS version 11.17.0; and CRDS context 1094 for NIRCam/F322W2 and NIRCam/F444W, and 1097 for MIRI/LRS. Our NIRCam reduction methods closely follow those used in previous \texttt{Eureka!} NIRCam spectroscopy analyses \citep{ahrer2022WASP39bERS,bell2023_methane,welbanks2023wasp107b}, and our MIRI/LRS reduction method generally followed the \texttt{Eureka!} v1 method described by \citet{bell2023_ers} and the \texttt{Eureka!} MIRI/LRS analysis of \citet{welbanks2023wasp107b}. \texttt{Eureka!} Control Files and \texttt{Eureka!} Parameter Files that can be used to reproduce this work are available for download (\url{https://doi.org/10.5281/zenodo.11168833}; \citealt{schlawin2024_wasp69b_zenodo}), but the important parameters are summarized below.

\texttt{Eureka!}'s Stages 1 and 2 make use of the \texttt{jwst} pipeline's \citep{bushouse2023jwstPipeline1p10p2} Stages 1 and 2 but allow for some modifications. For both NIRCam spectroscopy datasets, the only change to Stage 1 was increasing the jump rejection threshold to 6 to avoid excessive false positives when ramp fitting. For MIRI/LRS, we similarly increased the jump rejection threshold to 7 and also turned on the \texttt{firstframe} and \texttt{lasstframe} steps (which skip the first and last frames of each integration) as these frames are subject to excessive noise and generally increase the final noise level in the data. Finally, during the Stage 1 processing of the MIRI/LRS data we also applied \texttt{Eureka!}'s newly developed 390\,Hz noise removal and group-level background subtraction technique described in detail by \citet{welbanks2023wasp107b}; in general, this step removes the wavelength correlated noise in raw MIRI/LRS SLITLESSPRISM subarray data first pointed out by \citet{bouwman2023specPerformanceMIRI}. Our Stage 2 processing followed the \texttt{jwst} standard processing with the exception of turning off the \texttt{photom} and \texttt{extract1d} steps which are not required for time-series observations.
Later, we re-analyzed the data for checking the stellar absolute flux against a model.

Our Stage 3 processing of the NIRCam data closely followed the \texttt{Eureka!} NIRCam reductions of \citet{welbanks2023wasp107b}. In summary, we cropped the frames to only include the relevant pixels, masked outlier pixels, corrected for the curvature of the spectral trace, performed a linear column-by-column background subtraction using pixels $>$13\,px away from the source center, computed the position and width of the source for later use when fitting the lightcurves, and performed optimal spectral extraction \citep{horne1986optimalE} considering only the pixels within 5 pixels of the source center and using a cleaned median integration to compute our variance-weighted spatial profile. Our Stage 3 processing of the MIRI data also closely followed the \texttt{Eureka!} MIRI reduction of \citet{welbanks2023wasp107b} which differs only slightly from the NIRCam reduction steps. In particular, the differences from the NIRCam reduction steps include rotating the MIRI/LRS data to have wavelength increasing toward the right, manually applying an estimated gain of 3.1 e/DN \citep{bell2023firstLookWasp43,bell2023_ers}, using pixel indices 11--61 (excluding pixels within 9\,px of the spectral trace) to perform a column-by-column background subtraction, and only using pixels within 3\,px of the source center when performing optimal spectral extraction.

In Stage 4, we spectrally binned the data and sigma-clipped temporal outliers. For the NIRCam data we used 0.01\,$\mu$m spectral bins across 2.45--3.95\,$\mu$m for F322W2 and 3.89--4.97\,$\mu$m for F444W, while for the MIRI data we used coarser 0.25\,$\mu$m spectral bins from 5--12\,$\mu$m following the recommendation of \citet{bell2023_ers}. We did not see evidence of the ``shadowed region effect'' described by \citet{bell2023_ers}, and therefore include the 10.6--11.8\,$\mu$m data in our reduction. For all three observations, we performed 4-sigma clipping comparing each point to a smoothed version of the data computed with a 20-integration wide boxcar filter which helped to removed otherwise missed temporal outliers while ensuring not to mask the eclipse ingress or egress.

\subsubsection{\texttt{Pegasus}}\label{sec:pegasusRed}

We also reduced the NIRCam observations using the \texttt{Pegasus} pipeline (\url{https://github.com/TGBeatty/PegasusProject}). Our Pegasus reduction began with background subtraction on the \textsc{rateint} files provided by version 1.10.2 of the \texttt{jwst} pipeline using CRDS version 11.17.0. We began by applying a basic background subtraction to each \textsc{rateint} file by fitting a two-dimensional, second-order spline to each integration, covering the full 256$\times$2048 \textsc{rateint} images. For the spline fitting, we masked out image rows 5 through 75 to prevent the self-subtraction of light from the WASP-69 system. We then performed a single round of $3\,\sigma$ clipping on the unmasked portions of the image. Next, we fit individual splines to each amplifier region on the image using the unmasked, unclipped pixel values with a median box size of 20 pixels. This per-amplifier spline fitting was necessary to eliminate residual bias differences across the amplifier areas. We extrapolated the combined background spline for the whole image over the masked portions near WASP-69 and subtracted it from the original image values. Visual inspection of the \textsc{rateint} images showed that in roughly 2\% of the integrations the reference pixel correction failed for at least one of the amplifier regions, so after the spline fitting and subtraction we re-ran the reference pixel correction using \texttt{hxrg-ref-pixel} (\url{https://github.com/JarronL/hxrg_ref_pixels}). Finally, we attempted to remove some of the red-noise caused by the NIRCam readout electronics which is present along detector rows, by calculating the robust mean of each row using pixels from column 1800 onwards for the F322W2 images and up to column 600 for F444W (both chosen to avoid light from WASP-69), and then subtracting this mean from each row. Visually, this removed most of the horizontal banding typical for NIRCam \textsc{grismr} images.

We then extracted spectroscopic lightcurves from our background-subtracted images. To do so, we fit the spectral trace using a fourth-order polynomial and then used optimal extraction to measure the 1D spectrum in each image. We performed three rounds of iterative profile estimation for the optimal extraction routine, after which we judged the profile fit to have converged. Using the resultant 1D spectra, we extracted a broadband lightcurve from 2.45\,$\mu$m to 3.95\,$\mu$m at F322W2 and from 3.89\,$\mu$m to 4.97\,$\mu$m at F444W. For the spectroscopic lightcurves we subdivided each of these wavelength regions into individual 0.01\micron-wide spectral channels. During this extraction process, we linearly interpolated over each spectral column to account for partial-pixel effects in the 0.01\micron\ wavelength bins.

\subsection{Lightcurve Fitting}\label{sec:LCfitting}

\subsubsection{\texttt{tshirt}}\label{sec:tshirtLCfit}
We fit the lightcurves with a \texttt{starry} \citep{luger2019starry} lightcurve model, assuming a uniform intensity map characterized by an amplitude term only.
We use the probabilistic programming suite \texttt{pymc3} \citep{salvatier2016pymc3} and the \texttt{pymc3-ext} tools from \citet{foreman-mackey2021exoplanetJOSS} to calculate the posterior distributions of the model with No U-Turns sampling.
For the orbital parameter priors, we use the values from \citet{casasayas-barris2017sodiumWASP69b} listed in Table \ref{tab:props}.
For NIRCam, we bin the time series to a resolution of 300 equally spaced time bins at a cadence of 1.24 minutes to speed up model evaluation and also provide an empirical estimate of noise.
We estimate the errors by taking the standard error in the man of all integrations within each 1.24 minute bin and then taking the median of all bins' standard deviations and adopting this across all time bins.
edit1{For MIRI data, we trim the first 1000 integrations (ie. 45 minutes) to discard much of the initial ramp-up behavior \citep{bell2023firstLookWasp43} and bin the data to 100 equally spaced time bins}.

We have found that the measured flux of many different targets anti-correlates with NIRCam's detector housing temperature.
Figure \ref{fig:FPAHcorr} in Section \ref{sec:analysisDetails} shows the Focal Plane Housing Temperature and the out-of-transit and out-of-eclipse fluxes of many different exoplanet observations and how they anti-correlate.
While the mechanism is not exactly understood, we can model the flux as a linear function of the focal plane housing temperature deviation.
For both the short wavelength photometry and the long wavelength grism spectroscopy, we collect the detector housing temperature from the MAST  engineering database.
The short wavelength and long wavelength temperatures are accessible by querying the \texttt{IGDP\_NRC\_A\_T\_SWFPAH1} and \texttt{IGDP\_NRC\_A\_T\_LWFPAH1} telemetry mnemonics, respectively.
We find the temperature deviation by subtracting all temperatures by the median.
The temperature sensors have significant noise, so we smooth both the short wavelength and long wavelength housing temperatures first by fitting them with a 5th order polynomial and using this as an interpolation function for the temperature at each integration.
Finally, we model the NIRCam flux as a linear function of the smoothed temperature deviation.

\begin{figure*}
\gridline{
\fig{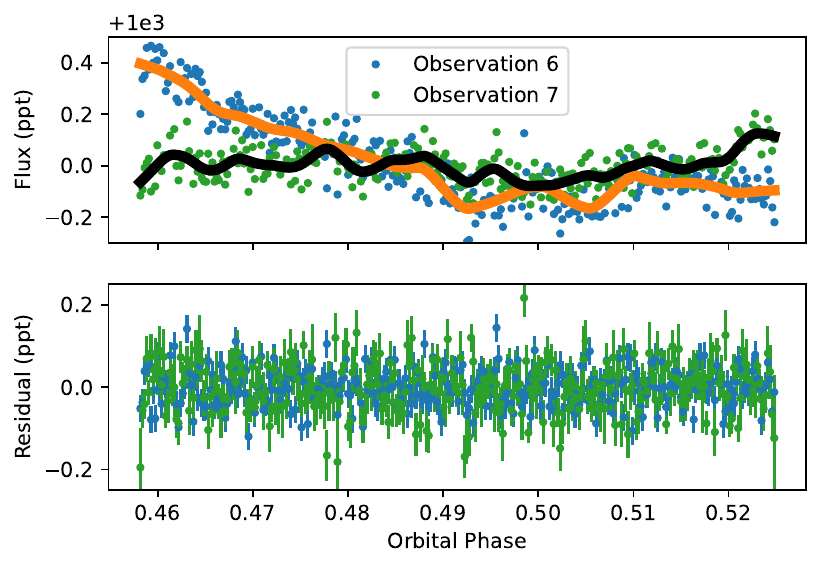}{0.49\textwidth}{Program 1185 Obs 6 and 7 F210M Eclipses, binned, with GP models}
\fig{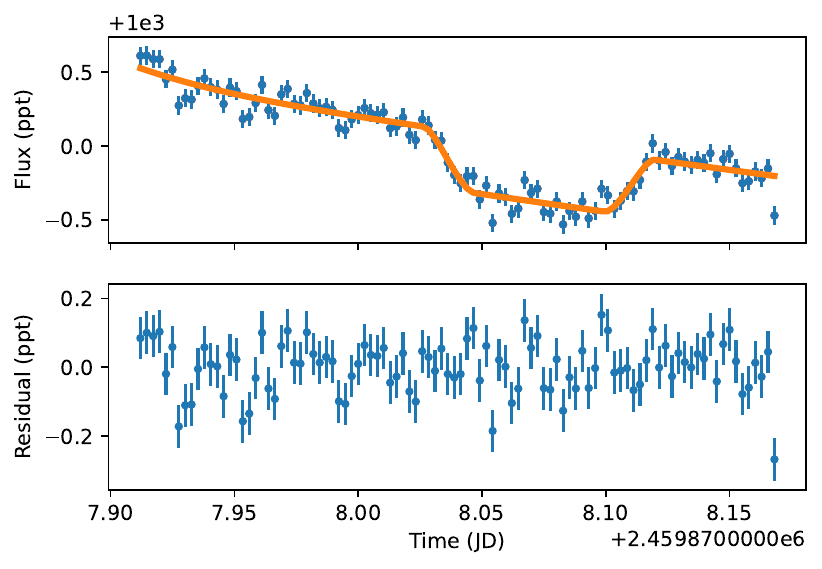}{0.49\textwidth}{Prog 1185 Obs 6 F322W2 Broadband}
          }
\gridline{\fig{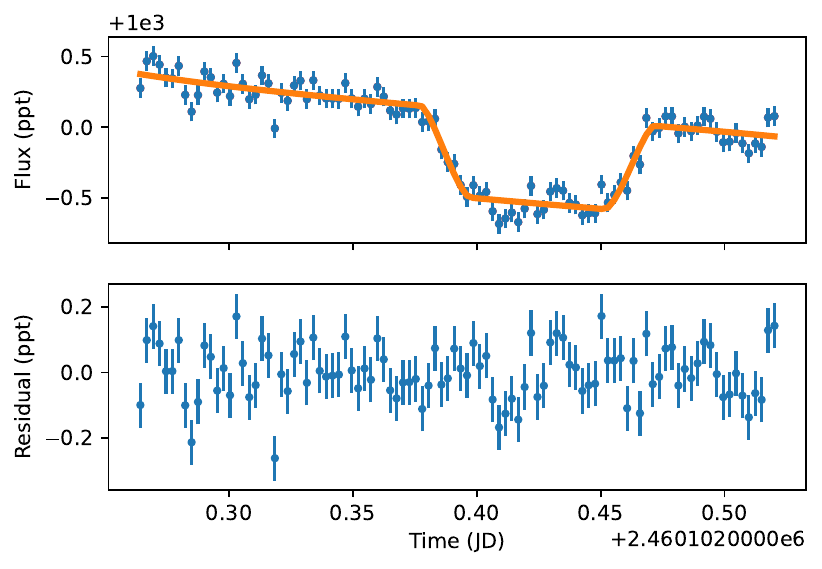}{0.49\textwidth}{Prog 1185 Obs 7 F444W Broadband}
\fig{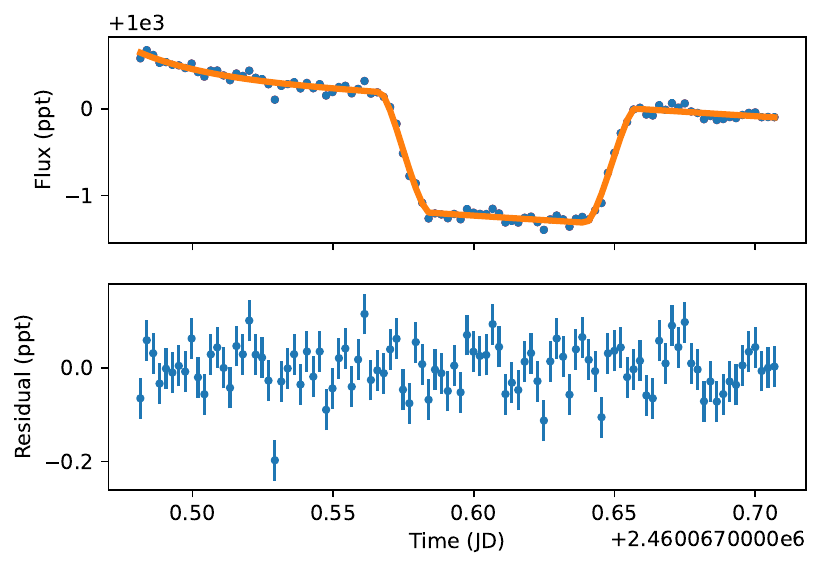}{0.49\textwidth}{Prog 1177 Obs 3 LRS}}
\caption{{\it Top Half-Panels:} Broadband eclipse Lightcurves with a Best-Fit Model. {\it Bottom Half-Panels:} Residuals of the model. \label{fig:bbLightcurve}}
\end{figure*}

In addition to the drift with focal plane housing temperature for NIRCam, we include a linear slope in time to account for other drifts such as stellar rotation.
We assume a prior slope of 0$\pm$1.6\%/hr for the linear drift.
For the MIRI data, we also include an exponential baseline function to fit for the ramp behaviors of MIRI LRS lightcurves \citep{bell2023firstLookWasp43}.
For the exponential ramp, we assume the behavior decays exponentially from the first integration with a timescale prior with a lognormal prior that has a timescale of 1 minute with a wide geometric mean of 2 and amplitude of 0.1\%.
Additionally, a strong level of correlated noise is visible in both NIRCam observations using the same F210M photomeric filter, as seen in Figure \ref{fig:bbLightcurve}.
Therefore, for the short wavelength F210M lightcurves only, we also model the time series with a Gaussian Process \citep[e.g.][]{gibsonGP}.
We use \texttt{celerite2} code \citep{foreman-mackey2018celerite} with a stochastically-driven damped harmonic oscillator kernel that has a fixed quality factor of Q=0.25 and a log-normal prior on the period of the oscillator $\rho$ that has a geometric mean equal to the duration of the observation.

We iteratively sigma clipped any fluxes that deviated by the model by more than 5~$\sigma$ and iterated 2 times to maximize the a priori probability using the \texttt{starry}, \texttt{pymc3} and \texttt{pymc3-ext} software suites.
The resulting maximum a priori lightcurve solutions are shown in Figure \ref{fig:bbLightcurve}.
This maximum a prior solution was used as an initial set of parameters for the Hamiltonian No UTurns Sampler (NUTS).
The sampling was tuned with 3000 steps and samples for another 3000 steps with two different chains.

We first found the posterior distribution of the broadband eclipse lightcurve.
We then fix the semi-major axis, inclination, period, transit time (linked to the eclipse time) at the broadband fit value for all spectroscopic lightcurve fits.
We allow the linear baseline and FPAH temperature deviation terms to be free parameters with the same priors as applied to the broadband.

We note that the short wavelength F210M lightcurves exhibited significant correlated noise, as seen in Figure \ref{fig:bbLightcurve}, upper left, especially for Observation 7.
Ignoring this noise and fitting with polynomial de-trending vectors results in significantly discrepant eclipse depths by 3.6$\sigma$.
However, when we use a GP fit to both observations, we find agreement to within 0.7$\sigma$.
The correlated noise seen in these observations is larger than found for HAT-P-14 b with the same defocused photometry mode \citep{schlawin2023SWNIRCamPerformance}.
The increased noise for WASP-69 b may be related to activity in the host star, which has been found to have spots and a rotation period of $\sim$24 days \citep{khalafinejad2021wasp69b_lowResHighRes,anderson2014wasp69wasp84wasp70Ab}.
Another possibility is JWST primary mirror flexures because the defocused photometry is highly sensitive to JWST's wavefront \citep[e.g.][]{schlawin2023SWNIRCamPerformance,mcelwain2023OTEdesignDevelPerformance}.
The longer wavelength spectroscopic observations (grism time series and MIRI LRS) show less of this correlated noise from either of these effects because stellar spot contrast decreases as function of wavelength and these modes are not as sensitive to JWST's wavefront.

\begin{figure*}
\gridline{\fig{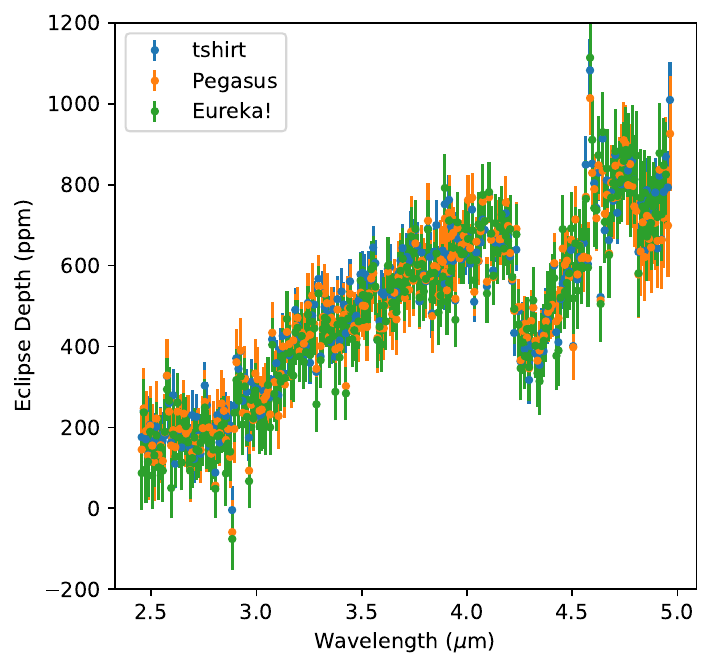}{0.49\textwidth}{NIRCam Emission Spectra From 3 Reductions}
          \fig{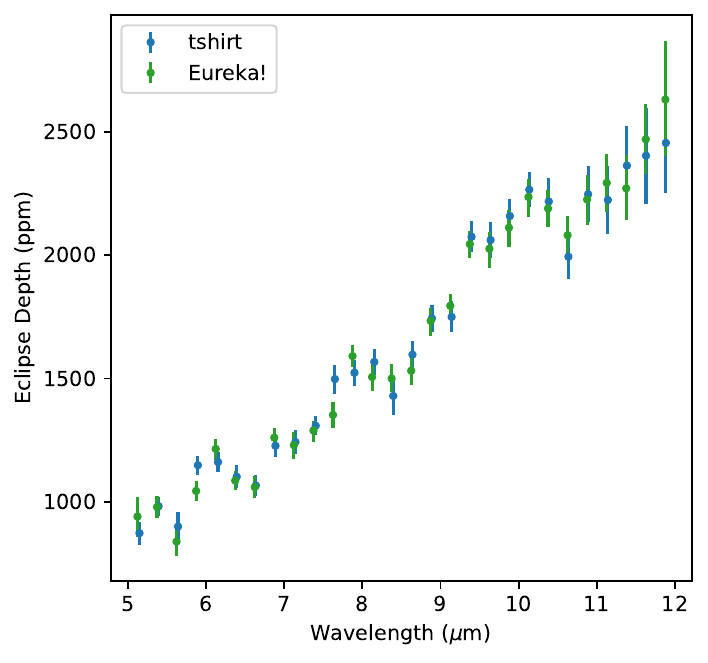}{0.49\textwidth}{MIRI LRS From 2 Reductions}
          }
\caption{{\it Left:} Eclipse spectrum with three independent extraction and lightcurve fitting methods on a common $\sim$0.010~\micron-spaced grid. All spectra agree within the $\sim$50-80 ppm errorbars.
{\it Right:} Eclipse Spectrum with MIRI with two independent extraction and lightcurve fitting methods \label{fig:wasp69EclipseSpec}.}
\end{figure*}

The resulting NIRCam emission spectrum of WASP-69 b is shown in Figure \ref{fig:wasp69EclipseSpec}.
There is a rise from 2.5 to 5.0~\micron\ expected for a $\sim$963 K zero-albedo equilibrium temperature planet orbiting a 4715 K star.
Additionally, there is a deep CO$_2$ absorption feature visible at 4.3~\micron, but no obvious CH$_4$ feature at 3.3~\micron.
A comparison with the \texttt{Eureka!} and \texttt{Pegasus} reductions (described below) is also shown in Figure \ref{fig:wasp69EclipseSpec}, which show consistency within 1~$\sigma$ errors, but some subtle 30-50 ppm offsets between reduction methods.

\subsubsection{\texttt{Eureka!}}\label{sec:eurekaLCfit}

As was done with \texttt{tshirt}, our \texttt{Eureka!} lightcurve fitting also used a \texttt{starry} \citep{luger2019starry} astrophysical model assuming a uniform map. For the orbital parameter priors, we fixed all parameters except $t_0$ to the values from \citet{casasayas-barris2017sodiumWASP69b}. For $t_0$, we used a Normal prior based on the posterior of \citet{casasayas-barris2017sodiumWASP69b} when fitting the MIRI/LRS broadband (5--12\,$\mu$m) lightcurve which resulted in an inferred $t_0=2455748.83473 \pm 0.00012$; we then adopted this new value and fixed $t_0$ for the spectroscopic fits to all lightcurves. We also assumed a stellar radius of 0.813\,$R_{\odot}$ \citep{casasayas-barris2017sodiumWASP69b} combined with $a/R_*$  to account for the light travel delay when computing the expected time of eclipse. For our systematic noise model, we used a polynomial in time (quadratic for NIRCam and linear for MIRI), a linear decorrelation against the spatial position and PSF-width of the spectral trace, and for MIRI an exponential ramp following the recommendations of \citet{bell2023_ers}. We also removed the first 800 integrations from the MIRI observations to reduce the impact of the strong initial exponential ramp. We also fitted an error inflation parameter which multiples our estimated uncertainties by a constant factor; this ensures a reduced chi-squared of 1.0 and avoids issues caused by incorrectly estimated gain values in Stage 3. For all our systematic noise models, we used minimally informative priors.

We used \texttt{pymc3}'s No U-Turns Sampler \citep{salvatier2016pymc3} to sample our posteriors using two independent chains with a target acceptance rate of 0.85. For the MIRI spectra we used 3000 tuning steps and 1500 posterior samples, while for the NIRCam spectra we used 4000 tuning steps and 3000 posterior samples. For all fits, the Gelman-Rubin statistic \citep{gelman1992inferenceFromIterativeSimulation} was at or below 1.01, ensuring the chains had converged. We then used the 16th, 50th, and 84th percentiles of the posterior samples to estimate the best-fit parameter values and their corresponding uncertainties. The fits to our NIRCam data showed minimal residual red noise, and the white noise in our residuals was $\sim$20--30\% above the estimated photon limit for F322W2 and $\sim$10--20\% above the estimated limit for F444W. Meanwhile, the MIRI data did exhibit significant residual red noise in some channels (especially between 6--8\,$\mu$m), and for MIRI the white noise level in our residuals ranged from $\sim$20--30\% above the estimated photon limit from 6--10\,$\mu$m with a gradual increase to 80\% above the limit longward of 10\,$\mu$m and a steep increase to 140\% above the limit shortward of 6\,$\mu$m (the cause of which could not be identified). To account for the impact of unmodelled red noise on our final spectra, we used the $\beta$ error inflation method developed by \citet{winn2008XO3b}; in particular, we computed $\beta$ at timescales spanning 22--32 minutes (within 5 minutes of WASP-69b's $\sim$27 minute ingress/egress duration) and multiplied our emission spectra uncertainties by the computed $\beta$ value. For most spectroscopic channels this inflated the uncertainties by $<$20\% but significantly increased the uncertainties of three channels (6.125, 6.875, and 7.75\,$\mu$m).

\subsubsection{Pegasus}\label{sec:pegasusLCfit}

Our \texttt{Pegasus} analysis used a \texttt{BATMAN} eclipse model \citep{kreidberg2015batman} to fit the spectroscopic lightcurves extracted via the \texttt{Pegasus} pipeline. We fixed the orbital parameters of WASP-69b to those measured in \citet{casasayas-barris2017sodiumWASP69b}, which left the free parameters in our spectroscopic lightcurve fitting to be the secondary eclipse depth and the slope and normalization of a background linear trend. We did not impose a prior on any of these parameters. We fit each spectral channel individually.

To fit the spectroscopic lightcurves we performed an initial likelihood maximization using a Nelder-Mead sampler followed by MCMC likelihood sampling. We used the maximum likelihood point identified by the Nelder-Mead maximization as the starting locus about which we initialized the MCMC chains. To perform the MCMC runs, we used the \texttt{emcee} Python package \citep{foreman-mackey2013emcee} using 12 walkers with a 2,000-step burn-in and then a 4,000-step production run for each spectral channel. We checked that the MCMC had converged by verifying that the Gelman-Rubin statistic was below 1.1 for each parameter in each spectral channel.

We additionally checked the goodness-of-fit and statistical properties of our eclipse modeling in each spectral channel. We did so by first verifying that the average of the per-point flux uncertainties in each channel's lightcurve matched the standard deviation of the residuals to the best fit eclipse model. We also computed the Anderson-Darling statistic for each channel's lightcurve residuals to check that the residuals themselves appeared Gaussian. We did not find statistically significant non-Gaussianity in the residuals to our spectroscopic lightcurve fits.

\subsection{Comparison of Extracted Spectra}\label{sec:specComparison}
A comparison of our three NIRCam reductions on a common $\sim$0.010\micron\ wavelength grid is visible in Figure \ref{fig:wasp69EclipseSpec}.
The \texttt{tshirt}, \texttt{Eureka!} and \texttt{Pegasus} reductions all agree well within the error bars of 50-80 ppm.
Even though each spectral extraction treated 1/f noise and there were differences in treatment of the systematic trends, the three reductions give very similar absorption features, slopes and overall depths.
A moving average of the observations showed agreement to better than the 50 ppm level.
Given the close agreement between reductions, we proceed with the \texttt{tshirt} reduction, which had the most 1/f noise mitigations and smallest broadband out-of-eclipse lightcurve scatter.
We also did two independent reductions of the MIRI LRS spectrum shown in Figure \ref{fig:wasp69EclipseSpec}.
Again, the agreement for both reductions is broadly consistent with only 2 points differing by 2.6 and 2.8 times the 1$\sigma$ uncertainties of the \texttt{Eureka!} reduction. These two wavelengths do not correspond to any strong opacity sources in our models.
We use the \texttt{Eureka!} data for our modeling because it has a longer heritage of analysis with MIRI LRS data \citep[e.g.][]{bell2024nightsideCloudsDisEqChemWASP43b,kempton2023reflectiveMetalRichGJ1214}.

\subsection{Absolutely Calibrated Stellar Spectrum}\label{sec:absCal}
\begin{figure*}
\gridline{\fig{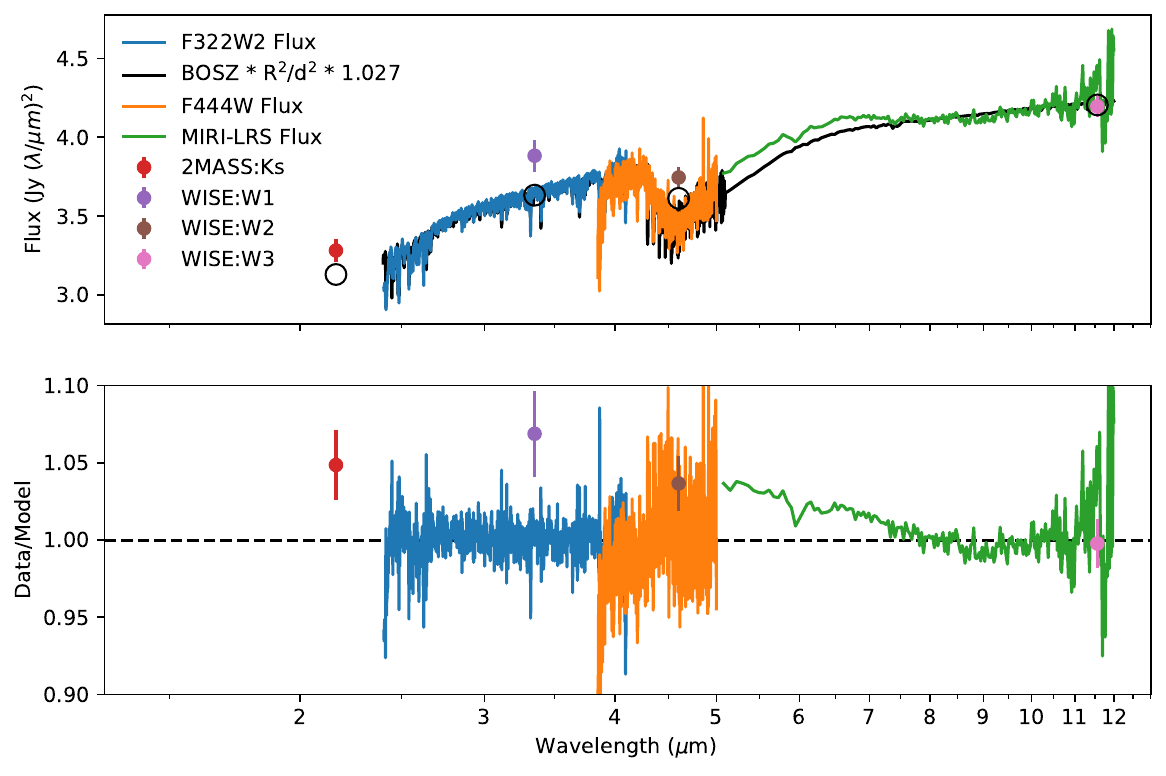}{0.69\textwidth}{}
          }
\caption{{\it Top:} Calibrated absolute flux for 3 observations of the host star WASP-69 A and a 4750 K BOSZ model \citep{bohlin2017bosz}.
Additionally, photometry from 2MASS K$_S$ and WISE bands 1-3 is shown as data points with error bars.
The black circles are synthetic photometry of the BOSZ model for those same photometric bands.
{\it Bottom:} The ratio of the data over the model shows closest agreement in the F322W2 filter (where a direct ratio with a calibrator was possible) and some 1-3\% discrepancies where stellar CO absorption is present.\label{fig:absFlux}.
No circumstellar debris disks are present above $\gtrsim 3\%$ flux levels that could dilute the planet's flux significantly.}
\end{figure*}

We also create an absolutely calibrated stellar spectrum across the NIRCam and MIRI wavelengths to find the intrinsic dayside planet flux as well as verify if a stellar model provides an accurate estimate for the star.
We first calculated calibrated stellar spectra and compared them to stellar models that most closely matched the stellar parameters from \citet{anderson2014wasp69wasp84wasp70Ab}: T$_{\rm *,eff}$ = 4700 $\pm$ 50 K, log(g$_*$)=4.5$\pm$0.15, [Fe/H] = 0.15 $\pm$ 0.08, R=0.813 R$_\odot$ and a Gaia DR3 distance of 50.29 $\pm$ 0.04 pc \citep{gaia2023DR3}.
For the BOSZ model \citep{bohlin2017bosz}, the parameters are T=4750 K, log(g)=4.5 and [Fe/H]=0.0.

For the NIRCam observations, we extracted stellar spectra of the solar-analog calibrator GSPC P-330E with the same extraction parameters as WASP-69 b to minimize errors due to aperture corrections and/or field-dependence differences with the NIRCam Wide Field Slitless calibration observations.
We used calibration program 1076 observations 1 and 2 for the F322W2 and F444W filters, respectively.
We used the CALSPEC model for GSPC P-330E (\texttt{p330e\_mod\_006.fits}) to convert from DN/s to physical flux units by dividing the stellar model by the observed count rates.

Finally, we multiplied this calibration factor by the observed count rate to calculate the calibrated NIRCam stellar spectrum of WASP-69 b.
Program 1076 observation 2 was take at a different field point that was separated by 94 pixels in the detector X direction from the science observations, so it necessitated interpolating the F444W calibrator to the same wavelengths, which was a shift of about 0.092~\micron.
An improved model for P-330E is being prepared \citep{rieke2023fainterGstars} that modifies the CALSPEC model by up to 1.5\% at 2~\micron\ but this is below the level of systematic errors, such as non-linearities in our measured absolute fluxes.
We also compared the F322W2 calibration to the value when extracting the spectrum with the \texttt{jwst} pipeline with the photometry step applied, an imaging flat field \texttt{jwst\_nircam\_flat\_0610.fits}, aperture correction \texttt{jwst\_nircam\_apcorr\_0003.fits}, and pixel area \texttt{jwst\_nircam\_area\_0054.fits} from CRDS and find agreement to within 3\% with the \texttt{jwst} pipeline correction being 3\% higher than the method with a ratio to GSPC P-330E.

For the MIRI observations, we re-ran \texttt{Eureka!}'s Stage 2 with the photometric calibration step turned on and then used \texttt{Eureka!}'s Stage 3 optimal extraction on the \texttt{\_cal.fits} calibrated data product using the exact same procedure we used to produce the MIRI lightcurve. The \texttt{Eureka!} optimal extraction allowed for better interpolation over bad pixels than a simple box extraction. Additionally, we analyzed the MIRI/LRS SLITLESS calibration observations of HD 167060, HD 106252, and HD 37962 using the exact same methods as for our WASP-69b science data to improve the flux-calibration of our observations.
We describe the use of these standard star observations below.

We compared the measured absolute fluxes of the star from before eclipse to stellar models, as shown in Figure \ref{fig:absFlux}.
While the measured fluxes from before eclipse include the planet emission, the planet contributes less than 0.3\% to the total flux so we ignore it at the level of calibration and systematic errors (several \%).
We use the stellar modeled flux at the photosphere of the star (i.e. $\pi$ times the intensity) and multiply it by the square of the ratio of the stellar radius \citep[0.813 R$_\odot$,][]{anderson2014wasp69wasp84wasp70Ab} to the Gaia DR3 distance \citep[50.2871 pc,][]{gaia2023DR3}.
We were able to confirm that a BOSZ model accurately predicted the spectrum to a 5\% level, but required an additional multiplicative factor to best match the data, which is 1.0266.
In other words, the combined flux multiplicative factor was 1.36399$\times 10^{-19}$ times the modeled flux at the stellar photospheric ``surface'', which is $\pi$ times the modeled intensity.
The multiplicative offset of 1.027 can either be due to calibration flux uncertainties or physical system parameter differences such as the distance to the system and radius.
Given 2\% uncertainties in the flux calibration of the NIRCam slitless grism mode and MIRI LRS mode \citep[][ accessed on 2024-02-01]{jdox2016}, this multiplication factor is within the calibration errors of both modes.
The strength of the CO absorption feature near 4.5~\micron\ is also slightly deeper by 1-3\%\ in the model as compared to the measured flux, either due to model systematics or differences in stellar abundances.
We also used our extracted and calibrated spectra for HD 167060, HD 106252 and HD 37962 to check for systematic calibration errors.
We used the average calibration factor (from Jy to DN/s) for these three stars using CALSPEC models and used this as an alternative calibration factor for WASP-69 b's calibration.
In this re-normalized spectrum, the drop from 5-7~\micron\ is still present, the residuals between 10-12~\micron\ are smaller and the overall spectrum is shifted upward by 5\%.
Given that the stellar model does not show any large ($>3\%$) discrepancies with the calibrated spectrum, we can rule out the possibility that debris disks can significantly dilute the spectrum of the planet at the measured precisions in $\sigma_{F_{\rm p}/F_*}$ (3-16\%, where $F_{\rm p}$ is the flux of the planet and $F_*$ is the flux of the star and $\sigma$ represents the uncertainty on the ratio of the two fluxes at the binned resolution).

We also checked our absolutely calibrated spectrum against 2MASS \citep{skrutskie06} and WISE \citep{wright2010wise} photometry.
The 2MASS and WISE photometry show slightly elevated fluxes as compared with the NIRCam grism spectra.
It is possible that there are residual non-linearity effects or the brighter-fatter effect \citep{plazas2018brighterFatter} that degrades the linearity correction of individual pixels.
A general analysis of the absolute calibration well below JWST requirements of 10\% accuracy for spectroscopy \citep{gordon2022absFlux1} is beyond the scope of this work.
However, given that there are no large deviations from the stellar model, no significant evidence for dust excesses, a small discontinuity between NIRCam F444W and MIRI LRS and future improvements to the JWST flux calibration ahead \citep{gordon2022absFlux1}, we proceed in using the stellar model and multiply it by $F_{\rm p}/F_*$ to derive the planet's dayside flux.

\section{General Properties of \texorpdfstring{WASP-69 \MakeLowercase{b's}} ~ Emission Spectrum}\label{sec:specGenProperties}

\begin{figure*}
\gridline{\vspace{-0.35in}\fig{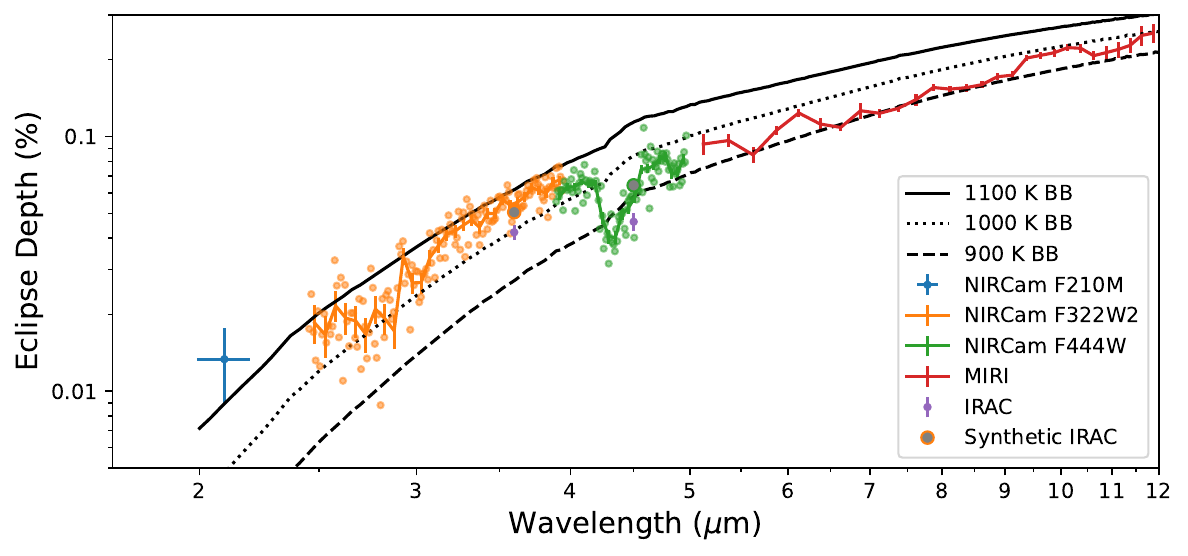}{0.553\textwidth}{}}
\gridline{\hspace{-0.07in}\vspace{-0.35in}\fig{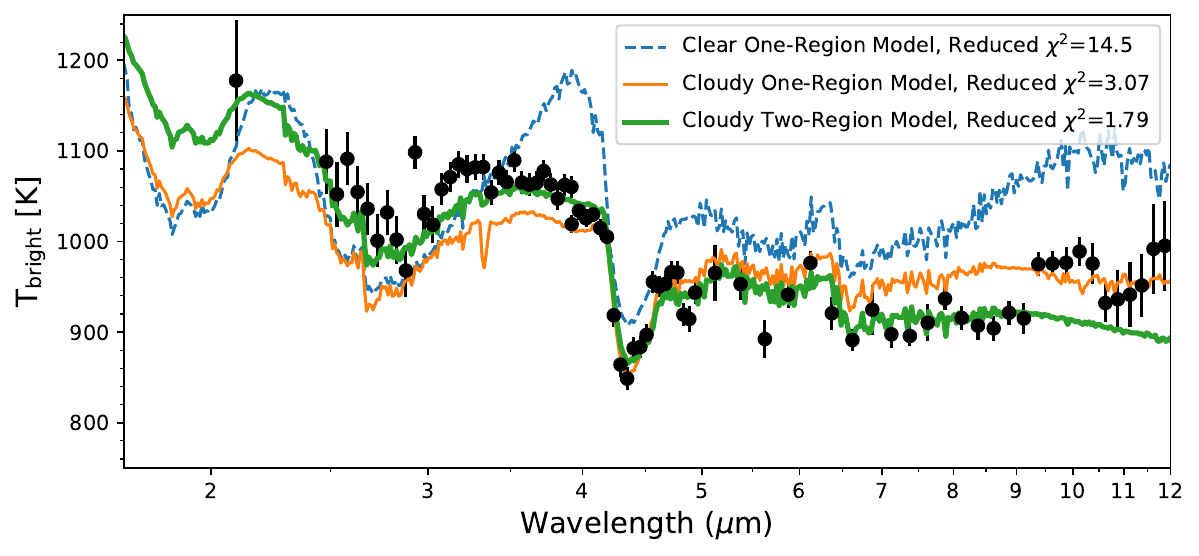}{0.56\textwidth}{}}
\gridline{\hspace{-0.1in} \fig{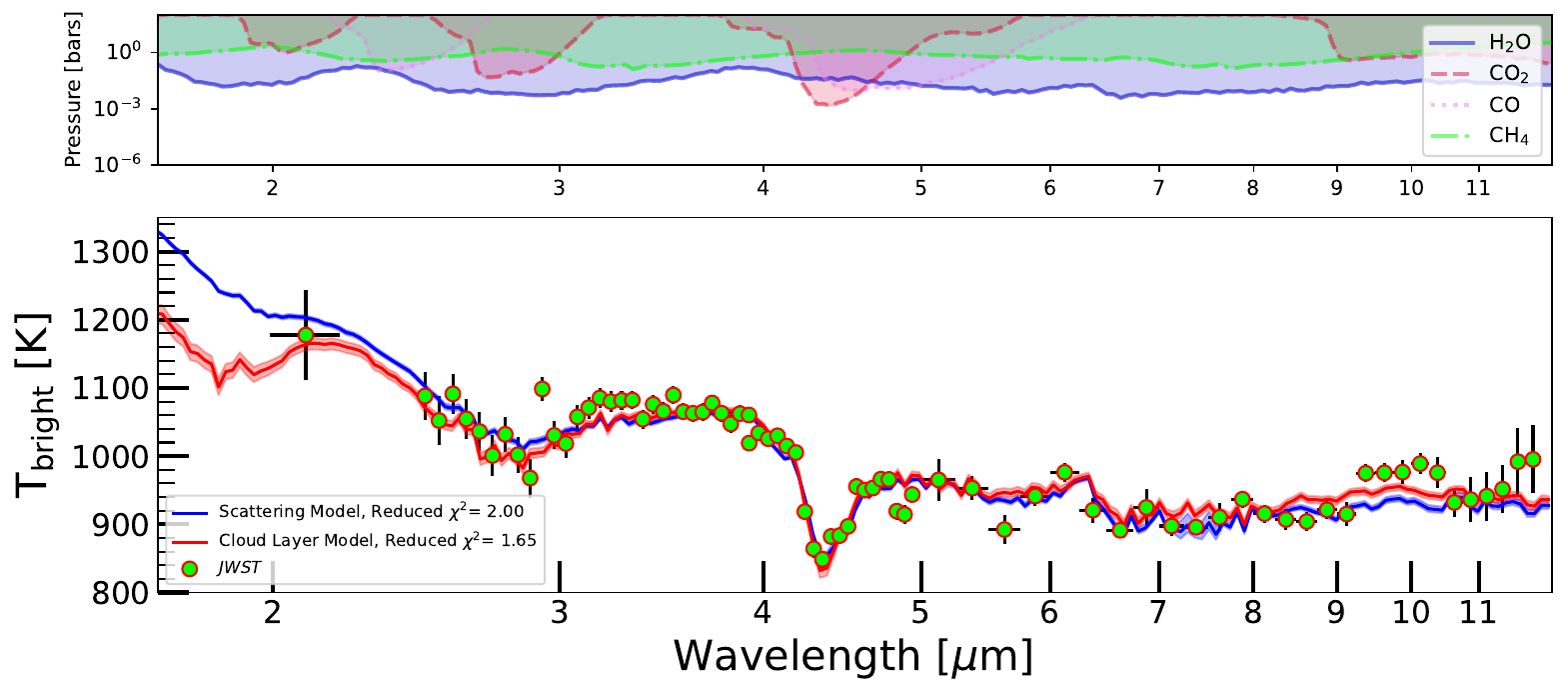}{0.55\textwidth}{}
          }
\caption{The emission spectrum of WASP-69 b shows significant absorption features from molecules in its atmosphere and a trend from high to low brightness temperature from 2.1~\micron\ to 12\micron.
{\it Top:} Binned emission spectrum of WASP-69 b in terms of planet-to-star flux for the NIRCam and MIRI data, colored by their respective filters.
The full higher resolution NIRCam data (which are used for atmospheric retrieval) are shown as points and the binned spectra are shown for illustrative purposes as lines with error bars.
The MIRI full resolution is the same as the binned one due to its intrinsically lower resolution.
Synthetic photometry is calculated for the IRAC 3.6 and 4.5 channels (gray circles) indicates larger eclipse depths than Spitzer published measurements \citep[purple points with error bars][]{wallack2019atmCompositionCoolGasGiants}.
Blackbody planetary spectra for example temperatures are shown as black lines.
{\it Middle:} The spectrum is poorly fit by a Clear 1D Homogeneous model (dashed blue line) where it over-predicts the flux at MIRI wavelengths (5-12~\micron), over-predicts the 3.9~\micron\ flux and under-predicts the shortest wavelengths (2.5-3.2~\micron).
It is necessary to include the effects of clouds (thin orange line) as well as dayside inhomogeneities (thick green line) to fit the spectrum of the planet.
{\it Bottom:} The brightness temperature spectrum shows absorption signatures of H$_2$O, CO$_2$ and CO.
The relative pressure level where each molecule has an optical depth of 1 is shown for each model in the style of \citet{mukherjee2024elfOwl}.
We plot 2 models (described in Section \ref{sec:retrievals}, which can alternatively fit the spectrum of WASP-69 b with either significant reflection (Scattering Model) or a high altitude cloud (Cloud Layer Model).\label{fig:wasp69TbSpec}}
\end{figure*}

Figure \ref{fig:wasp69TbSpec} shows the combined \texttt{Eureka!} MIRI and \texttt{tshirt} NIRCam emission spectrum, with NIRCam data binned to lower resolution for visualization purposes.
The bins for visualization are about 0.05~\micron\ for NIRCam, as compared to the original lightcurve bins of about 0.01~\micron\ with the MIRI LRS binning of 0.25~\micron\ already sufficiently low enough for visualization purposes.
Figure \ref{fig:wasp69TbSpec} also shows some representative blackbody planet spectra.
For these blackbody spectra, we multiply the Planck function by the planet-to-star radius ratio squared from Table \ref{tab:props} and divide this by the stellar intensity from our BOSZ stellar model \citep{bohlin2017bosz} with a stellar effective temperature of 4750 K, log(g)=4.5 log(cm s$^{-2}$), described in Section \ref{sec:absCal}.
The measured spectrum clearly is not well fit by a blackbody and shows absorption features and crosses from high to low blackbody temperature from short to long wavelengths.

We also show the Spitzer secondary eclipse values for the 3.6~\micron\ and 4.5~\micron\ bandpasses in Figure \ref{fig:wasp69TbSpec} \citep[421 $\pm$ 29, 463 $\pm$ 39;][]{wallack2019atmCompositionCoolGasGiants} and that they are shallower than the NIRCam spectroscopy.
We calculate the synthetic photometric eclipse depths using our JWST emission spectra and a photometric response curve interpolated to a common grid to compare the results between JWST and Spitzer.
We find synthesized eclipse depths of 542 $\pm$ 7 ppm and 644 $\pm$ 7 ppm for the 3.6~\micron\ and 4.5~\micron\ bandpasses respectively, assuming that each pixel is statistically independent.
In both bands, the JWST NIRCam eclipses are significantly deeper than the Spitzer values \citep{wallack2019atmCompositionCoolGasGiants} of 421 $\pm$ 29 ppm and 463 $\pm$ 39 ppm for the 3.6~\micron\ and 4.5~\micron\ bands, respectively.
For context, a previous comparison with the Spitzer IRAC 3.6~\micron\ eclipse depth and the synthetic NIRCam IRAC 3.6~\micron\ eclipse depth for WASP-39 b indicated consistency to within the 1~$\sigma$ uncertainty \citep{ahrer2022WASP39bERS}; however the Spitzer uncertainty on this transit depth is 176 ppm in \citet{sing2016continuum}, so the 120 and 180 ppm offsets we find in WASP-69 b would not be significant for WASP-39 b.
\citet{bean2023hd149026b} find a 4.4$\sigma$ discrepancy with the 3.6~\micron\ eclipse depth from \citet{zhang2018correlationPhaseOffsetTemperature}, ie 84 ppm $\pm$ 19 ppm for the hot Jupiter HD 149026~b.
Some of the difference we find between NIRCam and Spitzer IRAC for WASP-69 b could be due to red noise in the Spitzer lightcurves, which exhibited deviations from $1/\sqrt{N}$ statistics (where the noise for $N$ averaged data points is $1/\sqrt{N}$ times the non-averaged data).
This was observed for the second visits in both the 3.6~\micron\ and 4.5~\micron\ filters \citep{wallack2019atmCompositionCoolGasGiants}.
Another possibility is stellar or planet variability and/or a different set of systematics with JWST NIRCam versus Spitzer IRAC.

Given the large dynamic range of the eclipse depth (F$_{\rm p}$/F$_*$) across our full spectrum from 2~\micron\ to 12~\micron, it hard to see how individual gases affect the spectrum and see where atmospheric models perform well.
Instead, it is preferable for visualization purposes to compare the brightness temperature of the planet with the brightness temperature of atmospheric models because its dynamic range is much smaller.
We calculate the brightness temperature spectrum for WASP-69 b in the following manner: 
First, we average the intensity of the star from the BOSZ stellar model described in Section \ref{fig:absFlux} across each wavelength bin where the intensity is evaluated in wavelength space.
Second, we multiply this stellar intensity by the planet-to-star flux ratio and by the square of the star-to-planet radius ratio to evaluate the planet intensity.
Third, we invert the Planck function in wavelength space to calculate the associated blackbody temperature that corresponds to the planet intensity.
Fourth, we propagate intensity error intervals to brightness temperature error intervals by linearizing the inverse Planck function (with a first order Taylor expansion) and evaluating the width of the temperature uncertainty at its median intensity.
The brightness temperature (as well as the binning) are used for illustrative purposes to better view the measured spectrum and model over a wide dynamic range, but the original lightcurve wavelength bins (0.01 \micron\ bins, rounded to the nearest pixel for NIRCam spectroscopy, a filter bandpass from F210M and 0.25 \micron\ for MIRI LRS) were used to fit the models and perform retrievals using the eclipse depth (F$_{\rm p}$/F$_*$).

The brightness temperature spectrum of WASP-69~b is shown in Figure \ref{fig:wasp69TbSpec} (second panel from the top and bottom panel) for the binned version of the spectrum for visual clarity.
Additionally, we show the relative opacities for abundant and observable gases at the temperature of WASP-69 b in the second from the bottom panel.
The brightness temperature spectrum more clearly shows the spectral features of H$_2$O near 2.8{~\micron} and 6.7 to 8.0~\micron, and CO$_2$ near 4.3~\micron.
The red edge of the CO$_2$ feature also shows excess absorption due to CO near 4.7~\micron.
CH$_4$ is expected to be highly abundant in chemical equilibrium at the zero albedo full redistribution equilibrium temperature of WASP-69 b (963 K) for near-solar composition at 0.1 bar \citep{moses13}.
However, our data do not show any strong CH$_4$ features near 3.3~\micron\ nor 7.7~\micron.

\section{Atmospheric Retrievals}\label{sec:retrievals}
\subsection{Initial Fits with 1D Atmospheric Models for a Homogeneous Dayside}\label{sec:one-region}
We interpret the measured emission spectrum with theoretical atmospheric models beginning with the simplest assumptions.
We first attempted to fit the spectrum under the assumption that the disk-averaged emission from a planetary dayside can be represented by a single 1D atmospheric column, ie. a homogeneous dayside:
\begin{enumerate}
\item We initially do not include any aerosols in this model. We term this model the ``Clear One-Region Model.''\label{it:clear1DHomog}
\end{enumerate}
As will be described below, the ``Clear One-Region Model'' provides a poor fit to the data, so that is why we also include a cloud for the dayside following results of the transmission spectra that show evidence for aerosols \citep[e.g.][]{khalafinejad2021wasp69b_lowResHighRes,estrela2021aerosolsMicrobarWASP69b}.
\begin{enumerate}[resume]
\item In a second fit we term the ``Cloudy One-Region Model'', we assume a vertically uniform gray cloud opacity.
\end{enumerate}
We computed a grid of self-consistent radiative-convective equilibrium temperature-pressure (TP) profiles using the Extrasolar Giant Planet (EGP) code \citep{Marley&McKay99thermalStructure,Fortney+05comparitivePlanetaryAtmospheres,marley2015rev,thorngren2019intrinsicTemperature} for possible combinations of atmospheric metallicity of [M/H]=0.0, 0.5, 1.0, 1.5 and 2.0, C/O ratio of 0.25$\times$, 0.5$\times$, 1.0$\times$ and 2.0$\times$ solar C / O value, and the heat redistribution factor of $f_{\rm redist}=0.5, 0.6, 0.7, 0.8, 0.9, 1.0$.
The redistribution factor is $f_{\rm redist}=0.5$ in the case of full redistribution of absorbed stellar radiation over a solid angle of $4\pi$, whereas $f_{\rm redist}=1.0$ is dayside-only redistribution of absorbed stellar radiation to a solid angle of $2\pi$ (ie. a hemisphere).\footnote{The Dayside Effective Temperature is $T_{day} = T_* \sqrt{R_*/a} (1-A_{\rm B})^{1/4} (0.5 f_{\rm redis})^{1/4}$, where T$_*$ is the stellar effective temperature, $R_*$ is the stellar radius, $a$ is the semi-major axis and $A_{\rm B}$ is the Bond Albedo.}
We then use the open-source radiative transfer code \texttt{CHIMERA} \citep{Line2013} to compute the emission spectrum from the atmospheric RCE grid, where the molecular abundances at each pressure level are computed by a precomputed equilibrium chemistry table implemented by \texttt{CHIMERA} in default.
We used \texttt{PyMultinest} \citep{Buchner+14_pymultinest}, a python implementation of the \texttt{Multinest} tool \citep{feroz2009multinest}, to obtain posterior distributions for [M/H], C/O, and $f_{\rm redist}$.
The temperature at each pressure level for sub-grid point values of ${\rm [M/H]}$, C/O, and $f_{\rm redist}$ are interpolated from the RCE grid with the \texttt{RegularGridInterpolator} in Python scipy library, as  conducted in \citet{bell2023_methane}.
We assumed a uniform prior for each parameter and set the Nested Sampling live points to $500$.

As shown in the second from the top panel of Figure \ref{fig:wasp69TbSpec}, the Clear One-Region model tested here is unable to explain the emission spectrum of WASP-69b.
The best-fit spectrum has a \redChisq\ of 14.5.
The difficulty originates from the difference in brightness temperature between the NIRCam and MIRI bandpass.
The data reveal the average brightness temperature of $\sim1050~{\rm K}$ at $2.1$--$4~{\rm {\mu}m}$, which is much hotter than the temperature of $\sim950~{\rm K}$ at $>5~{\rm {\mu}m}$.
The Clear One-Region model underestimates the brightness temperature at $2.1$--$3.3~{\rm {\mu}m}$ and overestimates it at $>5~{\rm {\mu}m}$.
This result causes a dilemma for the one-region model in fitting the data: the atmosphere needs to be hotter to fit the planet's emission at $2.1$--$4$~\micron\ but it further worsens the discrepancy at $>5~$\micron.
The addition of a gray cloud opacity in the ``Cloudy One-Region Model'' shown in Figure \ref{fig:wasp69TbSpec} helps lower the flux at long wavelengths and at 3.9~\micron\ but still cannot achieve a high enough brightness temperature from 2.1 to 3.3~\micron\ and simultaneously low enough brightness temperature from 6 to 9~\micron.
In the following sections, we overcome the difficulty found here by investigating more detailed atmospheric properties.

\subsection{Models With Additional Parameters and Complexity}
\label{sec:typesOfRetrievals}
Given that our initial fit does not match the observed dayside spectrum of WASP-69~b well, we added additional complexity to the models.
We considered a variety assumptions about aerosols, reflected light, the temperature-pressure profiles and dayside inhomogeneities.
We used two different radiative transfer codes to simulate the emission spectra of WASP-69 b that each take as inputs temperature-pressure profiles and chemical abundances: PICASO \citep{batalha19refectedLightPICASO,mukherjee2023picaso3p0} and CHIMERA \citep{line2013chimera}.
The independent sets of radiative transfer models provide robustness to the results to a particular radiative transfer implementation.
We include the following types of reflection assumptions, aerosols and 3D effects, which are described in more detail later in this section:
\begin{enumerate}[resume]
\item A ``Cloudy Two-region model'' that includes two different  temperature-pressure profiles and sets of cloud properties between two geographically distinct regions.
Scattering is not significant.
(7 free parameters)\label{it:CHIMERA2TP}
\item A ``Scattering'' model that includes a wavelength-independent geometric albedo parameter as a free ``knob'' to the model, which can approximate the effects of aerosols without a specific assumption about their composition or sizes. This adds relatively more flux at short wavelengths than long wavelengths because the thermal contribution decreases at short wavelengths but the scattering term remains constant.  (10 free parameters) \label{it:PICASOrefl}
\item A ``Cloud Layer'' model that includes a distribution of silicate condensates that can emit at a different brightness temperature than the underlying molecular emission. There is no free parameter for the albedo like Model \ref{it:PICASOrefl}, but there is a modest level of scattering due to molecules and silicate condensates.\label{it:PICASOcloudLayer}
 (14 free parameters)

\end{enumerate}

For model \ref{it:CHIMERA2TP} we use the Extrasolar Giant Planet (EGP) code \citep{Marley&McKay99thermalStructure,Fortney+05comparitivePlanetaryAtmospheres,marley2015rev,thorngren2019intrinsicTemperature} with radiative transfer calculated by the CHIMERA code \citep[e.g.][]{line2013chimera}.
For models \ref{it:PICASOrefl} and \ref{it:PICASOcloudLayer} we use the PICASO code \citep[e.g.][]{batalha19refectedLightPICASO,mukherjee2023picaso3p0}.
We introduce more detailed model descriptions in what follows.


\subsubsection{Cloudy Two-Region Model}\label{sec:twocompModels}
It has been known that modeling exoplanet daysides with a single TP profile, as adopted in the One-Region models, has an inability to fit the emission spectrum if there is a strong temperature contrast in the dayside \citep{feng2016nonUniform,taylor20biasesInhomogeneousEmissionSpectra}.
In fact, the mock retrieval of \citet{feng2016nonUniform} demonstrated that a single TP profile cannot produce bright emission at $\sim2$--$3~{\rm {\mu}m}$ produced by a dayside with strong temperature contrast (see their Figure 4), which is reminiscent of the difficulty found in Section \ref{sec:one-region}. 
Fitting a homogeneous dayside model to planet with an {\it in}homogeneous dayside can also result in biased abundances and spurious molecular detections \citep{feng2016nonUniform,taylor20biasesInhomogeneousEmissionSpectra}.

To account for the dayside inhomogeneity, we introduce the Two-Region model that computes the emission spectrum using 2 TP profiles.
The setup of the model is largely the same as the one-region model presented in Section \ref{sec:one-region}.
We utilize the grid of radiative-convective equilibrium TP profiles used in the one-region models (see Section \ref{sec:one-region}), although we have extended the heat redistribution factor to $f_{\rm redist}=2$ because the dayside now has a local region that is hotter than the dayside average.
We split the dayside into hot and cool regions and compute the the emission spectrum from each region using the \texttt{CHIMERA}.
The observable spectrum is then computed as
\begin{equation}\label{eq:twoRegionMix}
    F_{\rm obs}=x_{\rm hot}F_{\rm p}(f_{\rm redist,hot},\kappa_{\rm cld,hot})+(1-x_{\rm hot})F_{\rm p}(f_{\rm redist,cold},\kappa_{\rm cld,cold}),
\end{equation}
where $x_{\rm hot}$ is the areal fraction of the hotter region of the dayside, and we have assigned different values of the heat redistribution factors $f_{\rm redist,hot}$ and $f_{\rm redist,cold}$ for each hot and cool regions.
$\kappa_{\rm{cld,hot}}$ and $\kappa_{\rm{cld,cold}}$ are the gray cloud opacities of the hot region and cold region, respectively.
Since the cloud properties can also be very different at each region due to the distinct TP profiles, we also independently assign the gray cloud opacity for hot and cold regions. 
We note that the gray cloud is treated as an purely isotropic scattering opacity source in \texttt{CHIMERA}.
The atmospheric metallicity and C/O ratio are expected to be horizontally uniform and thus are assumed to be the same in the hot and cool regions.
As in the One-Region models, we use \texttt{PyMultinest} with 400 live points to obtain the posterior distribution of 5 climate parameters: [M/H], C/O, $f_{\rm redist,hot}$, $f_{\rm redist,cold}$, and $x_{\rm hot}$ and 2 cloud parameters: $\kappa_{\rm redist,hot}$, $\kappa_{\rm redist,cold}$ for a total of 7 parameters.
We note that the model includes a gray cloud for computing the spectrum, but ignores the radiative effects of clouds on TP profiles.
For the priors on the temperature of the cold region, we restrict $f_{\rm redist,cold}$ to be between 0.5 and 1.0 (ie between full and and no heat re-distribution).
For the hot region, we set a wider prior on $f_{\rm redist,hot}$, from 0.5 and 2.0. The upper bound corresponds to the substellar point temperature in pure radiative equilibrium. This is higher than the no redistribution case discussed above, because the latter corresponds to the averaged dayside temperature, whereas the $f_{\rm redist,hot}=2$ bound corresponds to the maximum possible local temperature.

\subsubsection{Scattering Model}\label{sec:scatteringCloudy}
We use the open-source \texttt{PICASO} model \citep{batalha19refectedLightPICASO,mukherjee21cloudParameterizations} to model the planet's atmosphere within a 1D Bayesian retrieval framework.
The temperature-pressure profile is assumed to be in an analytic form of \citet{line2013chimera}, with five free parameters: the equilibrium temperature T$_{\rm eq}$ (simply a parameter and not necessarily consistent with the planet's equilibrium temperature), log(g), Planck mean infrared opacity $\kappa_{\rm IR}$, internal temperature $T_{\rm int}$ and the relative fraction of the second visible stream in the two-stream approximation $\alpha$.
We assume thermochemical equilibrium throughout the atmosphere in our retrieval framework. For a given $T(P)$ profile, atmospheric metallicity, and C/O ratio, the FASTCHEM chemical equilibrium model \citep{stock2022fastchem2} is used to generate the abundance profiles of gases like CO$_2$, CO, CH$_4$, NH$_3$, H$_2$O, etc. The solar composition elemental abundances from \citet{lodders2009abundances} are scaled for different metallicities and C/O ratios within the retrieval. The C/O ratio for a given metallicity is varied by scaling the C and O elemental abundances such that the total C+O remains unaltered. In addition to thermochemical equilibrium gases, we also retrieve on a constant (with altitude) SO$_2$ abundance to capture possibility of photochemically produced SO$_2$ \citep{tsai2022photochemistrySO2}.
This SO$_2$ abundance is not self-consistent with the chemical equilibrium chemical profiles, but it gives a clue whether photochemical calculations such as with VULCAN \citep{tsai2021VULCANcomparitiveStudy} may be necessary.
These chemical abundances along with the $T(P)$ profile are then used to generate the 1D emission spectrum of the planet with \texttt{PICASO}. 
To account for the horizontal temperature contrast, the model also adds a dilution parameter ($s_\mathrm{dilute}$), which is the fractional area of the planet's dayside that is emitted by this 1D model.
This smaller emitting area can compensate for the 2D structure of the planet and mitigate the biases in retrievals \citep{taylor20biasesInhomogeneousEmissionSpectra}.

In the scattering model retrieval setup, we include a constant reflected light term within our retrieval model in addition to the thermal component to fit the eclipse spectrum. The planet-to-star flux ratio spectrum in this case is defined as, 
\begin{equation}
    \dfrac{F_{\rm planet}(\lambda)}{F_{\rm star}(\lambda)} = A_{\rm g}\dfrac{R_{\rm p}^2}{a^2} + \dfrac{F_{\rm planet,thermal}(\lambda)}{F_{\rm star}(\lambda)}
\end{equation}
where $A_{\rm g}$ is the geometric albedo of the planet, R$_{p}$ is the planet radius, and $a$ is the semi-major axis. We choose the $A_{\rm g}$ term to be wavelength-independent to simply asses the amount of starlight reflection that would be needed to explain the near-infrared data, even though such a wavelength-independent reflected light term is not very realistic and we do not include the effect that scattering can have on radiative transfer. 
For the prior on the geometric albedo, we put in a strict upper limit of 0.67 on $A_{\rm g}$. We use this upper limit because if the $A_{\rm g}$ is greater than 0.67, then the Bond Albedo of the planet would be higher than 1 if the planet is assumed to be Lambertian. We also note that this retrieval setup can violate energy conservation if the thermal component plus the reflected light component are larger than the total incident energy on the planet.
We sample the posterior distribution with the Dynasty code \citep{speagle2020dynesty}.

\subsubsection{Cloud Layer Model}\label{sec:cloudLayerModel}

We use a second model setup with the same PICASO framework, equilibrium chemistry, parameters for the metallicity, C/O ratio, and analytic \citet{line2013chimera} Temperature-Pressure profile but with a cloud deck.
We also use the Dynasty sampler \citep{speagle2020dynesty}.
Our cloud deck setup attempts to fit the data without an arbitrary free Geometric Albedo that allows high values of reflected light.
Instead, we simulate the (small) contribution to scattering light from molecular Raleigh scattering and condensate clouds and the (larger) effects on radiative transfer of planet thermal emission that the clouds can have.
We include a parametric form of condensate cloud deck in this retrieval setup, which can be used with any cloud species' optical properties.
The cloud is parameterized assuming a log-normal cloud particle size distribution, where $r_{\rm mean}$ is the {\it geometric} mean (and also the median) of the particle size distribution and $\sigma$ is the {\it geometric} standard deviation the particle size distribution as in \citet[]{ackerman2001cloudPrecipitation} Equation 9.
We assume that $r_{\rm mean}$ and $\sigma$ remain the same for all cloudy atmospheric layers and calculate the layer-by-layer cloud optical depth $\tau_{\rm{cld}}$, asymmetry parameter $g_0$, and single scattering albedo $\rm{w}_0$ using Mie scattering calculations. 
These layer-by-layer Mie properties are then scaled using the normalization factor--$ndz$. This lets us calculate the cloud Mie properties at the base of the cloud deck. We fit for a base pressure -- $P_{\rm base}$ deeper than which the cloud optical depth is zero. At pressures smaller than $P_{\rm base}$, the layer-by-layer optical depth of the cloud deck is calculated using $r_{\rm mean}$ and $ndz$ is scaled using the scaling factor,
\begin{equation}
    f = e^{-{f_{\rm sed}}z/H}    
\end{equation}
where $z$ is altitude relative to the cloud deck base. We set the reference altitude $z$ arbitrarily and the scale height $H$ to be a constant throughout the atmosphere and use $f_{\rm sed}$ as a free fitting parameter. We fit for the five free cloud parameters $log(r_{\rm mean})$, $\sigma$, $P_{\rm base}$, ${f_{\rm sed}}$, and $ndz$ in this setup.
We also include a dilution parameter ($s_\mathrm{dilute}$) to approximate the 3D effects of day-to-terminator temperature contrasts, as for the Scattering model described in Section \ref{sec:scatteringCloudy}.

We apply our cloud layer retrieval setup using optical 
properties of Enstatite (MgSiO$_3$) \citep{scott1996forsterite}.
We initially tried other sulfide compositions such as Na$_2$S but found unrealistically high temperatures that deviated significantly from radiative-convective equilibrium grids.

\subsection{Retrieval Results}\label{sec:retrievalResults}
\subsubsection{Two-Region Retrieval}\label{sec:twoRegRetrieval}
\begin{figure*}[t]
\centering
\includegraphics[clip,width=0.99\hsize]{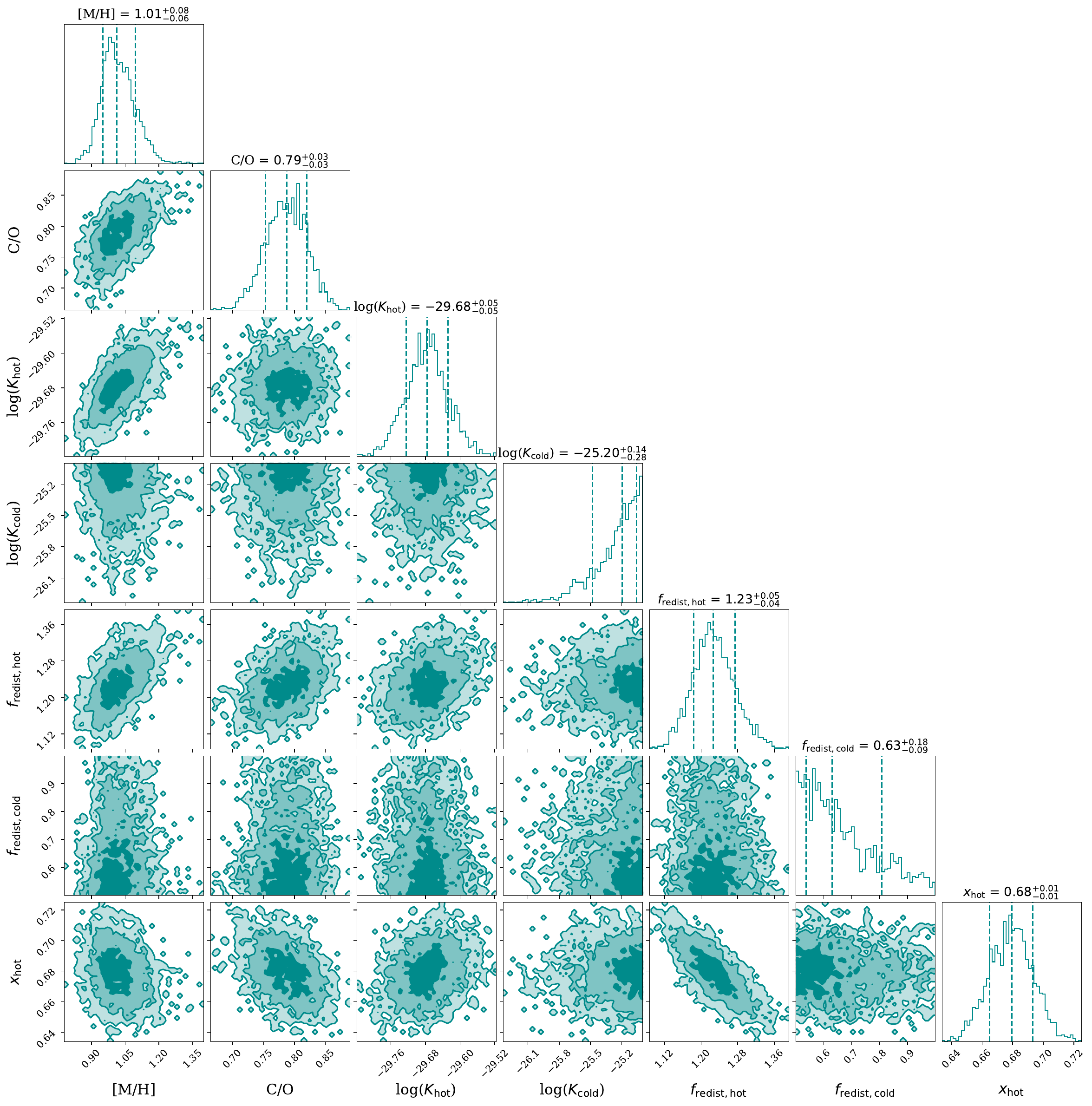}
\caption{Retrieved Posterior Distribution for the Two-Region Model.
The physical constraints of the model give tight bounds on the metallicity and C/O ratio of the planet.
The hotter region fills about 68\% of the dayside in this retrieval.
}\label{fig:twoRegionRetrieval}
\end{figure*}

\begin{figure*}[t]
\centering
\includegraphics[clip,width=0.99\hsize]{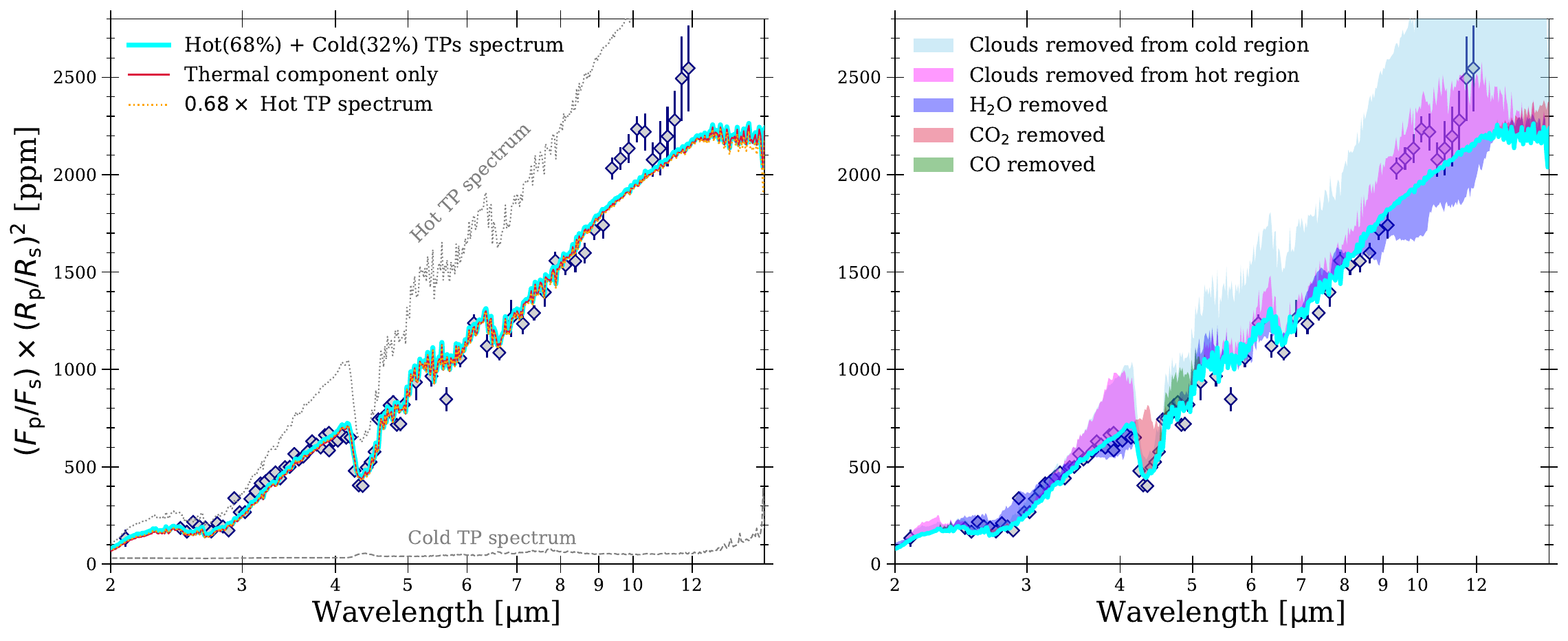}
\caption{{\it Left:} The Two-Region Model can fit the spectrum with a hot temperature-pressure profile that covers 68\% of the dayside area and a cooler cloudier region that contributes negligibly to the planet flux.
The binned data are shown as gray diamonds with error bars.
The individual spectra from the hot component (gray dotted line) and cold component (gray dashed line) are combined via Equation \ref{eq:twoRegionMix} to give a best fit spectrum (thick cyan line).
A model with no reflected light and only thermal flux (thin red line) and the hot component's emission only (dotted orange line) show that reflected light and the cold component's emission are both negligible compared to the thermal emission from the planet's hot region.
This also means that the dilution parameter used in the Cloudy Layer and Scattering models should give a very similar results as two separate regions.
{\it Right:} Our measured spectrum shows signatures of H$_2$O, CO$_2$ and CO molecules as well as clouds to both suppress the emission from the cold region and the 4~\micron\ and 9--11~\micron\ emission from the hot region.
We remove one molecule at at time and plot the change in spectrum as a shaded region spanning from the best fit model (cyan line) to the molecule-removed and cloud-removed spectra (color-shaded regions).
The binned data data (gray diamonds) are the same as on the left plot.
We note that removing H$_2$O can decrease the flux as compared to the best-fit model because it increases the single scattering albedo and thus decreases the emissivity of the photosphere. \label{fig:twoRegionSpectrum}}
\end{figure*}

The introduction of a dayside inhomogeneity greatly improves the model ability to explain the data.
Figure \ref{fig:wasp69TbSpec} shows the median emission spectrum of the two-region model.
While One-Region models over-predicts the long wavelengths and under-predict the short wavelengths, the Two-Region model better fits the overall shape of the spectrum  (see Figure \ref{fig:wasp69TbSpec} and \ref{fig:twoRegionSpectrum}).
This result demonstrates that it is critical to accounting for the temperature contrast on the day side, as suggested by \citet{feng2016nonUniform} and \citet{taylor20biasesInhomogeneousEmissionSpectra}.
Figure \ref{fig:twoRegionRetrieval} shows the posterior distributions of the two-region model.
We retrieved a super-solar metallicity of [M/H] = 1.01$^{+0.08}_{-0.06}$ and moderately super-solar C/O ratio of $0.79^{+0.03}_{-0.03}$.
It should be noted that the uncertainties of the parameters retrieved here are likely underestimated, as our two-region model relies on the TP profiles interpolated from the 1D RCE grid, which has less flexibility to change the TP profile compared to parameterized TP profiles. 
It is also interesting to mention that the cloud opacity retrieved for the cold region is much higher than that for the hot region, which indicates a strong cloudiness contrast on the dayside.

To further elaborate on the results of the two-region model, the left panel of Figure \ref{fig:twoRegionSpectrum} shows the median emission spectrum along with individual emission spectra from the hot and cold regions.
The data can be well fitted by the hot region with an area fraction of $68\%$.
A comparison between the median spectrum with the diluted hot region spectrum further reveals that the emission from the cold region has negligible impacts on the total thermal emission from the dayside, which is attributed to thick clouds in cold regions.
Thus, our result indicates that the overall shape of the observed spectrum is controlled by the emission from the hot regions where it is hotter and less cloudy as compared to the cold regions.
We also note that the median spectrum is invariant if we omit the reflected light component, indicating that the reflected light is negligible in the Two-Region model.
Given that the cold region's flux is negligible due to thermal scattering in the atmosphere, the temperature on the cold region largely unconstrained by the data.
The $f_{\rm redist,cold}$ parameter pushes up against our prior of 0.5 (full redistribution of heat) as shown in Figure \ref{fig:twoRegionRetrieval}.
Thus, we only find an upper limit on the redistribution factor and the data do not probe the deeper layers as will be discussed in the contribution functions described in Section \ref{sec:cloudProperties}.
That being said, the temperatures implied by $f_{\rm redist}$ (1220$\pm$12 K in the hot region and $\sim$1030 K in the cold region) are roughly consistent but slightly below the temperature contrast in a General Circulation Model for WASP-69 b in a cloudless simulation \citep{mehta2024inprep}.

The observed spectrum shows the absorption features of several molecules.
As shown in the left panel of \ref{fig:twoRegionSpectrum}, the spectrum clearly shows the absorption feature of CO$_2$ at ${\sim}4.3~{\rm {\mu}m}$ that is a strong indicator of a high metallicity atmosphere \citep[e.g.,][]{moses11,ers2023wasp39b_CO2}.
The spectrum also indicates the presence of CO that is needed to explain the relatively low flux at ${\sim}4.6~{\rm {\mu}m}$.
The H$_2$O absorption and cloud opacity have comparable impacts and control the overall shape of the spectrum.
On the other hand, the spectrum does not show any noticeable CH$_4$ features. 
As will be discussed in Section \ref{sec:compositionSummary} , the atmospheric TP profile inferred from the observed spectrum leads to an equilibrium CH$_4$ abundance of $\la{10}^{-6}$ at $P<{10}^{-1}~{\rm bar}$, which is orders of magnitude lower than the abundances of CO and H$_2$O.
This makes the CH$_4$ feature unnoticeable even without the disequilibrium quenching from the deep hot atmosphere.
However, this would not be obvious under the assumption of a homogeneous model that is closer to the zero-albedo full-redistribution equilibrium temperature of 963 K for WASP-69 b, which would contain more significant CH$_4$ absorption.

The observed spectrum indicates the presence of clouds in both the hot and the cold regions.
The hot region needs to be veiled by a moderate amount of clouds; otherwise, the spectrum around $4~{\rm {\mu}m}$ becomes too bright to explain the observation.
The cold regions should be veiled by clouds much thicker than those in the hot regions to suppress the emission from the cold region.
The model fails to explain the faint emission at $5~{\rm {\mu}m}$ if the cold regions have clear atmospheres.
We will discuss the potential cloud compositions in the hot and cold regions in Section \ref{sec:cloudProperties}.

Our two-region model still struggles to explain the sudden increase of the observed emission at ${\ga}9~{\rm {\mu}m}$.
Since removing clouds from the hot region leads to thermal emission comparable to the observed value at ${\ga}9~{\rm {\mu}m}$, the data potentially indicate that clouds in the hot region begin to be transparent at ${\sim}9~{\rm {\mu}m}$ through Mie scattering, which cannot be modeled by our gray scattering clouds in the two-region model.
If this is true, the clouds in the hot region would mainly consist of ${\sim}(9/2\pi)$= 1.4~\micron\ cloud particles.
Another intriguing possibility is the sudden decrease in the cloud's single scattering albedo at ${>}9~{\rm {\mu}m}$, which we will further discuss in Section \ref{sec:cloudProperties}.

\begin{figure*}
\gridline{\fig{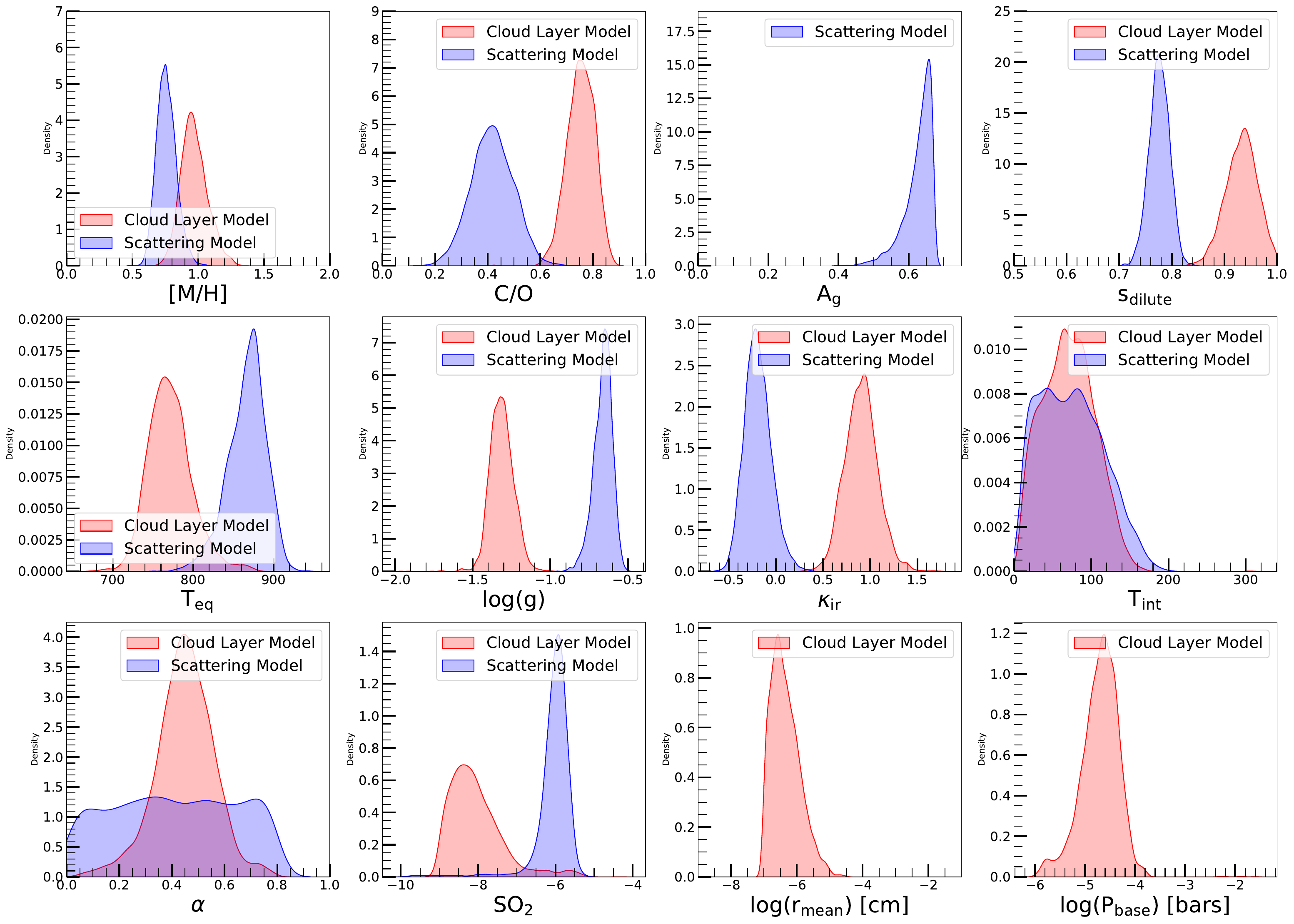}{0.99\textwidth}{}}
\caption{Posterior densities of retrieved atmospheric parameters from PICASO for the scattering model (red) and the cloud layer model (blue).
The modeled parameters (from left to right and down are the atmospheric log metallicity relative to solar composition ([M/H]), absolute carbon-to-oxygen ratio (C/O), geometric albedo ($A_g$), scattering model only), dayside area dilution ($s_\mathrm{dilute}$), Five Temperature-Pressure Profile parameters from from \citet{line2013chimera}: the $T_{eq}$ parameter (not consistent with the planet's equilibrium temperature, described in the text), $log(g)$, $\kappa_{IR}$, the internal heat flux temperature $T_{int}$,$\alpha$, and finally a free-floating SO$_s$ abundance (SO$_2$).
Stair-step 2D representations of these posteriors can be found in Figures \ref{fig:cornerScatteringModel} and \ref{fig:cornerCloudLayer}.
\label{fig:picasoPosteriors}}
\end{figure*}

\subsubsection{Scattering Model Retrieval}\label{sec:scatRetrieval}
We show the posterior distribution of the Scattering Model spectrum in the bottom panel of Figure \ref{fig:wasp69TbSpec}.
The posterior distributions on different parameters like atmospheric metallicity, C/O ratio, dilution factor, and geometric albedo obtained from this retrieval setup is shown in the panels of Figure \ref{fig:picasoPosteriors} in blue. This retrieval setup constrains the atmospheric metallicity of the planet to be supersolar at with [M/H] = +0.75$\pm$0.15 dex and the absolute C/O ratio at 0.42$\pm$0.16.
The dilution parameter is significantly below 1.0, indicating that the emission is dominated by a central region of higher brightness approximated by this 1D model.
There is also a peak in the posterior abundance of SO$_2$ near a mixing ratio of $10^{-6}$, which likely improves the fit near 7.5~\micron, but it has a long tail extending to lower mixing ratios.
Given that there is no significant absorption feature in the brightness temperature spectrum at 7.7~\micron\ we do not consider this significant evidence for SO$_2$.

In our scattering model setup, we find that the retrieved geometric albedo pushes right up against our prior of 0.67 in order to fit the spectra at short wavelengths, as shown in Figure \ref{fig:picasoPosteriors}.
The retrieved very high reflected component is much higher than the albedo inferred in other hot Jupiter atmospheres  which are, for example, 0.096 $\pm$ 0.02 for HD 209458 b \citep{brandeker2022cheopsAlbedoHD209458b} and 0.076 $\pm$ 0.02 for HD 189733 b \citep{krenn2023hd189733bAlbedoCHEOPS}, as measured by CHEOPS secondary eclipse, with a range of 0.02 to 0.27 for 1500 K to 1700 K a sample of hot Jupiters \citep{adams2022opticalAlbedosSixHotJupiters}.
Recently, an ultra-hot (1980 K) Neptune LTT 9779 b was found to have a large geometric albedo of 0.80$^{+0.10}_{-0.17}$ from its CHEOPS optical secondary eclipse depth \citep{hoyer2023LTT9779b}, and even at 1980 K, the thermal contribution contributed negligibly ($<$10) ppm for a 115 ppm eclipse.
However, a Neptune-mass planet like LTT 9779 b may have a different composition than a warm Jupiter like WASP-69 b so it is unknown if Jupiter-mass planets can attain such high geometric albedos.
The $T(P)$ profile constrained from this retrieval is shown along with the abundances of key molecular absorbers of CH$_4$, CO, CO$_2$ and, H$_2$O using blue in Figure \ref{fig:retrievedProfiles}.

\subsubsection{Cloud Layer Retrieval}\label{sec:cloudLayerRetrieval}
\begin{figure*}
\gridline{\fig{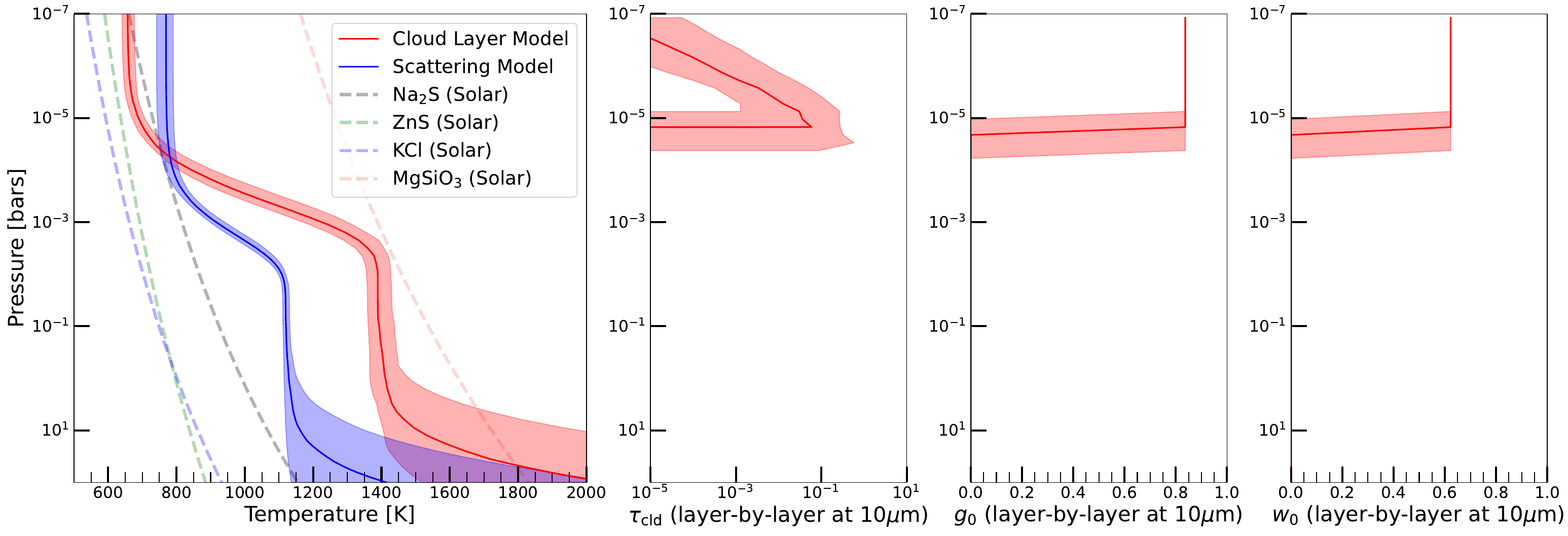}{0.99\textwidth}{}}
\gridline{\fig{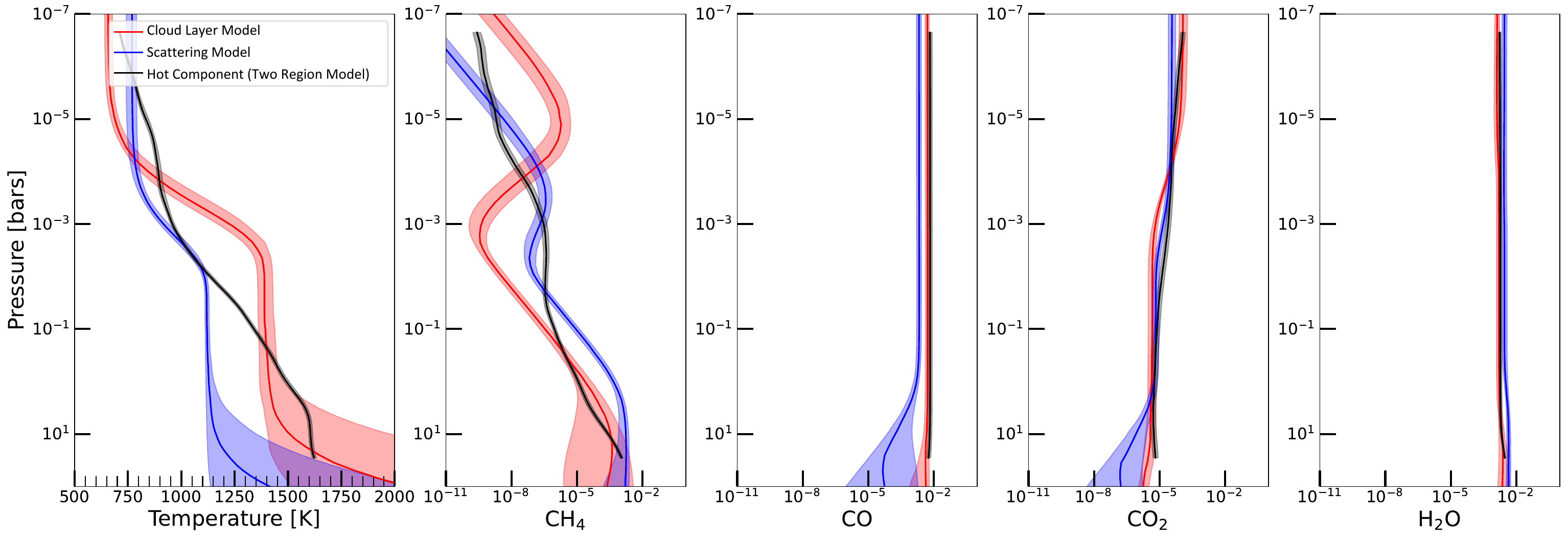}{0.99\textwidth}{}
          }
\caption{Temperature-Pressure Profile (upper left) and cloud properties for the silicate cloud model \label{fig:retrievedProfiles}.
The TP profile for the Cloud Layer model crosses the condensation curve for Na$_2$S (dashed gray line) at very high altitudes and MgSiO$_3$ (dashed pink line) at very deep altitudes, while the Scattering model crosses neither of these cloud candidate's condensation curves.
While the Cloud Layer crosses these two condensation curves, Na$_2$S has inefficient nucleation rates from theoretical models \citep{gao2018microphysicsGJ1214clouds} and MgSiO$_3$ would require extreme verticle mixing to be lofted to the high altitudes needed to match the observed brightness spectrum.
The upper right 3 panels show the optical properties of the retrieved silicate clouds: optical depth $\tau_{cld}$, asymmetry $g_0$ and single scattering albedo $w_0$ at a reference wavelength of 10~\micron.}
\end{figure*}

The fit to the brightness temperature spectra obtained with this setup is shown in red in Figure \ref{fig:wasp69TbSpec}.
The posterior distributions of the parameters are shown in Figure \ref{fig:picasoPosteriors}.
This setup estimates the planet's atmospheric metallicity to be +0.96$\pm$0.2 above Solar and the absolute C/O ratio to be 0.75$\pm$0.1, as seen in Figure \ref{fig:picasoPosteriors}.
Thus, the compositional constraints depend on the assumptions about aerosols and the level of reflection from WASP-69 b.
As with the Scattering retrieval described in Section \ref{sec:scatRetrieval}, there is a peak in the SO$_2$ posterior, but it is at very low abundances where the 7.7~\micron\ feature does not show up significantly in the spectrum.
Our constraints on the layer-by-layer cloud optical depth, asymmetry parameter, and single scattering albedo at a wavelength of 10~\micron\ along with the constrained $T(P)$ profile is shown in Figure \ref{fig:retrievedProfiles}.
The retrieval prefers a silicate cloud deck near $\sim$ $10^{-4.5}$ bar pressure level extending up to $10^{-6}$ bar with a peak cloud optical depth close to 0.1.
The retrieved particle size is very small with a median radius of 10$^{-6}$ cm.

Figure \ref{fig:retrievedProfiles} shows the retrieved temperature-pressure profiles for the Cloud Layer model.
The temperature-pressure profile crosses the saturation pressure curves for Na$_2$S and MgSiO$_3$ (enstatite).
This indicates that Na$_2$S can form a cloud at a similar altitude as the inferred cloud deck base $10^{-4.5}$ bar.
However, Na$_2$S does not have the right optical properties to work in our Cloud Layer model, which requires a high absorption coefficient at MIRI wavelengths ($5$ to $12$\micron) as compared to short wavelengths.
MgSiO$_3$, on the other hand, has the right optical properties \citep[e.g.][]{taylor2021thermalScattering} to explain the low ($\sim$930 K) brightness temperatures at the MIRI wavelengths without significantly absorbing the NIRCam wavelengths $2$ to $5$ \micron.
However, our retrieved TP profiles for even the warmest Cloud Layer model does not cross the MgSiO$_3$ condensation curve shown in Figure \ref{fig:retrievedProfiles}, except at very low altitudes (10 bars or deeper), so MgSiO$_3$ particles would likely rain down from observable pressures.
The enstatite particles would need to be lofted to the retrieved 10$^{-4.5}$ to $10^{-6}$ bar pressure level from below 10 bars.
This would require an extreme level of vertical mixing (such as inferred for WASP-107 b to loft high altitude silicate cloud particles \citep{dyrek2024silicatesWASP107}).

\begin{figure*}
\gridline{\fig{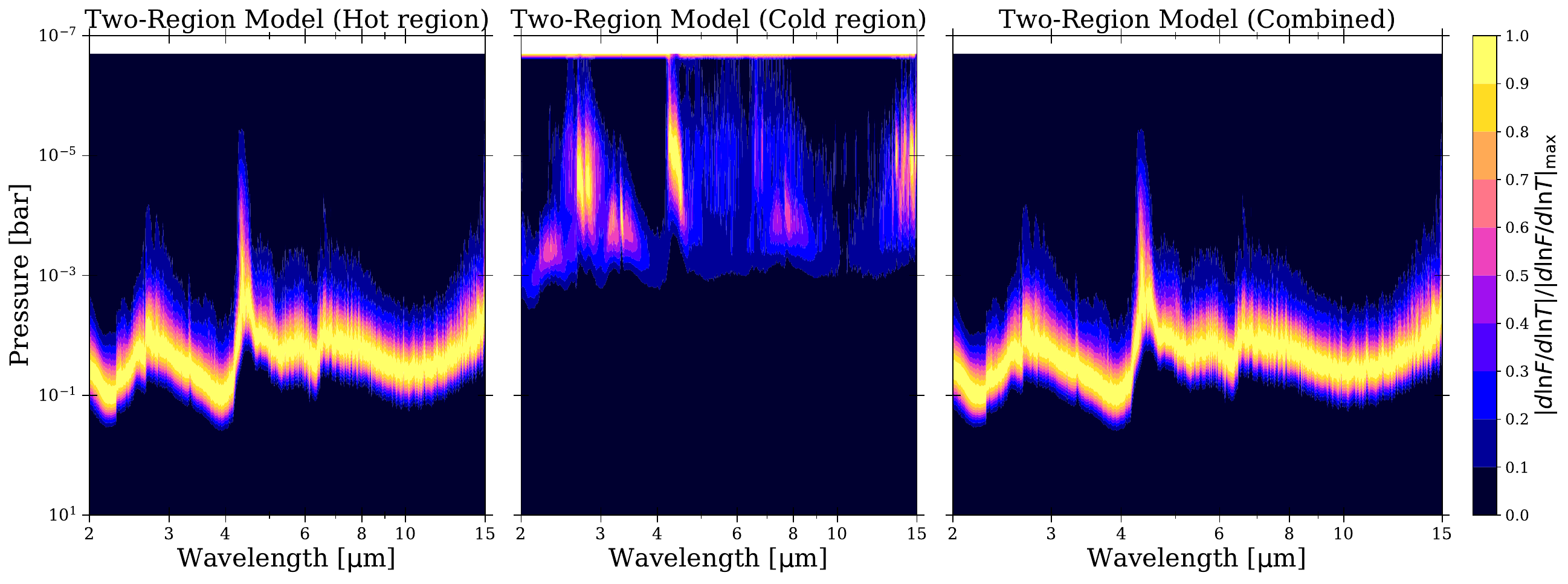}{0.99\textwidth}{}}
\gridline{\fig{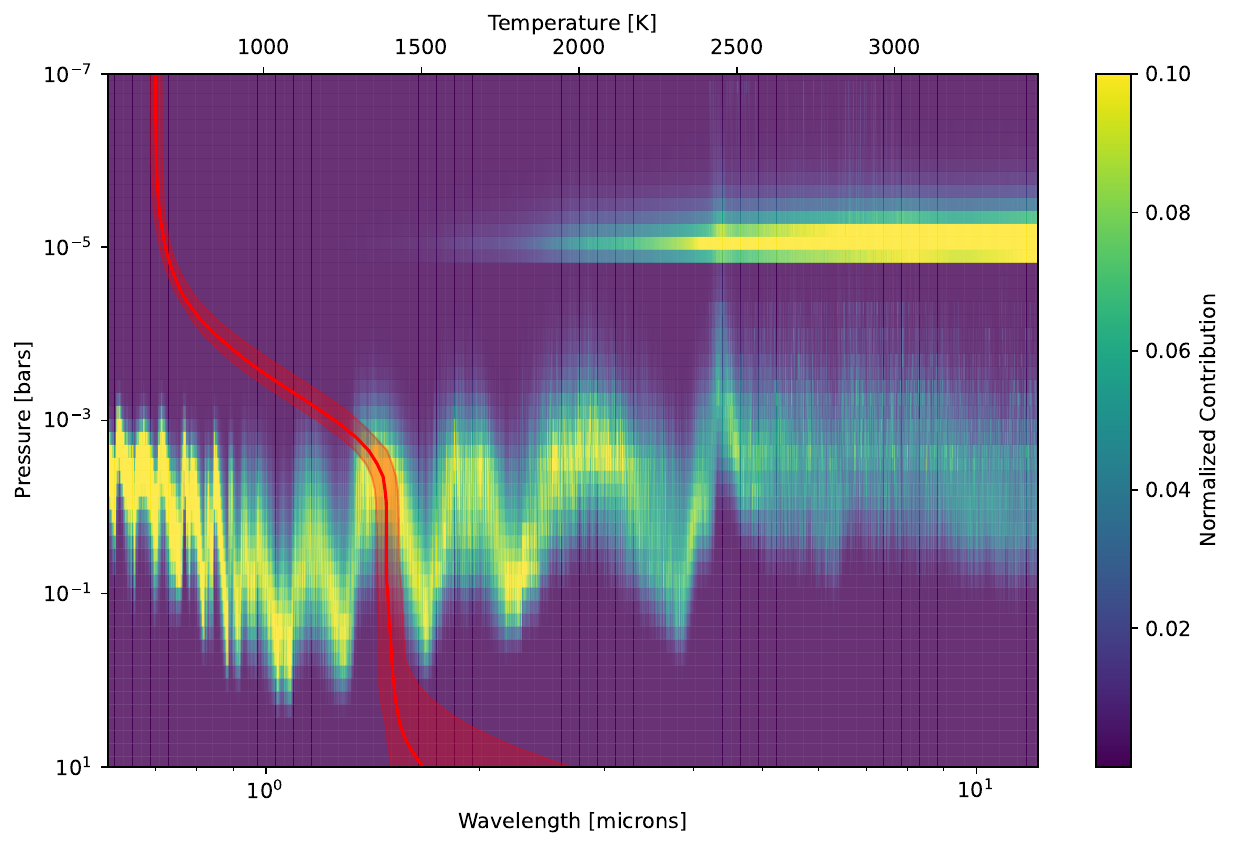}{0.49\textwidth}{Cloud Layer Model}}

\caption{Contribution functions for the Two Region model (top row) and the Cloud Layer model (bottom plot).
The individual regions' contributions are shown for the Two Region model (top left two plots) as well as the combined full dayside (right plot).
A high altitude ($10^{-4}$ to $10^{-5}$ bar) cloud dominate the longer wavelength emission (with smaller brightness temperatures) while the warmer lower layers dominate at short wavelengths (with larger brightness temperatures) in both models.
The posterior retrieved Temperature-Pressure Profile for the CHIMERA model (bottom middle) is shown as a red curve with the temperature at the top axis.
\label{fig:contributionFunctions}}
\end{figure*}

The normalized contribution per wavelength is plotted in Figure \ref{fig:contributionFunctions}, which shows that the cloud dominates the emission at longer wavelengths (thus explaining the lower brightness temperature) while gas near 10$^{-1.5}$ bar dominates the emission at short wavelengths (thus explaining the higher brightness temperature).
MgSiO$_3$ clouds produce this wavelength dependence because the absorption coefficient of these silicates is much larger at wavelengths longer than 4~\micron\ than at short wavelengths \citep[e.g.][]{taylor2021thermalScattering}.
The cloud emission at long wavelengths has a lower brightness temperature because it is emitted by a cooler upper layer.
The gas emission at short wavelengths has a higher brightness temperatures because it comes from a warmer layer below the cloud deck.
We also include reflected light from Rayleigh scattering by molecular gas and silicate clouds, but their Albedo is negligible ($\sim$10 ppm or A$_g \approx 0.08$) compared to the thermal component or the high geometric Albedo inferred from the Scattering Model (A$_g \approx $\planetGAlbedo) described in Section \ref{sec:scatteringCloudy}.

\subsubsection{Summary of Model Retrieval Results}\label{sec:retrievalSummary}

\begin{deluxetable*}{l|llll|cccc}[t!]
\tablecaption{Summary of retrieved atmospheric metallicity and C/O ratio from each model setup. The most plausible scenarios are the Two-Region (Cloudy) and Cloud Layer retrievals\label{tab:model_summary}}
\tablecolumns{4}
\tabletypesize{\scriptsize}
\tablewidth{0pt}
\tablehead{
\colhead{Model} &
\colhead{Retrieval Code} &
\colhead{Dayside Geometry} &
\colhead{TP Profile} &
\colhead{Reflected Light} &
\colhead{Metallicity [M/H]} &
\colhead{C/O ratio} &
\colhead{\redChisq} &
\colhead{log$Z$}
}
\startdata
One-Region (Clear) & CHIMERA & Homogeneous & RCE Grid & Molecular & $1.30^{+0.02}_{-0.02}$ &  $0.11^{+0.02}_{-0.01}$ & 14.5 & $-588$\\
One-Region (Cloudy) & CHIMERA & Homogeneous & RCE Grid & Molecular & $0.96^{+0.03}_{-0.04}$& $0.65^{+0.03}_{-0.03}$ &3.07& $-439$\\
Two-Region (Cloudy) & CHIMERA & 2 TP & RCE Grid & Molecular & $1.01^{+0.08}_{-0.06}$ &$0.79^{+0.04}_{-0.04}$ &1.79 &  $-255$ \\
Scattering & PICASO & Diluted & \citet{line2013chimera} & Constant A$_G$ & 0.75$^{+0.15}_{-0.15}$ & 0.42$^{+0.16}_{-0.16}$ & 2.00 & $-253$ \\
Cloud Layer & PICASO & Diluted & \citet{line2013chimera} & Silicate & 0.96$^{+0.20}_{-0.17}$ & 0.75$^{+0.19}_{-0.10}$ & 1.65 & $-256$\\
\enddata
\end{deluxetable*}

The 5 models considered in this work are summarized in Table \ref{tab:model_summary}, with the major assumptions, derived abundances, reduced chi-squared (\redChisq) and Bayesian Evidence (log(Z)).
The 1D homogeneous models that were originally explored in Section \ref{sec:one-region} under-predict the short wavelength planet flux and over-predicts the long wavelength planet flux thus giving large \redChisq\ values.
The difference in log Bayesian evidence between the 1D Cloudy Homogeneous model and the Two-Region model is 184, corresponding to a 19~$\sigma$ difference.
The three models that best fit the data (with a \redChisq\ of 2.0 or less) are the the Scattering Model, the Two-Region Model and the Cloud Layer Model with \redChisq\ of 2.0, 1.79 and 1.65 respectively.
The Bayesian evidence favors the Scattering model, but the extremely high Geometric Albedo of \planetGAlbedo\ is likely unrealistic.
The Two-Region model and Cloud Layer models, by contrast have negligible stellar reflected light, as shown in Figure \ref{fig:twoRegionSpectrum}.
While we regard the Scattering Model as less plausible, it is still consistent with existing JWST data and can be used to assess the robustness of other conclusions like the composition of WASP-69 b's atmosphere inferred from the Cloud Layer and Two-Region Models.
We further discuss the synthesized inferences about WASP-69 b's cloud properties and atmospheric composition across all models in Section \ref{sec:discussion}.

\section{Eclipse Mapping}\label{sec:eclipseMapping}

\begin{figure*}
\gridline{
\fig{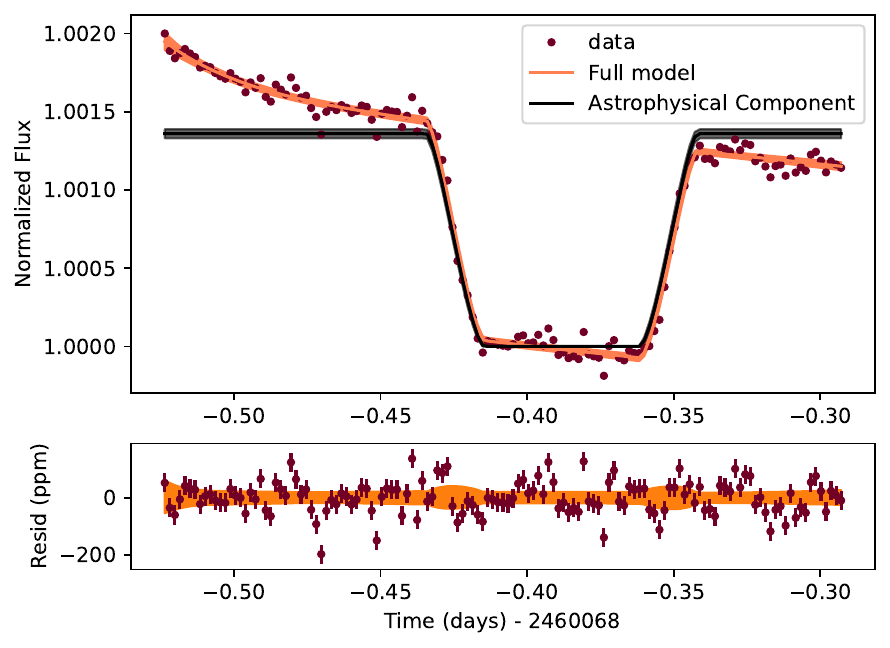}{0.49\textwidth}{Lightcurves for Uniform Map}
\fig{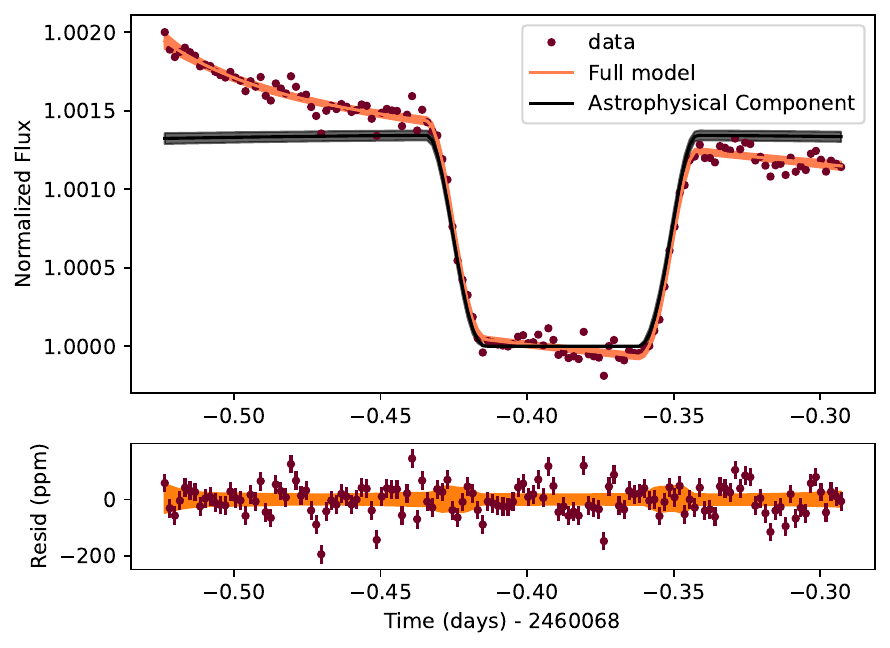}{0.49\textwidth}{Lightcurves Non-Uniform Map}
          }
\gridline{
\fig{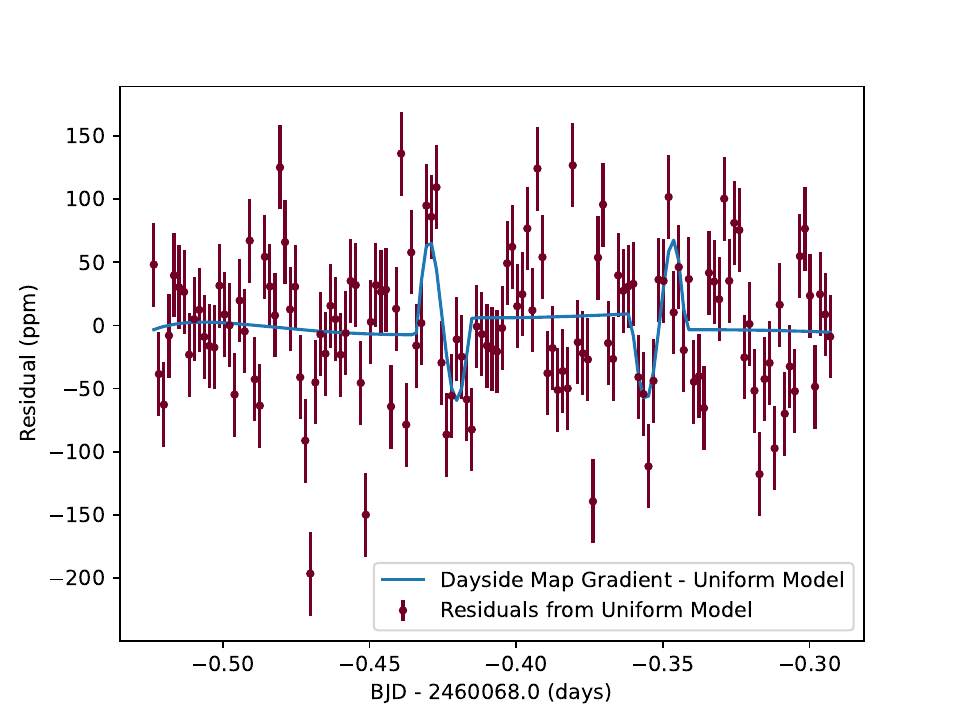}{0.49\textwidth}{Residuals and Model Differences}}
\caption{MIRI LRS broadband lightcurves and Residuals for Lightcurve Fits to the Uniform Map as well as the model difference between non-uniform map and the uniform map.
The non-uniform map allows for a gradient from the sub-stellar point to the terminator through the Y$_{2,0}$ and Y$_{1,0}$ spherical harmonics.
The Bayesian Information Criterion favors the non-uniform map (BIC=184.1 versus BIC=191.1), but higher signal to noise is needed to robustly map WASP-69 b to high confidence \label{fig:eclipseMapLC}.}
\end{figure*}

\begin{figure*}
\gridline{
\fig{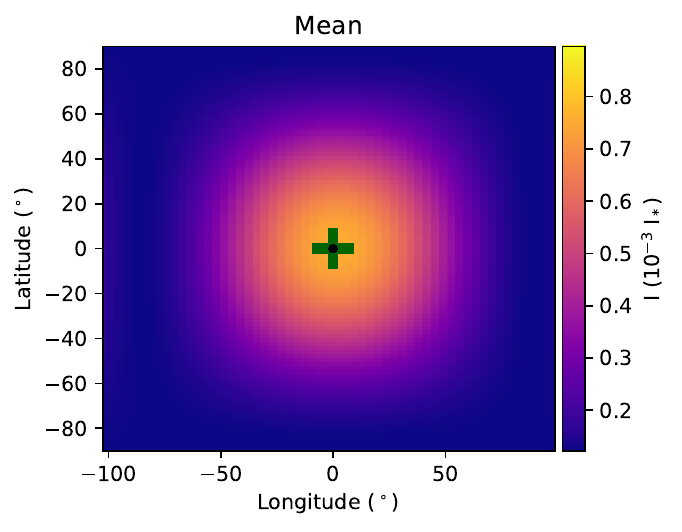}{0.49\textwidth}{}
\fig{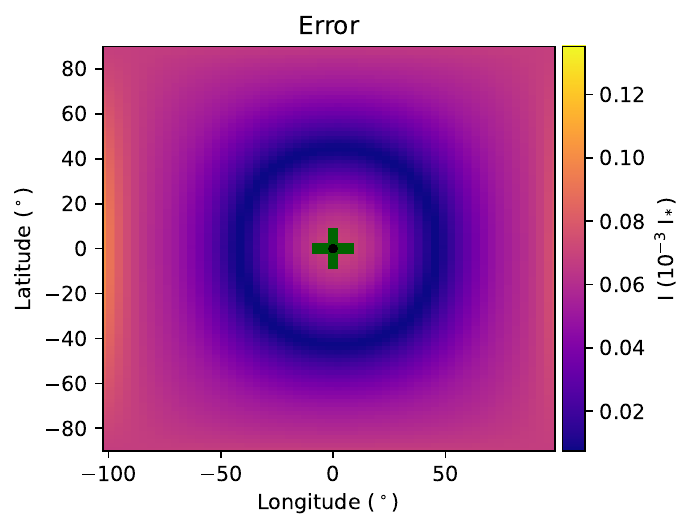}{0.49\textwidth}{}
          }
\caption{Posterior maps for the non-uniform map fits to WASP-69 b's dayside. The maps only include the Y$_{2,0}$ and Y$_{1,0}$ spherical harmonics to sense temperature gradients, but do not reveal location information.
The mean map ({\it Left}) and Error ({\it Right}) favor a dayside brightness distribution that is centrally concentrated and inefficient at redistributing heat.
The plus sign shows the peak brightness, which is forced to be at the substellar point by the limited number of spherical harmonics. \label{fig:eclipseMapStats}.}
\end{figure*}

Motivated by the inhomogeneous dayside inferred by both the Two-Region retrieval and the Scattering Model's dilution parameter, we investigated whether the lightcurves show further and independent evidence of an inhomogeneous dayside.
Eclipse mapping \citep[e.g.][]{williams2006eclipseMapping,rauscher2009eclipseMapping,deWit2012eclipsemap189,majeau2012eclipsemap189,coulombe2023broadbandThermalWASP18b}, can use the star as a spatial scanner to map some non-uniform features of the planet.
We used the broadband MIRI lightcurve, which had the highest signal-to-noise eclipse signal to map the dayside of the planet.

First, it is necessary to use the highest available precision orbital parameters to decrease the correlations of these parameters with maps.
This is achieved by fitting the publicly available TESS lightcurve from sector 55 taken between 2022-08-05 and 2022-09-01.
The posterior parameters we find are listed in Table \ref{tab:props}.

We model the MIRI LRS lightcuve a with \texttt{starry} \citep{luger2019starry} and include a systematic trend for the lightcurves.
The lightcurve is modeled as
\begin{equation}\label{eq:eclipseMapModel}
F(t) = F_a(t) \left( 1 + C e^{-(t-t_{\rm{min}})/\tau} \right) (A + B x),
\end{equation}
where $F_a(t)$ is the astrophysical variation modeled by \texttt{starry}, $t$ is the time, $t_{\rm{min}}$ is the minimum (ie. start) time, $\tau$ is an exponential time constant, $C$ is the exponential amplitude constant and $A$ and $B$ are polynomial baseline terms.
$x$ is the scaled time
\begin{equation}
x= 2 (t - t_{mid})/(t_{max} - t_{min}),
\end{equation}
where t$_{mid}$ is the mid-time.

We fit the broadband MIRI LRS lightcurve with a uniform map, a spherical harmonic degree 1 map and a spherical harmonic degree 2 map and a fourth fit with the spherical harmonics corresponding to dayside brightness gradients only.
We sample the posterior with No-U-Turns sampling with \texttt{pymc3} \citep{salvatier2016pymc3} and the \texttt{pymc3-ext} tools from \citet{foreman-mackey2021exoplanetJOSS}.
All of our model fits results show that there is excess noise as compared to the photon and read noise estimate.
The best-fit (maximum a-priori) spherical harmonic degree 2 lightcurve model has a \redChisq\ = 1.05, which is smaller than the Uniform model, which has \redChisq\ = 1.11, where \redChisq\ is the reduced-chi-squared metric.
We included an error inflation factor that was fit as a hyper-parameter for the uniform fit and then that same uncertainty is used in cross-model comparisons.
We find that the theoretical photon and read noise error should be 33 ppm, whereas the inflated error is 56 ppm. The ratio of the inflated to theoretical error is on the higher end as compared with other MIRI LRS lightcurves, but less than the commissioning target L169-9 b \citep{bouwman2023specPerformanceMIRI}.=
We also test whether the extra degrees of freedom in the spherical harmonic degree 2 lightcurve are justified with the Bayesian Information Criterion (BIC) and find that it is 214.5 for the spherical harmonic degree 2 versus 191.1 for a uniform map, which favors uniform map due to its 8 fewer free parameters.
We repeated an eclipse mapping fit but with only two free parameters in the spherical harmonic map besides the overall amplitude: Y$_{2,0}$, Y$_{1,0}$, which create a gradient from the sub-stellar point to the terminator - see \citet{schlawin2023eclipseMappingLongTermDrifts} Figure 5 or \citet{luger2019starry} Figure 1, for example.
This time the non-uniform map has \redChisq\ = 0.99 and a BIC=184.0, so both statistics favor a non-uniform map with dayside temperature gradients.
The spherical harmonic degree 1 map, on the other hand, is not statistically preferred over the Uniform map.

To better visualize the residuals, we also show the difference between the two best-fit models and the residuals for a uniform eclipse map fit in Figure \ref{fig:eclipseMapLC}, bottom.
The lightcurve for the non-Uniform explains some of the 50-100 ppm deviations in the residuals at ingress and egress.
While the evidence for a non-Uniform map is still at a similar level as the excess noise in the data, the MIRI LRS broadband lightcurve supports the possibility that a non-Uniform dayside model may explain the spectrum.

Figure \ref{fig:eclipseMapStats} shows the posterior map distributions for the spherical harmonic degree 2 fits to the lightcurves.
The maps suggest significant day-to-terminator contrasts on the planet.
We also analyzed the lightcurve independently with the eigenmapping method \citep{rauscher2018moreInformativeMapping} using the same systematic model as described in Equation \ref{eq:eclipseMapModel} using the \texttt{ThERESA} mapping code in its single-wavelength implementation \citep{challener2021ThERESA}.
The BIC preferred 2 eigenmap components up to 2nd order spherical harmonics.
The \texttt{ThERESA} maps also significantly favored a non-Uniform dayside map over a uniform one by $\Delta$BIC of 53.6.
The excess noise in the lightcurves beyond photon and read noise means that caution is warranted when interpreting the maps.
Transit observations (planned for JWST Cycle 2 GO program 3712 and Cycle 3 GO program 5924) and a full phase curve will better measure if this high contrast is indeed needed for the planet because the transmission spectrum will constrain the properties around the terminator (ie longitudes of $\pm$ 90 degrees).
The black points in Figure \ref{fig:eclipseMapStats} are the locations of peak brightness for random sample posterior draws from the \texttt{pymc3} No U-Turns sampling of the lightcurve.
The location of peak brightness is consistent with the substellar point (5$\pm 14 \degree$,-5$ \pm 24 \degree$).
The large variance in the peak brightness longitude is due in part to the baseline we fit to the data in Equation \ref{eq:eclipseMapModel}.
Non-flat baselines can decrease the precision of the m=1 spherical harmonic terms, which constrain the peak brightness longitude \citep{schlawin2023eclipseMappingLongTermDrifts}.
The ThERESA analysis of the same lightcurve indicates an eastward hotspot shift of $\sim20\degree$ with less uncertainty in the ThERESA maps due to the fewer number of mapping terms and free parameters, but an improved orbital ephemeris is needed to better constrain the hotspot shift.
This ephemeris will be improved with the upcoming JWST transit observations of WASP-69 b.
As with confirming the central concentration of brightness from the eclipse maps, a full phase curve would better constrain the longitude of WASP-69 b's hotspot.

\section{Discussion}\label{sec:discussion}
\subsection{Energy Balance}\label{sec:energyBalance}
\begin{figure*}
\gridline{\fig{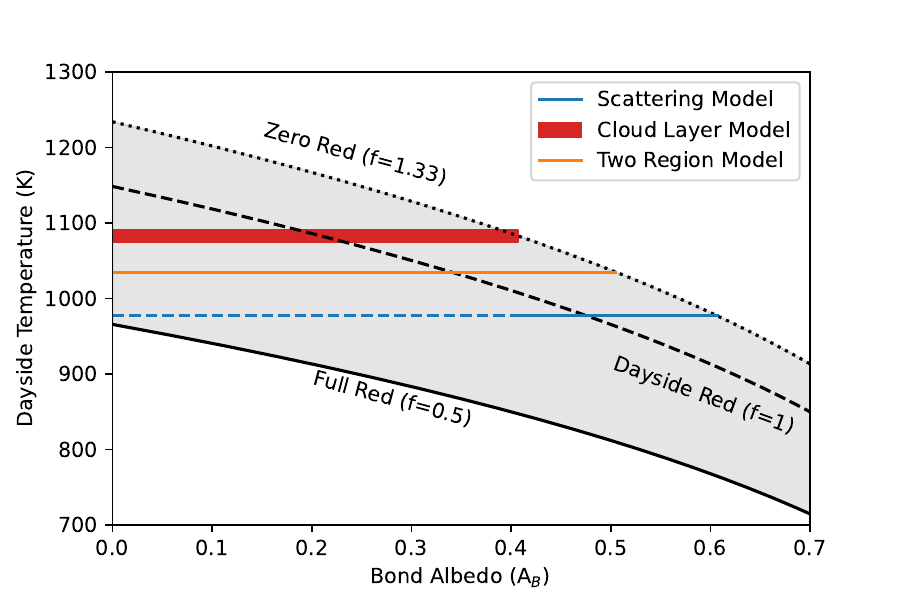}{0.5\textwidth}{}
          }
\caption{The dayside temperature from our JWST spectrum indicates that heat is not fully re-circulated around the planet.
The allowed ranges of dayside temperature from full redistribution of heat around the planet (solid line)
 to zero redistribution of heat (dotted line) for a given Bond Albedo are bounded within the gray region.
The integrated thermal flux from the planet extrapolated to 0~\micron\ and $\infty~\micron$ is calculated for our better-fitting models (thick red line, an orange line and and thin blue line).
The high geometric albedo of the Scattering model (blue) which pushes against the prior of 0.67 strongly indicates a high Bond albedo so we show values below 0.4 with a dashed line.
The Scattering Model barely conserves energy at the highest Geometric Albedo and would result in a very extreme dayside-to-nightside temperature contrast.\label{fig:energyBalance}.}
\end{figure*}

The total planet emission as measured over the NIRCam and MIRI wavelength ranges can be used to estimate the dayside temperature, which constrains the Bond albedo and the heat recirculation efficiency of the planet.
For the energy budget, we assume that internal heat flux is negligible for this circularized hot Jupiter planet, which is also supported by our internal temperature $T_{\rm int}$ of less than 200 K shown in Figure \ref{fig:picasoPosteriors}.
We calculate the dayside effective temperature of the planet for the Scattering model, Cloud Layer Model and Two-Region models, which were the best fits in terms of \redChisq.
We integrate the thermal component of a best-fit model spectrum from 0 to $\infty$, which requires extrapolation beyond our measured wavelengths from 2.0~\micron\ to 12~\micron.
We then find the temperature of a spherical homogeneous blackbody radiator the size of the planet that has the same flux.

This dayside effective temperature can be compared to the incoming radiation from the host star to derive a Bond albedo.
However, the day/night heat recirculation efficiency cannot be measured with only an eclipse lightcurve of WASP-69~b.
A full phase curve would be needed to measure the global effective temperature, as has been done for GJ 1214 b \citep{kempton2023reflectiveMetalRichGJ1214}.
We therefore show the combined constraints on both the Bond albedo ($A_B$) and heat redistribution efficiency in Figure \ref{fig:energyBalance}.
We mark the the allowed dayside temperatures in the gray region in Figure \ref{fig:energyBalance}, which are bounded by full heat redistribution ($f=0.5$ or $\varepsilon=1.0$ in the parameterization of \citet{cowan2011statisticsOfAlbedoAndRecirculation}) and no heat redistribution ($f=\frac{4}{3}$ or $\varepsilon = 0$).

We examine the constraints on the redistribution efficiency, model-by-model from their dayside temperature.
The Scattering Model spans a wide range of possible redistribituion efficiencies that depend on the Bond Albedo.
However, the high constant-wavelength Geometric Albedo inferred from the Scattering model \planetGAlbedo\ or higher indicates that the Bond Albedo is also likely large.
For a Lambertian sphere that scatters equally at all wavelengths, the Geometric Albedo is $\frac{2}{3}$ times the Bond Albedo \citep{heng2021geometricAlbedos}.
In all four of the Solar System's giant planets, the infrared geometric albedo is less than the Bond Albedo \citep{dePater2001PlanetaryScienceBook}.
If the Bond albedo of WASP-69 b is indeed larger than \planetGAlbedo\ (ie the right hand side of the blue line in Figure \ref{fig:energyBalance}, the redistribution efficiency would have to be near $f=1.33$ (ie. no redistribution of heat).
However, it is possible (with a preference for back-scattering starlight), that the geometric Albedo at some wavelengths exceeds the Bond albedo.
In the Solar System giant planets, the visual Geometric Albedo is on average $\sim\frac{2}{3}$ times the Bond Albedo.
Therefore, we mark all Bond Albedos below 0.4 with a dashed line for the Scattering model.
Given the likely high Bond Albedo for the planet, we conclude that the Scattering Model's dayside temperature implies an inefficient heat redistribution for the planet (with $f > 0.87$ or $\varepsilon < 0.56$ in the parameterization of \citet{cowan2011statisticsOfAlbedoAndRecirculation}).
The Bond Albedo can be much smaller (near 0.0) for the Two Region and Cloud Layer models shown in Figure \ref{fig:energyBalance} (ie the left hand side of the Figure).
However, the dayside temperatures inferred from these models are significantly warmer, so even for a Bond Albedo of 0.05, the heat redistribution must be inefficient  ($f > 0.7$ or equivalently $\varepsilon < 0.76$ ) for the Two Region Model and ($f > 0.8$ or equivalently $\varepsilon < 0.64$) for the Cloud Layer Model.
Thus, in all of the models we fit to the dayside spectrum of WASP-69 b, full redistribution of heat is ruled out and some significant day-to-night temperatures are expected for the planet.

Dynamical models can help inform whether inefficient heat redistribution and high Bond Albedo (thinner blue line in Figure \ref{fig:energyBalance}) are plausible for WASP-69 b.
In models of zero albedo equilibrium temperatures of 1061 K with no clouds or hazes, circulation is efficient with temperature contrasts of less than 100 K between the day and night sides at 1 mbar pressures, but clouds can significantly enhance the day-to-night temperature contrast at the same pressure to 500 K as well as alter the photosphere altitude \citep{roman2021clouds3DthermalStructures}.
Therefore, depending on the clouds in an atmosphere, there can be a large variety of day-to-night contrasts and recirculation efficiencies possible for a planet.
The phase curve amplitudes of similar temperature gas giant planets in short orbits measured by the Spitzer space telescope have also shown a wide variety of longitudinal brightness distributions are possible at 1100-1500 K equilibrium temperatures, potentially influenced by clouds, with some cooler planets like WASP-34 b showing significant day/night contrasts \citep{may2022spitzerPhaseCurves}.
WASP-43 b also shows evidence for nightside clouds and a day-to-night brightness temperature that varies between 1520 and 860 K for the two hemispheres \citep{bell2024nightsideCloudsDisEqChemWASP43b}.
This is similar for our emerging picture of WASP-69 b, that there are significant aerosols in the atmosphere that can enhance the day-to-night contrast of the planet.
Furthermore, the inferred dayside inhomogeneity from the Two Region model (Section \ref{sec:twoRegRetrieval}), dilution factor in the Scattering model (Section \ref{sec:scatRetrieval}) and tentative results of eclipse mapping (Section \ref{sec:eclipseMapping}), all point to a central concentration of high temperature gas as compared to the limbs of the planet, which would be expected for inefficient heat redistribution.

\subsection{WASP-69 b's Atmospheric Composition}\label{sec:compositionSummary}

All of the models explored in this work give posterior distributions that are enriched in heavy elements as compared to solar composition.
This largely comes from the constraints on the dominant molecular absorbers in the observed 2~\micron\ to 12~\micron\ region: H$_2$O, CO and CO$_2$.
Even the lowest metallicity retrieval from the Scattering model has [M/H]= 0.75$\pm$ 0.15.
The two models which give the lowest \redChisq\ and (the Cloud Layer and Two-Region models) give consistent compositional constraints within 1$\sigma$, in addition to showing similar cloud effects described in Section \ref{sec:cloudProperties}.
Despite different assumptions about the 3D geometry of the planet and the temperature-pressure profile(s), the Cloud Layer and Two-Region retrievals overlap well in inferred chemical enrichment over solar of [M/H]=0.96$^{+0.20}_{-0.17}$ and [M/H]=1.01$^{+0.08}_{-0.06}$, respectively.
Similarly, the posterior C/O ratios overlap with values of 0.75$^{+0.19}_{-0.10}$ and 0.79$^{+0.04}_{-0.04}$ respectively.
The Cloud Layer model has broader compositional constraints because of the more flexible temperature-pressure profile whereas the Two-Region model uses fixed profiles based on radiative-convective equilibrium calculations.
We did not include radiative feedback from the clouds in the Two-Region model so the very small standard deviations in the posterior metallicity and C/O ratio likely under-estimate model uncertainty.
Therefore, we recommend adopting a metallicity of \planetMetallicity\ and a C/O ratio of \planetCtoO\ (from the Cloud Layer model) for 1$\sigma$ (68\%) posterior intervals for modeling and comparative planetology.

We also note that the Scattering Model is not completely ruled out (and has the largest log(Z) or Bayesian evidence in Table \ref{tab:model_summary}) so WASP-69 b's composition may be less metal-enriched than the Two-Region and Cloud Layer retrievals indicate.
While we consider this a less likely scenario, it is worth considering the possibility that WASP-69 b's atmosphere has a metallicity of  \planetMetallicityScattering\ solar and a C/O ratio of \planetCtoOScattering\ for 1$\sigma$ (68\%) posterior intervals.
The Scattering Model can be ruled out or confirmed by measuring the optical reflection from WASP-69~b, which diverges strongly from the Cloud Layer and Two-Region models below wavelengths of 2~\micron, as shown in Figure \ref{fig:wasp69TbSpec}.
CHEOPS (archived), WFC3 UVIS (planned) or JWST NIRISS SOSS lightcurves of WASP-69 b can help answer this question at short wavelengths.
Until those data are analyzed and published, we carry forward these two possible compositions for WASP-69 b.

WASP-69 b's host star also shows enrichment relative to solar composition with [Fe/H]=0.144$\pm$0.077 \citep{anderson2014wasp69wasp84wasp70Ab} and another determination of [Fe/H]= 0.29 $\pm$ 0.04 \citep{sousa2021sweetcat}.
Machine-learning characterization of the spectrum indicates [Fe/H]=0.41 $\pm$ 0.03, [C/H] = 0.27 $\pm$ 0.05 and [O/H]=0.18 $\pm$ 0.07 \citep{polanski2022chemAbund25JWSThostStars}.
If indeed WASP-69 A has a super-solar C/O ratio, this could explain the slightly super-solar C/O ratio for WASP-69 b using the Cloud Layer and Two Region models.
More characterization of the star should be performed to better constrain its metallicity and heavy element enrichment.
If the star is super-solar in heavy element enrichment of [Fe/H]=0.29 $\pm$ 0.04 from \citet{sousa2021sweetcat}, WASP-69 b may only be $\sim$0.7 dex (ie. 5$\times$) above stellar composition.
This would place WASP-69 b more similar in chemical enrichment to Jupiter ($\sim$5$\times$) than Saturn ($\sim$10$\times$) relative to the Sun \citep{atreya2022saturn}.

The brightness temperature spectrum plotted in Figure \ref{fig:wasp69TbSpec} does not show any obvious signs of a 7.7~\micron\ SO$_2$ feature, that can be very prominent due to photochemical production of SO$_2$ \citep{dyrek2024silicatesWASP107}.
The Two Region model fits the data well in terms of both Bayesian Evidence and \redChisq\ and this model assumes chemical equilibrium and thus essentially no SO$_2$.
The Cloud Layer retrieval has a long tail in the posterior mixing ratio down 10$^{-10}$ and the Cloud Layer model retrieves a mixing ratio of 10$^{-8}$, where the feature size becomes small.
Thus, we do not see the significant abundances of SO$_2$ inferred in the hot Jupiter WASP-39 b of 10$^{-5.2}$ to 10$^{-5.7}$ \citep{rustamkulov2022nirspecPrism,alderson2022wasp395bG395H}.
One reason could be the moderately high C/O ratio of \planetCtoO\ inferred for WASP-69 b in the Cloud Layer and Two Region models, which decreases the expected SO$_2$ abundances from photochemical models \citep{tsai2022photochemistrySO2}.

Our observations are potentially sensitive to absorption features from CH$_4$ near 3.3~\micron\ and 7.8~\micron, but there are no such features in the emission spectrum of WASP-69 b, as shown in Figure \ref{fig:wasp69TbSpec}.
This is well modeled in our Scattering, Cloud Layer and Two-Region models, which assume chemical equilibrium.
At the high inferred temperatures of these models ($\gtrsim$1000 K at 10$^{-3}$ bar), the CH$_4$ abundance is low enough that H$_2$O absorption dominates and obscures any CH$_4$ features.
Note in Figure \ref{fig:wasp69TbSpec} (second panel from the bottom) that CH$_4$ doesn't become optically thick until pressures deeper than 10$^{-1}$ bar, which is deeper than where H$_2$O becomes optically thick.
In cooler planets, such as WASP-80 b (850 K), methane presents a strong feature that was observed in absorption \citep{bell2023_methane}.
If the terminator and nightside of WASP-69 b is significantly cooler than the dayside, as inferred by our retrievals, CH$_4$ will become a more significant absorber in WASP-69 b's transmission spectrum.
Our inferred moderately large C/O ratio of \planetCtoO\ from the Cloud Layer and Two Region models would significantly enhance the CH$_4$ abundance in chemical equilibrium.
Continued analysis of existing spectra \citep{guilluy2022fiveMoleculesWASP69b,khalafinejad2021wasp69b_lowResHighRes} and future JWST measurements (e.g. GO programs 3712 and 5924) will provide better insight into the chemical processes shaping WASP-69 b's atmosphere and CH$_4$, now that the dayside abundances have been constrained in this work.

One other consideration in our abundance inferences from atmospheric retrievals is that there may be systematic offsets in the JWST emission spectra.
We do not expect significant non-linearity effects in our data because the difference between the in-eclipse and the out-of-eclipse (0.1\% to 0.25\% in this case for MIRI, 0.02 to 0.08\% for NIRCam) do not appreciably change the detector well filling factor.
However, we did experiment with a retrieval that allows for the MIRI LRS data spectrum to have an offset relative to the NIRCam data.
We used the Cloud Layer model with an MIRI offset with an allowed range of -500 ppm to +500 ppm.
This retrieval results in a bimodal posterior distribution with offsets of -80 ppm and +40 ppm that have significantly different abundances.
The -80 ppm offset resulted in a similar metallicity as the Cloud Layer, Scattering and Two Region models ([M/H] $\approx$ 1.0, with a moderately lower C/O ratio (C/O$\approx 0.6$).
The higher MIRI offset resulted in a very extended cloud and an order magnitude higher metallicity.
Therefore, if our MIRI data have a systematic offset at all wavelengths, the inferred composition of the planet can change.
Fortunately, we do not find significant evidence for offsets in our independent data reductions shown in Figure \ref{fig:wasp69EclipseSpec}, with a mean MIRI LRS offset between \texttt{Eureka!} and \texttt{tshirt} of 2 ppm, despite larger differences at individual data points.

\subsection{Aerosol Properties}\label{sec:cloudProperties}

Both the Cloud Layer and Two-Region Models show evidence for clouds that most significantly affect the long wavelength radiation, as seen in Figure \ref{fig:contributionFunctions}.
The Cloud Layer model has a combination of this infrared-absorbing cloud deck located at 10$^{-4.5}$ to 10$^{-6}$ bar and gaseous deeper emission from 10$^{-1}$ to 10$^{-3}$ bar with a homogeneous (but slightly diluted) 1D model.
The Two-Region model instead fits the data with separate temperature-pressure profiles and gray scattering clouds so that the hot region is dominated by the molecular features near 10$^{-1}$ to 10$^{-3}$ bars, and cloud scattering suppresses the flux at 4.0~\micron\ and 7 to 9~\micron.
The cold region of the Two-Region model has negligible emission because of the combination of cold temperature and thermal scattering from clouds \citep[e.g.][]{taylor2021thermalScattering} that strongly suppresses the planet emission.
In either case, the averaged dayside properties are best explained with the inclusion of clouds, and the clouds help reduce the flux at long wavelengths.
The next best model in terms of \redChisq\, the Scattering model, also would require aerosols to produce the significant reflection inferred by the geometric albedo parameter A$_G \approx$\planetGAlbedo.
The Raleigh scattering of molecules only provides a geometric albedo of A$_G \lesssim 0.1$, so this would require reflective aerosols in the atmosphere of WASP-69 b to achieve the inferred geometric albedo.
Thus, in all three of the best models with a \redChisq\ $\lesssim2.0$, we infer the presence of aerosols in the dayside of WASP-69 b.

\subsubsection{Potential Mineral Cloud Compositions}\label{sec:CloudComp}
We now discuss the possible compositions of aerosols that could exist in WASP-69 b's dayside, first considering cloud formation from commonly assumed condensed minerals \citep[e.g.][]{Zhang20_review,gao2021aerosolsExoplanetAtmospheres}.
As shown in Figure \ref{fig:retrievedProfiles}, the retrieved temperature-pressure profiles from the Cloud Layer model crosses the the saturation vapor pressure curves for Na$_2$S and MgSiO$_3$ and the Scattering model crosses the curve for Na$_2$S within 1~$\sigma$.
While the Cloud Layer model uses MgSiO$_3$ (enstatite) optical properties \citep{scott1996forsterite} to fit the observed dayside spectrum of WASP-69 b, the inferred cloud that extends from 10$^{-4.5}$ to 10$^{-6}$ bar is far higher than the $\sim$10 bar pressure where MgSiO$_3$ cloud is expected to form.
The situation is the same for other silicates like quartz (SiO$_2$), which condenses at a similar temperature \citep{grant2023quartzWASP17b} and forsterite (Mg$_2$SiO$_4$), which condenses at even higher temperatures  \citep[e.g.,][]{Morley+12_SVP,wakeford2017condensateClouds}.
This situation is reminiscent of WASP-107b, for which a recent JWST-MIRI transmission spectrum indicates the presence of silicate clouds at high altitudes based on a 9-11~\micron\ feature despite its cool equilibrium temperature of 740 K.
At these low temperatures, silicate clouds would be expected to rain down to the deep atmosphere \citep{dyrek2024silicatesWASP107}.
An extreme lofting mechanism, such as vigorous vertical mixing \citep{gao2018microphysicsGJ1214clouds,Ormel&Min19} and/or nonspherical cloud particles with high porosity \citep{Ohno+20fluffyaggregate,Samra+20_mineral_snowflake}, would be needed to elevate these particles from the $\sim10~{\rm bar}$ to the 10$^{-4.5}$--10$^{-6}~{\rm bar}$ pressure levels inferred by our model retrievals for aerosols in WASP-69 b.

Besides silicate clouds, salt clouds such as KCl have been predicted to form efficiently thanks to their low surface energy which drastically enhance the nucleation rate \citep{gao2018microphysicsGJ1214clouds,Lee+18_Nucleation,gao2020aerosolsSilicatesAndHazes}.
However, as shown in Figure \ref{fig:retrievedProfiles}, the atmospheric temperature profile inferred from the dayside emission spectrum is too hot to form KCl clouds.
Instead, KCl would have to be formed in the cooler terminator and nightside of the planet and horizontally transported to the dayside.

As discussed in Section \ref{sec:cloudLayerRetrieval}, the Cloud Layer model crosses the saturation vapor pressure curve of Na$_2$S at a similar pressure level as the inferred cloud. 
Nucleation theory suggests that Na$_2$S, ZnS, and MnS nucleate very inefficiently due to their high surface energy, leading to the predction that those sulfide clouds are absent in exoplanet atmospheres \citep{gao2018microphysicsGJ1214clouds,gao2020aerosolsSilicatesAndHazes}. 
However, one should keep in mind that the nucleation theory involves many uncertainties, such as the contact angles of the nucleating embryo on the condensation nuclei, and Na$_2$S clouds may form if there are condensation nuclei suitable for Na$_2$S nucleation \citep[for discussion, see][]{arfaux2023hazeCloudMicrophysicsWASP39b}.

To investigate the possibility of Na$_2$S clouds, we attempted to fit the spectrum of WASP-69 b with an alternative model with the same configuration as the Cloud Layer model described in Section \ref{sec:cloudLayerModel} but with Na$_2$S's optical properties \citep{montaner1979opticalConstantsNa2S}.
This Na$_2$S Cloud Layer model required a temperature-pressure profile with about twice the temperature from a radiative-convective equilibrium calculation and thus was not a physically plausible solution.
One other challenge for Na$_2$S is that WASP-69 b's transmission spectrum shows Na atomic features \citep{casasayas-barris2017sodiumWASP69b,langeveld2022surveyNaHighRes,khalafinejad2021wasp69b_lowResHighRes}, which should be suppressed for significant levels of Na$_2$S cloud formation.
Thus overall, there are a few candidate cloud particle compositions that could form directly on WASP-69 b's dayside at high altitudes but do not nucleate efficiently (Na$_2$S and MnS) and there are more readily-formed cloud particle candidates (KCl and silicates) that could form in colder regions or deep in the dayside, respectively, and  transported to the observed dayside pressures.
Photochemical haze particles can also form in conditions like atmosphere of WASP-69 b but are expected to create a temperature inversion not observed in WASP-69 b's dayside.
We discuss photochemical hazes in Section \ref{sec:photoChem}.

We next consider atmospheric features of candidate cloud particles that may explain some of the features in WASP-69 b's dayside spectrum that are not fully explained by the Two Region, Scattering and Cloud Layer models.
Inspection of Figure \ref{fig:wasp69TbSpec} reveals that the MIRI LRS spectrum shows a jump in brightness temperature at a wavelength of 9.4~\micron.
This is not the same wavelength as the known instrument artifact, the shadow-region effect, from 10.6 to 11.8~\micron\ \citep{bell2023_ers}, nor is there any evidence for a shadow-region effect in the lightcurves.
Two possibilities for this model-data disagreement are either an absorption feature from WASP-69 b near 8.6~\micron\ against a rising continuum or a change in the optical properties of aerosols at 9.4~\micron.
Evidence for SiO$_2$ (Quartz) has been found in the hot exoplanet WASP-17 b from its 8.6~\micron\ absorption feature \citep{grant2023quartzWASP17b}.
WASP-69 b also has an apparent 8.6~\micron\ absorption feature compared to the Scattering and Cloud Layer models as visible in Figure \ref{fig:wasp69TbSpec}.
Unfortunately, SiO$_2$ is condensed at temperatures above our inferred temperature-pressure profile for WASP-69 b, except at deep pressures more than 10 bar, so it would require extreme vertical mixing to reach the high altitudes near 10$^{-4.5}$, as with the enstatite and forsterite discussed above.
Another possibility is that the scattering properties of a cloud (like in the Two-Region model) change abruptly with wavelength near 9.4~\micron.
The single scattering albedo of MgSiO$_3$ also drops off steeply from near 1.0 at wavelengths of 8~\micron.
If another type of particle dropped in its single scattering albedo near 9.4~\micron, we expect that it would fit the spectrum of WASP-69 b well.
Thus, we encourage future studies of particles that can exist in solid form on the dayside temperatures and pressures of WASP-69 b and also have an 8.6~\micron\ absorption feature or changes in single scattering albedo near 9.4~\micron.

\subsubsection{A Lack of Photochemical Haze?}\label{sec:photoChem}
It has been suggested that hydrocarbon photochemical hazes tend to produce temperature inversions in the atmospheres of exoplanets \citep{morley2015superEclouds,Lavvas&Arfaux21photochemHazesStructure,Arfaux&Lavvas22_HJs,Steinrueck+23tempStructurePhotochemHaze}.
Our observed emission spectrum disfavors the presence of such a temperature inversion, at least in the hot region that controls the spectral shape, because the CO$_2$ feature unambiguously appears in absorption rather than in emission. 
We also independently computed an emission spectrum by taking into account the haze's radiative feedback using the TP profiles computed by the \texttt{EGP} code coupled with a two-moment microphysical haze model \citep{Ohno&Kawashima20superRaleighSlopes} and found that the addition of hazes tends to worsen the model fit for haze optical constants tested.
Namely, we used Titan-like haze compositions \citep{Khare+84opticalConstantsTholins} and soot \citep{Lavvas&Koskinen17} .
The lack of hydrocarbon hazes could be consistent with the apparent lack of CH$_4$ in the dayside spectrum and several modeling studies that predicted the decline in the haze abundance at $T_{\rm eq}\ga 950~{\rm K}$ due to the conversion of CH$_4$ into CO \citep{morley2015superEclouds,Kawashima&Ikoma19,gao2020aerosolsSilicatesAndHazes}. 
On the other hand, WASP-69 b is known to show a steep spectral slope in optical wavelengths \citep{murgas2020rayleighScatteringWASP69b,estrela2021aerosolsMicrobarWASP69b}, and our results may challenge previous studies that attribute the optical slope to photochemical hazes \citep[e.g.,][]{Lavvas&Koskinen17,Ohno&Kawashima20superRaleighSlopes}.
However, we note that the impacts of photochemical hazes on the TP profile should be sensitive to the actual optical properties of exoplanetary hazes, which have been highly uncertain to date.
Further understanding of the haze optical properties, such as by experimental studies \citep{Corrales+23photochemHazesCtoO,He+23organichazesWaterRich}, would be needed to quantitatively investigate to what extent photochemical hazes might exist in the atmosphere of WASP-69b.

\section{Conclusions}\label{sec:conclusions}

We analyzed a 2~\micron\ to 12~\micron\ JWST emission spectrum of WASP-69 b and find the following:
\begin{itemize}[noitemsep]
\item We observe molecular features in absorption, so the temperature profile decreases with altitude
\item We observe features of of CO$_2$, CO and H$_2$O in the atmosphere, but no features of CH$_4$. 
\item The shortest wavelengths (less than 3~\micron) observed in NIRCam have unexpectedly high brightness temperatures above 1000 K as compared to the longest wavelengths (more than 5~\micron)
\item There is some marginal evidence of an inhomogeneous dayside for WASP-69 b from the MIRI broadband eclipse maps.
\end{itemize}

We model the emission spectrum with a variety of assumptions about the atmosphere and find the following:
\begin{itemize}[noitemsep]
\item Cloudless 1D homogeneous chemical equilibrium models cannot fit our observed spectrum well and imply that additional complexity is needed to explain WASP-69 b's dayside spectrum.
\item Aerosols are present in WASP-69 b's dayside atmosphere because they are needed in our models to fit the shape of the planet's 2~\micron\ to 12~\micron\ spectrum, especially the high brightness temperature at short wavelengths as compared to long wavelengths and the lack of a strong peak in brightness near 4.0~\micron.
\item We find three models that can explain WASP-69 b's spectrum 1) a Scattering Model with a free parameter for the geometric albedo, possibly due to aerosols 2) A Cloud Layer Model that has a high altitude silicate cloud layer and 3) A Two-Region model that has an inhomogeneous dayside that is a combination warm and cool components that both contain gray aerosols.
The Scattering model requires unexpectedly high albedos of \planetGAlbedo, which is less plausible.
\item The abundances of CO$_2$, CO and H$_2$O point to super-solar concentrations of heavy elements in the atmosphere.
Considering the close agrement between the Cloud Layer and Two-Region Models, the metallicity of the atmosphere is enriched by \planetMetallicity \ compared to solar composition and the C/O ratio is \planetCtoO, but we cannot completely rule out the Scattering model, which has an atmospheric metallicity of \planetMetallicityScattering\ and a C/O ratio of \planetCtoOScattering.
All of these compositional intervals are at 68\% confidence.
\item The lack of CH$_4$ in the dayside spectrum is expected from the retrieved temperature-pressure profiles, even in chemical equilibrium without requiring chemical quenching because the temperatures are high.
\item The inferred heat redistribution efficiency factor for the planet is $f \gtrsim 0.7$ where 0.5 is full redistribution of heat spherically around the planet and 1.0 is redistribution to the dayside hemisphere only.
Equivalently $\varepsilon < 0.76$ in the parameterization of \citet{cowan2011statisticsOfAlbedoAndRecirculation}.
This is true for the cooler temperature Scattering model given its high implied Bond Albedo and the higher temperature Two-Region and Cloud Layer models, which rule out full redistribution of heat at all Bond Albedos.
\item The Two Region retrieval and Scattering Model retrieval suggest there is inhomogeneity in the dayside and that the emission is dominated by a warm component. This, combined with the inefficient heat redistribution factor and tentative eclipse mapping signal all point to a planet that has a hot spot at the substellar point.
\item The lack of a temperature inversion rules out absorptive Titan-like hazes that could warm upper altitudes in the planet's atmosphere
\end{itemize}

Together, we find multiple different models and retrievals give evidence that aerosols are responsible for shaping the spectrum of WASP-69 b.
Either these aerosols are highly reflective, as implied in the Scattering Model, or they shape the MIRI wavelenghts from 5~\micron\ to 12~\micron\ as compared to NIRCam short wavelengths.
The Scattering Model's reflective scenario requires an extremely high Geometric Albedo of \planetGAlbedo\ or higher (assumed to be constant with wavelength), which is unlikely.
The high geometric albedo (equivalent to an 80 ppm eclipse) can be tested with short wavelength eclipse observations such as CHEOPS, HST WFC3 UVIS or JWST NIRISS SOSS.
The TESS observations favor low geometric Albedos but the eclipse depth uncertainty $\sim$ 50 ppm does not significantly constrain the inferred Geometric Albedo, whereas the CHEOPS archive includes 19 eclipses with a larger aperture telescope.
Aerosols are also needed in our favored models: the Cloud Layer model, which includes high altitude silicates and the Two Region Model, which includes a gray scattering opacity that mutes some features in the hot region and drastically decreases the emission from the cold region.
While silicates like the MgSiO$_3$ (enstatite) composition assumed in the Cloud Layer model have the right kind of optical properties to explain most of the spectrum of WASP-69 b, they condense at far deeper pressures (10 bar) compared to their inferred high altitudes $10^{-4.5}$ to $10^{-6}$ bar.
Ongoing searches for particles that form efficiently at 800-1000 K temperatures in a metal-enriched atmosphere and also change in single scattering albedo near 9.4~\micron\ are encouraged.

Several lines of evidence point to a hot spot near WASP-69 b's substellar point.
First, the Two-Region model and Scattering models show evidence of an inhomogeneous dayside that is dominated by warm emission.
Second, eclipse map from the MIRI LRS broadband light tentatively favors a central concentration in brightness over a uniform dayside.
Third, the inferred heat redistribution efficiency of all of our retrievals rules out a full global redistribution of heat.
Inefficient heat redistribution is also seen in General Circulation Models of exoplanets that include the feedback effects of clouds \citep[e.g.][]{roman2021clouds3DthermalStructures,bell2024nightsideCloudsDisEqChemWASP43b}.

WASP-69 b is at an interesting temperature (T$_{\rm{eq}}$ = 963 K at zero albedo and full redistribution) where the chemical transition from a CO-dominated to CH$_4$ dominated atmosphere is expected to occur.
Probes of the terminator and nightside of WASP-69 b, with its inferred large day-to-night temperature contrast will constrain the physics of atmospheres across these temperature boundaries on the same planet.
We note that our retrieved C/O ratio from the Cloud Layer and Two Region models of \planetCtoO\ as well as our inferred temperature gradients on the planet imply that there is plenty of Carbon and cool enough temperatures for a large CH$_4$ signature in WASP-69 b's transmission spectrum.
CH$_4$ has been already detected in high resolution cross correlation \citep{guilluy2022fiveMoleculesWASP69b} and will be further constrained in JWST GO programs 3712 and 5924.
800-1000 K is also a temperature range where specific aerosols may form in planet atmospheres.
We find that evidence for aerosols in WASP-69 b and comparisons with planets both warmer and cooler will help illuminate what kind of chemical and aerosol transitions occur generally in this temperature regime.

\begin{acknowledgements}
Funding for E Schlawin is provided by NASA Goddard Spaceflight Center.
The authors are grateful for support from NASA through the JWST NIRCam project though contract number NAS5-02105. T.J.B.~and T.P.G.~acknowledge funding support from the NASA Next Generation Space Telescope Flight Investigations program (now JWST) via WBS 411672.07.05.05.03.02.
This research has made use of the SIMBAD database, operated at CDS, Strasbourg, France;
and NASA's Astrophysics Data System Bibliographic Services.
We respectfully acknowledge the University of Arizona is on the land and territories of Indigenous peoples. Today, Arizona is home to 22 federally recognized tribes, with Tucson being home to the O'odham and the Yaqui. Committed to diversity and inclusion, the University strives to build sustainable relationships with sovereign Native Nations and Indigenous communities through education offerings, partnerships, and community service.  
This work benefited from the 2023 Exoplanet Summer Program in the Other Worlds Laboratory (OWL) at the University of California, Santa Cruz, a program funded by the Heising-Simons Foundation and NASA.
Thanks to Jeroen Bouwman for some tips on analyzing MIRI LRS data.
This publication makes use of data products from the Wide-field Infrared Survey Explorer, which is a joint project of the University of California, Los Angeles, and the Jet Propulsion Laboratory/California Institute of Technology, funded by the National Aeronautics and Space Administration.
Numerical computations were in part carried out on PC cluster at Center for Computational Astrophysics, National Astronomical Observatory of Japan.
\end{acknowledgements}

%

\vspace{5mm}
\facilities{JWST(NIRCam), JWST(MIRI)}

The JWST data presented in this article were obtained from the Mikulski Archive for Space Telescopes (MAST) at the Space Telescope Science Institute. The specific observations analyzed can be accessed via \dataset[DOI: 10.17909/v2v9-k243]{https://doi.org/10.17909/v2v9-k243}.


\software{astropy \citep{astropy2013}, 
          \texttt{photutils v0.3} \citep{bradley2016photutilsv0p3},
          \texttt{matplotlib} \citep{Hunter2007matplotlib},
          \texttt{numpy} \citep{vanderWalt2011numpy},
          \texttt{scipy} \citep{virtanen2020scipy},
          \texttt{starry} \citep{luger2019starry},
          \texttt{batman} \citep{kreidberg2015batman},
          \texttt{ThERESA} \citep{challener2021ThERESA},
          \texttt{MC$^3$} \citep{cubillos2017correlatedNoiseExoplanetLC},
          \texttt{pymc3} \citep{salvatier2016pymc3},
          \texttt{celerite2} \citep{foreman-mackey2018celerite},
          \texttt{emcee} \citep{foreman-mackey2013emcee},
          \texttt{webbpsf} \citep{perrin2014webbpsf},
           }



\appendix

\section{Illustrative Forward Models}\label{sec:illustrativeForwardModels}
\begin{figure*}
\gridline{
\fig{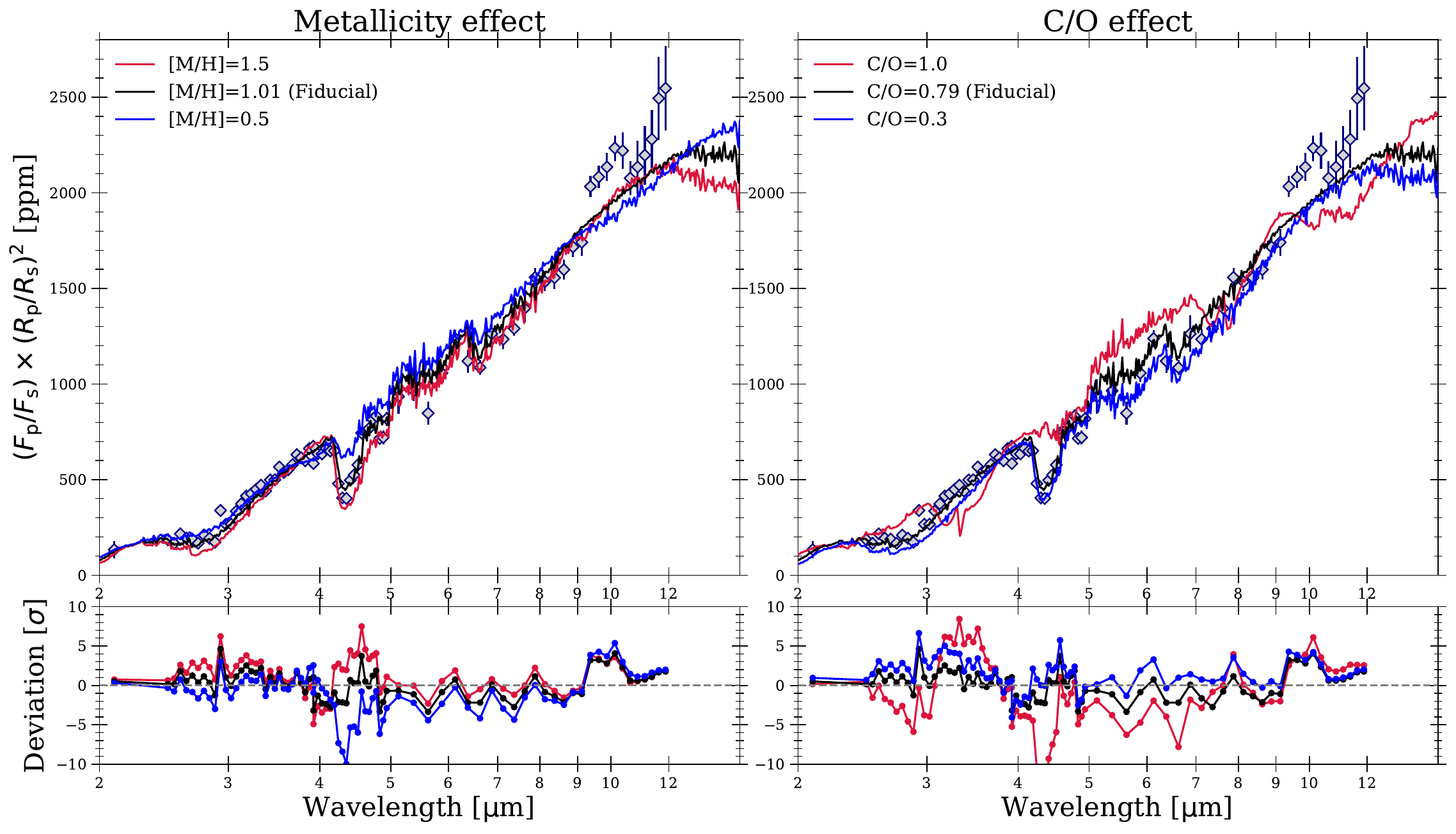}{0.99\textwidth}{}}
\caption{The abundance of WASP-69 b is constrained largely by the H$_2$O feature at 2.8~\micron, the lack of a CH$_4$ feature at 3.3~\micron, the CO$_2$ feature at 4.3~\micron\ and the CO feature at 4.5 to 5.0~\micron.
{\it Upper Left:} Three forward models are shown for 3 different atmospheric enrichments of [M/H]=1.5, [M/H]=1.01 (the best-fit value) and [M/H]. In these three examples, the relative abundances of heavy elements, including the C/O ratio is held constant.
{\it Lower Left:} The residuals of the models at three different atmospheric enrichments shows the most significant offsets are due to H$_2$O near 2.8~\micron\ and CO$_2$ at 4.3~\micron.
{\it Upper Right:} Three forward models are shown for 3 different C/O ratios of 1.0, 0.79 (best-fit value) and 0.3. These three models preserve the overall abundance of heavy elements.
At high C/O ratios, CH$_4$ is expected to be a significant absorber and causes a pronounced 3.3~\micron\ feature, not observed in WASP-69 b.
{\it Lower Right:} High C/O ratios are ruled out because there is no strong 3.3~\micron\ CH$_4$ feature and low C/O ratios are disfavored because they have large abundances of H$_2$O and make a deeper 2.8~\micron\ feature as well as shrink the size of the 4.3~\micron\ CO$_2$ absorption trough.
\label{fig:abundanceForwardModels}}
\end{figure*}

Given the many parameters and model assumptions in this work, described in Section \ref{sec:retrievals}, it is illustrative to show forward model spectra and residuals for representative models to further understand which parts of the spectrum drive our atmospheric inferences.
We first begin with the fiducial Two-Region model described in Bullet point \ref{it:CHIMERA2TP}, Section \ref{sec:twocompModels} and Section \ref{sec:twoRegRetrieval}.
The key atmospheric abundance parameters from which we report results of the paper are the atmospheric metal enrichment and the ratio of Carbon-to-Oxygen.
These abundance values allow comparison to Solar System planets as well as planet formation and evolution models \citep[e.g.][]{mordasini2016planetFormationSpec}.
As see in Figure \ref{fig:abundanceForwardModels}, this model's metallicity and C/O are largely constrained by the strength of the H$_2$O, CO$_2$ and CO absorption, as well as the lack of a strong CH$_4$ feature.

We also explored whether Sulfur species might be constrained in WASP-69 b and could improve the fit of the One-Region model described in Section \ref{sec:one-region} without invoking both clouds and an inhomogeneous dayside.
While our retrievals did not include the Sulfur-to-Hydrogen ratio as a free parameter, we consider the effects of varying two significant Sulfur-bearing species: H$_2$S and SO$_2$.
We find that increased or reduced H$_2$S abundance cannot significantly improve the 1-region model fit.
We find that non-equilibrium SO$_2$, such as produced by photochemistry \citep{tsai2022photochemistrySO2}, can improve the fit around 7.7~\micron.
However, it does not significnaly improve the generally high NIRCam flux at 3.5~\micron\ and generally low MIRI flux from 5 to 9~\micron.

\begin{figure*}
\gridline{
\fig{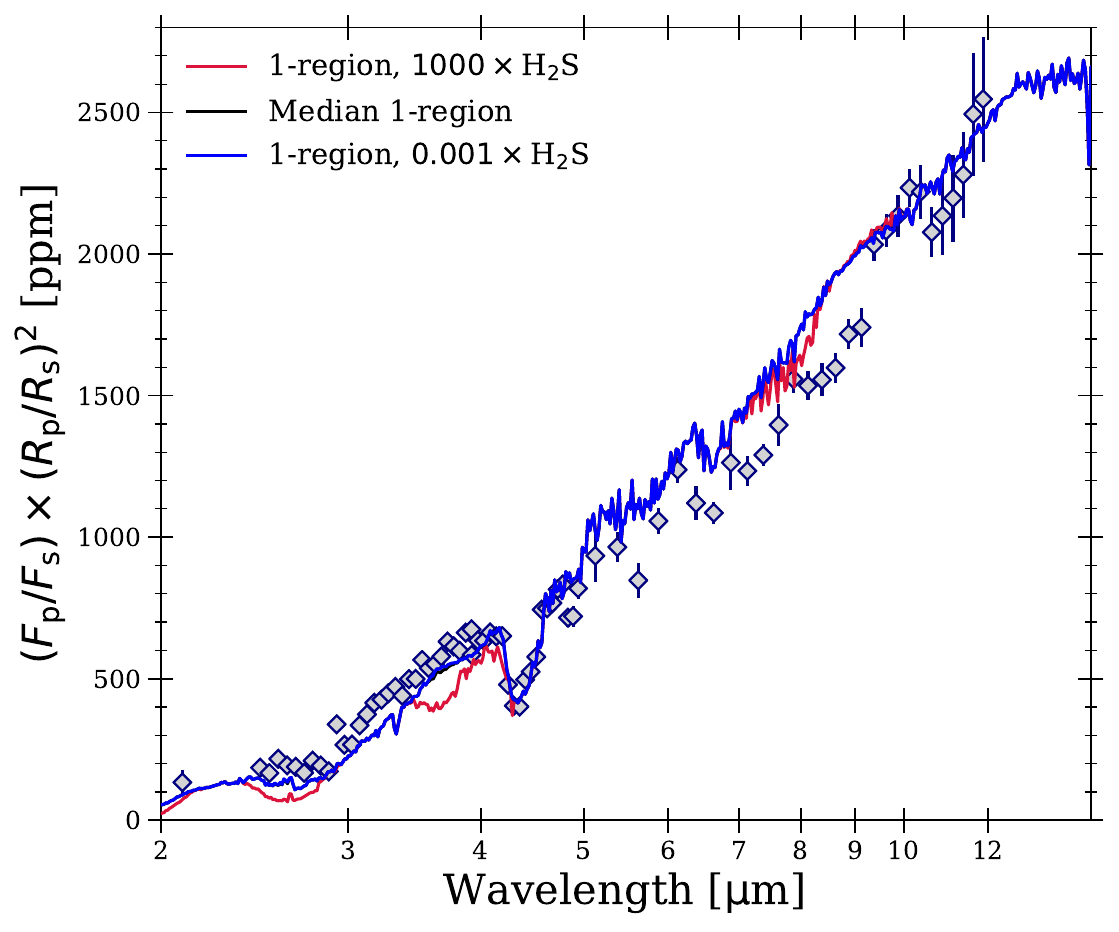}{0.48\textwidth}{Models of Varying H$_2$S Abundance}
\fig{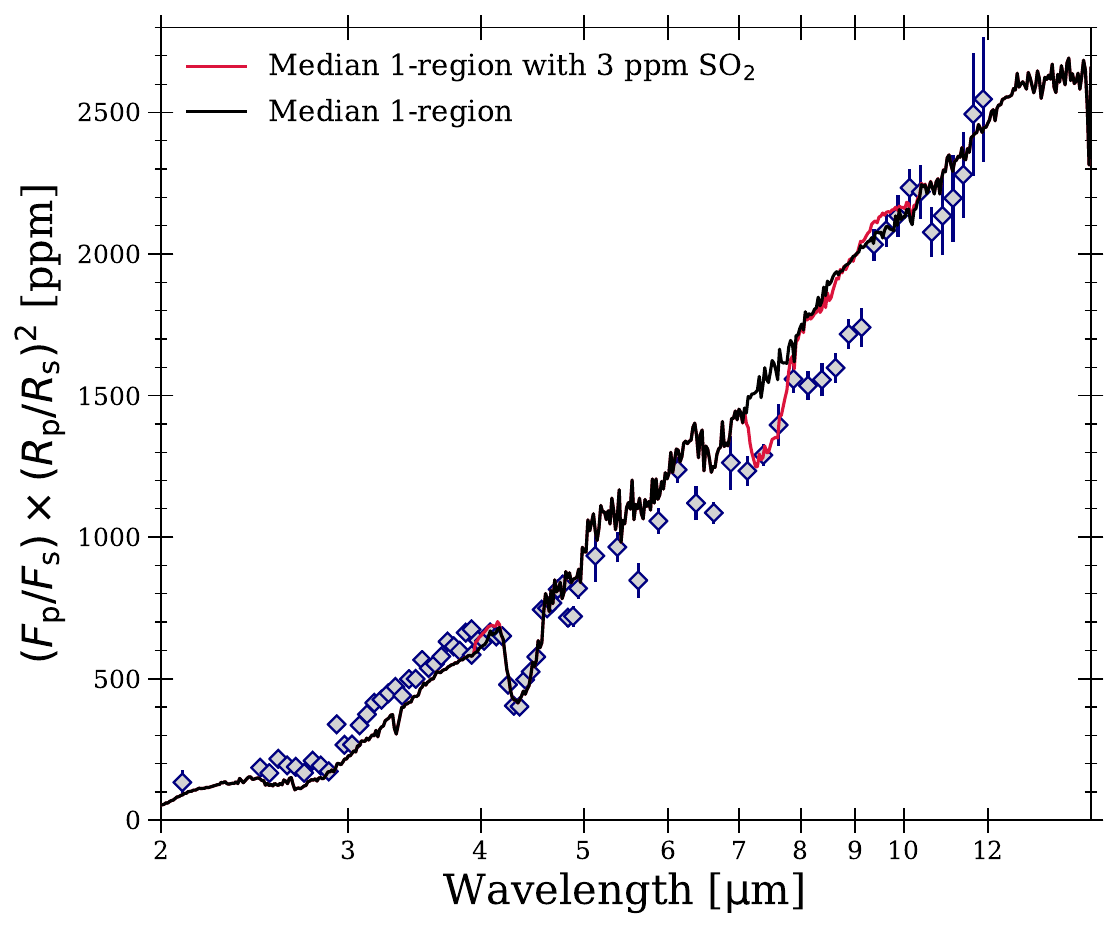}{0.48\textwidth}{Models of Varying SO$_2$ Abundance}}
\caption{
{ Left:} We experimented whether enhanced or decreased H$_2$S could help fit the data better without requiring a temperature gradients on the Dayside.
We show the Single-Region model with equilibrium H$_2$S for reference (black) as well as highly enhanced and highly diluted abundances of H$_2$S, neither of which can significantly improve the model as compared to the data.
{\it Right:} Using the same single region forward model, we also vary the SO$_2$ abundance to explore another Sulfur-bearing molecule. While extra SO$_2$ can help fit the data from 7.3 to 7.7~\micron, it cannot raise the 2.8~{\micron} flux and lower the 6 to 9~\micron\ flux to match the data as well as the Two-Region model.
\label{fig:sulfurForwardModels}}
\end{figure*}

\section{Posterior Distributions}\label{sec:postDistributions}
\begin{figure*}
\gridline{
\fig{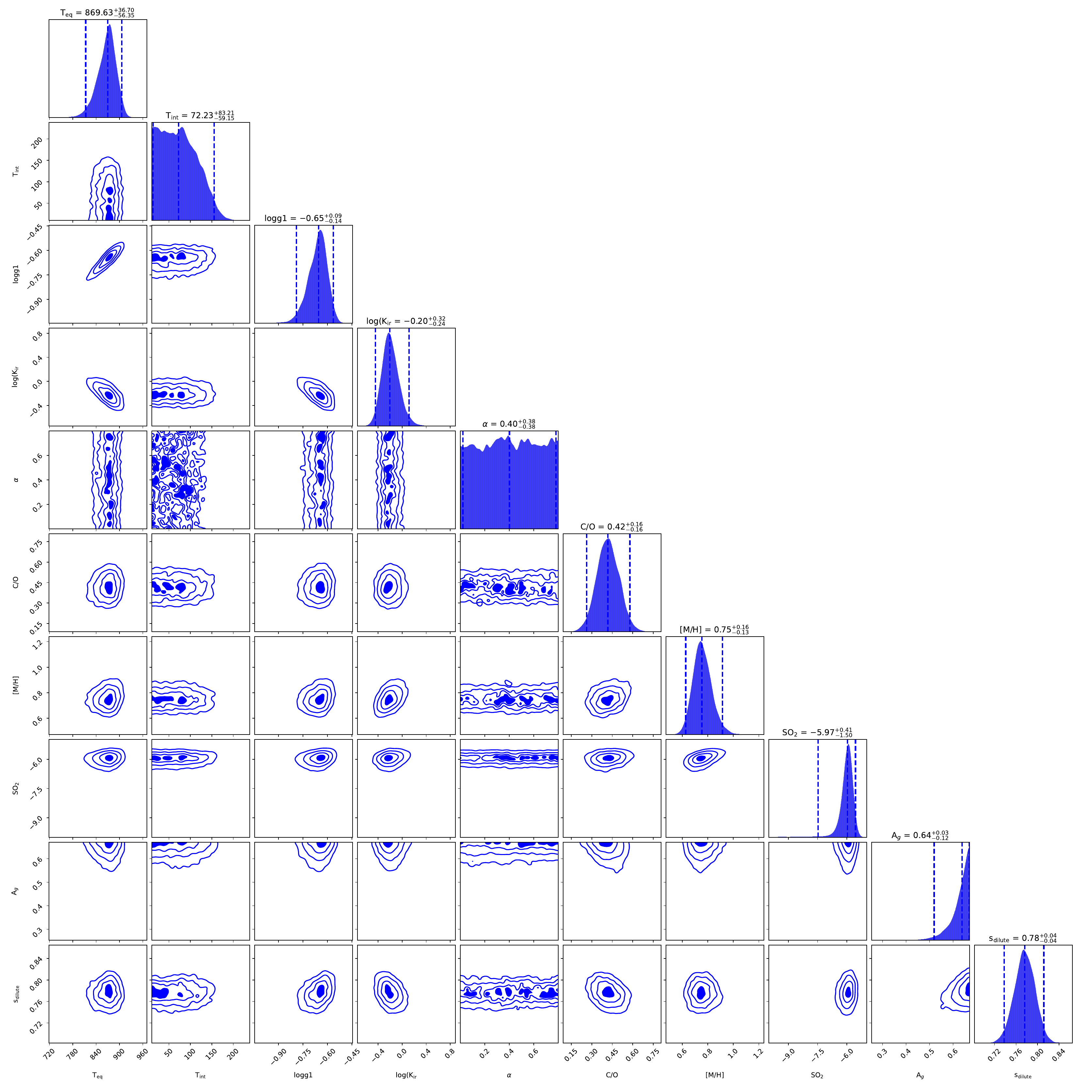}{0.99\textwidth}{}}
\caption{
Two-dimensional posterior distributions of the Scattering model  (Bullet Item \ref{it:PICASOrefl}, Sections \ref{sec:scatteringCloudy} and \ref{sec:scatRetrieval}) .
The parameters are Temperature-Pressure Profile parameters: $T_{eq}$ (not necessarily consistent with planet's equilibrium temperature), $T_{int}$, $log(g)$, $\kappa_{IR}$ and $\alpha$, the Carbon to Oxygen ratio C/O, the atmospheric metallicity [M/H], the free SO$_2$ abundance SO$_2$, the constant geometric albedo $A_g$ and the area fraction of the model emission compared to the full planet disk $s_{dilute}$.
\label{fig:cornerScatteringModel}}
\end{figure*}

\begin{figure*}
\gridline{
\fig{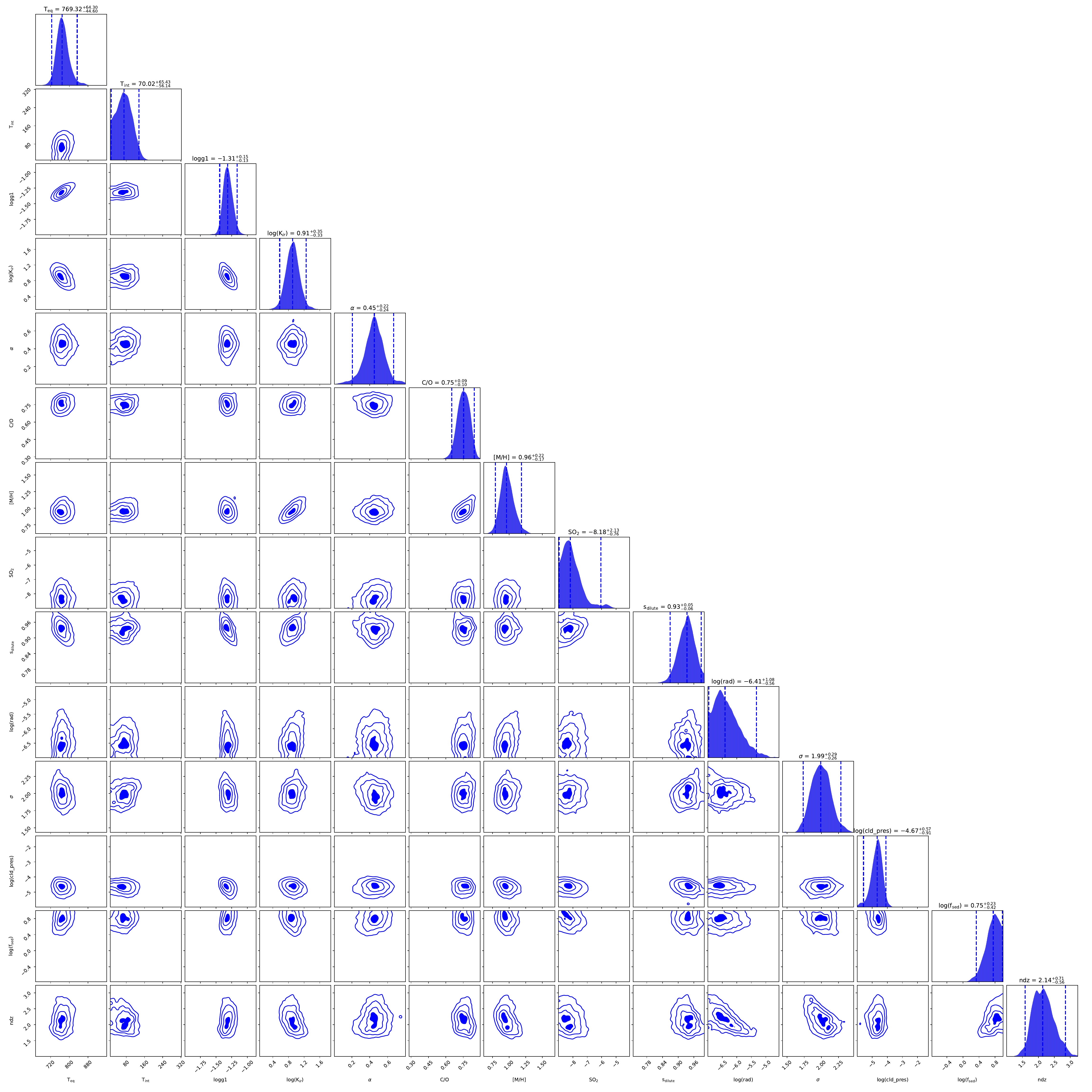}{0.99\textwidth}{}}
\caption{
Two-dimensional posterior distributions of the Cloud Layer model (Bullet item \ref{it:PICASOcloudLayer}, Sections \ref{sec:cloudLayerModel} and \ref{sec:cloudLayerRetrieval}).
The parameters are Temperature-Pressure Profile parameters: $T_{eq}$ (not necessarily consistent with planet's equilibrium temperature), $T_{int}$, $log(g)$, $\kappa_{IR}$ and $\alpha$, the Carbon-to-Oxygen ratio C/O, the atmospheric metallicity [M/H], the free SO$_2$ abundance SO$_2$, the area fraction of the model emission compared to the full planet disk $s_{dilute}$, and the cloud parameterization parameters: the logarithm of the geometric mean radius \texttt{log(rad)}, the geometric standard deviation of radius $\sigma$, the logarithm of the cloud base pressure in bars \texttt{log(cld\_pres)}, the log of the sedimentation efficiency log(${f_{\rm sed}}$) and the normalization factor $ndz$.
\label{fig:cornerCloudLayer}}
\end{figure*}

For completeness, we also include the full stair-step 2D representations of the Scattering model (Bullet Item \ref{it:PICASOrefl}, Sections \ref{sec:scatteringCloudy} and \ref{sec:scatRetrieval}) and Cloud Layer model (Bullet item \ref{it:PICASOcloudLayer}, Sections \ref{sec:cloudLayerModel} and \ref{sec:cloudLayerRetrieval}).
The Stair-Step plot of the Scattering  Model is Shown In Figure \ref{fig:cornerScatteringModel}.
The Stair-Step plot of the Cloud Layer Model is Shown In Figure \ref{fig:cornerCloudLayer}.

\section{Extra Analysis Details}\label{sec:analysisDetails}

\begin{figure*}
\gridline{
\fig{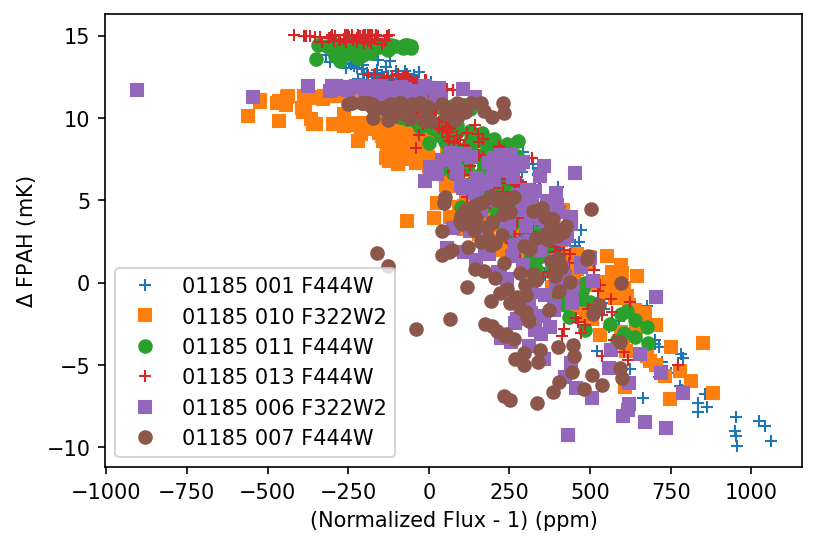}{0.49\textwidth}{}}
\caption{The Detector Housing Temperature is anti-correlated with flux across many different NIRCam observations, so we use it as means to de-trend the data.
Here, the X axis shows the change in the wavelength-integrated flux from the median in parts per million for all out-of-transit and out-of-eclipse integrations for 6 different example observations in the MANATEE program number 1185.
The Y axis shows the change in Focal Plane Array Housing (FPAH) as measured by the telemetry keyword IGDP\_NRC\_A\_T\_LWFPAH1.
Each observation is listed with the program number, observation number and which long wavelength filter was used respectively.
The observations describe in this paper are 006 and 007, as listed in Table \ref{tab:observations}.
\label{fig:FPAHcorr}}
\end{figure*}

\begin{figure*}
\gridline{\fig{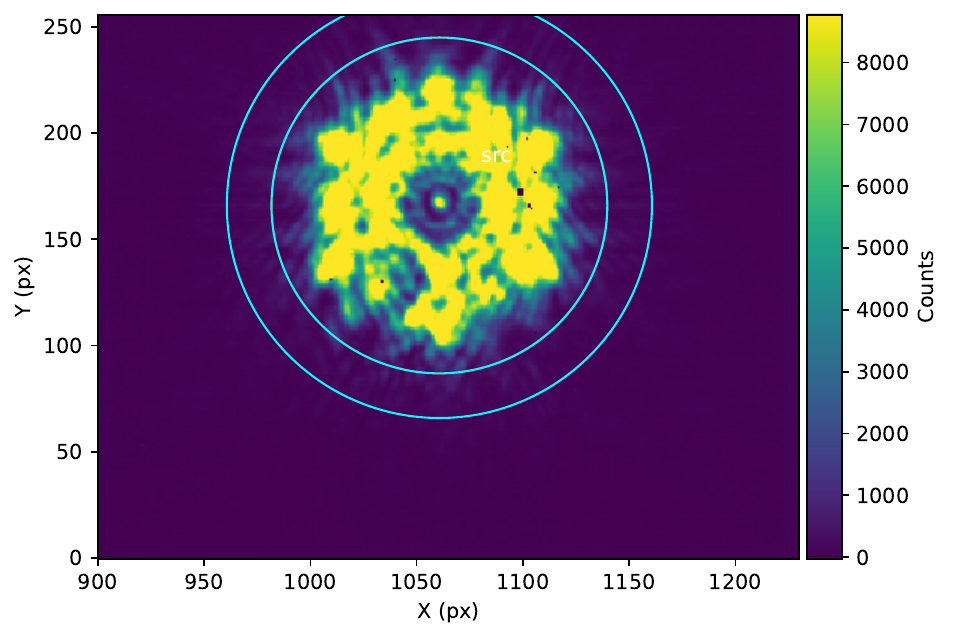}{0.49\textwidth}{Photometric Aperture}
          }
\caption{Extraction Aperture. The inner and outer circles are the background annulus radii and the source aperture is the same as the inner background annulus radius \label{fig:apertures}}
\end{figure*}

All NIRCam grism time series in the MANATEE program show a gradual decrease in flux over the duration of the exposure.
While the exact mechanism of this decline is not fully understood, we have identified an anti-correlation with detector housing temperature, as measured by the IGDP\_NRC\_A\_T\_LWFPAH1, available in the MAST engineering database.
Figure \ref{fig:FPAHcorr} shows the change in temperature versus the change in flux for out-of-transit and out-of-eclipse data in the MANATEE program 1185.
We note that many different observations of different brightnesses show this anti-correlation and therefore use the focal plane housing temperature as a de-trending vector.
Another specific data reduction visual is the aperture used for photometric extraction of the short wavelength data, which is shown in Figure \ref{fig:apertures}.

\bibliographystyle{apj}
\bibliography{this_biblio}



\end{document}